\documentclass{aa}  

\usepackage{graphicx}
\usepackage{txfonts}
\usepackage{multirow,bigdelim}
\usepackage{natbib}
\usepackage{url}
\usepackage{amsmath}
\usepackage[ table ]{xcolor}
\usepackage{ctable}
\usepackage{ulem}
\bibpunct{(}{)}{;}{a}{}{,} 
\def\kms{\hbox{km\,s$^{-1}$}}
\def\cmt{\hbox{cm$^{-2}$}}
\def\cmth{\hbox{cm$^{-3}$}}
\def\tex{\hbox{$T_\mathrm{ex}$}}
\def\tkin{\hbox{$T_\mathrm{kin}$}}
\def\tbr{\hbox{$T_\mathrm{b}$}}
\def\hdens{\hbox{$n(\mathrm{H_2})$}}
\def\tenfift{\hbox{$\times$10$^{15}$}}
\def\tenfort{\hbox{$\times$10$^{14}$}}
\def\tensixt{\hbox{$\times$10$^{16}$}}
\def\tensevt{\hbox{$\times$10$^{17}$}}

\definecolor{Gray}{gray}{0.85}

\begin{document}

   \title{Exploring the molecular chemistry and excitation in obscured luminous infrared galaxies}

   \subtitle{An ALMA mm-wave spectral scan of NGC~4418}

   \author{F. Costagliola\inst{1,3}
          \and
          K. Sakamoto\inst{2}
          \and
          S. Muller\inst{3}
          \and
          S. Mart\'in\inst{4}
          \and
          S. Aalto\inst{3}
          \and 
          N. Harada\inst{2}
          \and
          P. van der Werf\inst{5}
          \and
          S. Viti\inst{6}
          \and 
          S. Garcia-Burillo\inst{7}
          \and
          M. Spaans\inst{8}
          }

   \institute{Istituto de Astrof{\'i}sica de Andaluc{\'i}a (IAA-CSIC), Glorieta de la Astronom{\'i}a, s/n, E-18008, Granada, Spain, \email{costagli@iaa.es}
   \and Academia Sinica, Institute of Astronomy and Astrophysics, P.O. Box 23-141, Taipei 10617, Taiwan
   \and Chalmers University of Technology, Onsala Space Observatory, SE-439 92 Onsala, Sweden
   \and Institut de RadioAstronomie Millim\'etrique, 300 rue de la Piscine, Domaine Universitaire, 38406 Saint Martin d'Hères, France   
   \and Leiden Observatory, Leiden University, 2300 RA Leiden, The Netherlands
   \and Department of Physics and Astronomy, University College London, Gower Street, London WC1E 6BT, UK 
   \and Observatorio Astronómico Nacional (OAN)–Observatorio de Madrid, Alfonso XII 3, 28014 Madrid, Spain 
   \and Kapteyn Astronomical Institute, University of Gr\"oningen, PO Box 800, 9700 AV Gr\"oningen, The Netherlands
   }


 
  \abstract
   {Extragalactic observations allow the study of molecular chemistry and excitation under physical conditions which may differ greatly from what found in the Milky Way. The compact, obscured nuclei (CON) of luminous infrared galaxies (LIRG) combine large molecular columns with intense infrared (IR), ultra-violet (UV) and X- radiation and represent ideal laboratories to study the chemistry of the interstellar medium (ISM) under extreme conditions.}
   {To obtain for the first time a multi-band spectral scan of a LIRG, in order
to derive the molecular abundances and excitation, to be compared to other Galactic and extragalactic environments.}
   {We obtained an ALMA Cycle~0 spectral scan of the dusty LIRG NGC~4418, spanning a total of 70.7~GHz in bands 3, 6, and 7. We use a combined local thermal equilibrium (LTE) and non-LTE (NLTE) fit of the spectrum in order to identify the molecular species and derive column densities and excitation temperatures. We derive molecular abundances and compare them with other Galactic and extragalactic sources by means of a principal component analysis.}
   {We detect 317 emission lines from a total of 45 molecular species, including 15 isotopic substitutions and six vibrationally excited variants. Our LTE/NLTE fit find kinetic temperatures from 20 to 350~K, and densities between 10$^5$ and 10$^7$~\cmth. The spectrum is dominated by vibrationally excited HC$_3$N, HCN, and HNC, with vibrational temperatures from 300 to 450~K. We find that the chemistry of NCG~4418 is characterized by high abundances of HC$_3$N, SiO, H$_2$S, and c-HCCCH and a low CH$_3$OH abundance. A principal component analysis shows that NGC~4418 and Arp~220 share very similar molecular abundances and excitation, which clearly set them apart from other Galactic and extragalactic environments. 
   }
   {Our spectral scan confirms that the chemical complexity in the nucleus of NGC~4418 is one of the highest ever observed outside our Galaxy. The similar molecular abundances observed towards NCG~4418 and Arp~220 are consistent with a hot gas-phase chemistry, with the relative abundances of SiO and CH$_3$OH being regulated by shocks and X-ray driven dissociation.  The bright emission from vibrationally excited species confirms the presence of a compact IR source, with an effective diameter $<$5~pc and brightness temperatures $>$350~K. The molecular abundances and the vibrationally excited spectrum are consistent with a young AGN/starburst system. We suggest that NGC~4418 may be a template for a new kind of chemistry and excitation, typical of compact obscured nuclei (CON). Because of the narrow line widths and bright molecular emission, NGC~4418 is the ideal target for further studies of the chemistry in CONs.}


   \maketitle
%

\section{Introduction}

Extragalactic chemistry is a field that is quickly expanding leading to new, powerful diagnostic tools for the star-forming and active galactic nuclei (AGN) activity in galaxies \citep[e.g., ][]{meier14,viti2014,martin2015}. The extreme environments found in some extragalactic objects provide the opportunity of studying the properties of the interstellar medium (ISM) beyond the typical conditions found in the Milky way. Shocks, stellar- and AGN radiation, dust shielding, and cosmic rays strongly impact the chemistry and excitation of the molecular ISM. Establishing the chemical and physical conditions of the molecular gas becomes a particularly important identification tool when the activity itself is buried in dust. 

A clear case is offered by the compact obscured nuclei (CON) of  IR-luminous (LIRGs) and ultraluminous galaxies \citep[ULIRGS, e.g., ][]{sanders_96}. These galaxies radiate most of their energy as thermal dust emission in the infrared and constitute the dominant population among the most luminous extragalactic objects. Observations at mid-IR and millimeter wavelengths suggest that they may play a crucial role in galaxy evolution, representing the early obscured stages of starburst galaxies and AGNs \citep[][]{spoon07,nascent,sakamoto2013,costagliola2013}. Because of the high amounts of gas and dust (N(H$_2$)>10$^{24}$~\cmt) and the extreme obscuration (A$_\mathrm V$>100~mag) the nature of the activity cannot be reveled by standard optical, IR or even X-ray observations. This makes it essential to identify new diagnostic tools in the mm/submm window, which can probe deeper in the column of dust.

Until recently, research in this field has followed mainly two paths: molecular line-ratio surveys, and single-band spectral scans. Line-ratios surveys have mostly focused on the brightest transitions of dense-gas tracers such as HCN, HNC, and HCO$^+$, trying to compare the observed values to chemical models of photon-dominated regions (PDRs), X-ray dominated regions (XDRs), and hot-cores \citep[e.g., ][]{imanishi07,loenen07,krips08,baan08,costagliola11,viti2014}. However, the line ratios of such bright tracers show only small variations, of the order of a few, and their theoretical interpretation is still controversial \citep[e.g.,][]{kohno2001,aalto07,meijerink07,costagliola11}. Since these molecules are ubiquitous in the dense ISM and are excited under a variety of physical conditions, the observed trends may be due to opacity and excitation effects, rather than a different chemistry. 

Unbiased spectral scans are powerful tools to find new tracers, more sensitive to the physical conditions of the gas. However, wide-band molecular line surveys have been carried out only for a very limited number of extragalactic objects \citep[e.g.,][]{martin06,martin2011,muller_pks,aladro_1068}. While these observations have made a leap forward in the study of the chemical complexity of galaxies, most of them have been limited by the narrow frequency range covered, which does not allow a complete analysis of the molecular excitation.
In order to compare chemical models with the observations, a better handle on the abundances of the molecular species is needed. This can be achieved only by observing multiple transitions in different frequency bands, sampling a wide range of excitation states. Here we present the first multi-band spectral scan of the obscured luminous infrared galaxy NGC~4418 obtained with the Atacama Large Millimeter/submillimeter Array (ALMA) observatory\footnote{\url{http://www.almaobservatory.org/}}.

\subsection{NGC~4418 : The prototypical obscured LIRG}

The LIRG NGC~4418 ($L_{\rm IR}=10^{11}$~L$_\odot$) has the optical morphology of an early-type spiral and was first detected as a bright source by the IRAS satellite. Lying at $D$=34~Mpc, the galaxy is part of an interacting pair, with the companion being an irregular blue galaxy roughly two arc-minutes (16~kpc) to the southeast. 

The optical spectrum of NGC~4418 has been described by \cite{roche86} as {\it unremarkable}, with only faint emission from S[II] and H$_{\rm\alpha}$ on a strong continuum, and it does not hint at the presence of a bright IR source. This is explained by mid-IR observations \citep{spoon_01, roche86}, which show a deep silicate absorption at 10~${\rm\mu}$m, one of the deepest ever observed, corresponding to roughly 100 magnitudes of visual extinction. Whatever is powering the strong IR flux of NGC~4418, it must be hidden behind a thick layer of dust, which makes it undetectable at optical wavelengths.

The high IR luminosity requires a compact starburst or an AGN to heat up a large column of dust. However, because of the high obscuration, the direct investigation of the nuclear region is a challenging task, and the nature of the energy source is still unclear.  Observations with the {\it Chandra} X-ray satellite by \cite{maiolino03} show a flat hard X-ray spectrum, which would be an indication of a Compton-thick AGN, but the photon statistics are too limited to be conclusive. The absence of a clear X-ray signature may imply either that the galaxy is powered by star formation alone, or that the obscuring column is so high that most of the X-ray emission cannot escape its nuclear region.

Additional evidence of nuclear activity in NGC~4418 comes from near- and mid-IR observations.  High-resolution {\it HST } near-infrared and Keck mid-infrared images by \cite{evans03} show that the galaxy has a 100-200 pc optically thick core,  with a high IR surface brightness. The observed spectral energy distribution implies a dust temperature of 85 K, which, when compared with the total IR flux, implies the presence of an optically thick source of no more than 70~pc across. 

In two recent papers \citep{sakamoto2013,costagliola2013} we used the Submillimeter Array (SMA) and the Multi-Element Radio Linked Interferometer Network (MERLIN) to directly probe the nucleus of NGC~4418 at mm and radio wavelengths with sub-arcsecond resolution. These studies confirm the existence of a $\sim$20~pc (0.$''$1) hot dusty core, with a bolometric luminosity of about 10$^{11}$ L$_\odot$, which accounts for most of the galaxy luminosity. The high luminosity-to-mass ratio ($L/M\simeq$500 L$_\odot$ M$_\odot^{-1}$) and luminosity surface density (10$^{8.5\pm0.5}$ L$_\odot$ pc$^{-2}$) are consistent with a Compton-thick AGN to be the main luminosity source. Alternatively, an extreme (SFR$\simeq$100 M$_\odot$ yr$^{-1}$), young ($\leq$5 Myr), compact starburst could also reproduce the properties of the inner core. 

Combined observations with MERLIN and the European Very Long Baseline Interferometry Network (EVN) by \citet{varenius2014} show that 50\% of the total radio emission at 5~GHz originates from eight compact sources in the inner 0.$''$2 of the nucleus. The four brightest sources lie inside a 0.$''$1 circle centered on the position of the 860~$\mu$m continuum from \citet{sakamoto2013}. These sources have an average surface brightness which is close to the limit of what can be produced by well-mixed thermal/non-thermal emission from any surface density of star formation. \citet{varenius2014} suggest that the radio emission could be explained by super star clusters with intense star formation with some contribution from an AGN.

Herschel PACS observations by \citet{galfonso2012}, and our combined MERLIN/SMA study \citep{costagliola2013} reveal the presence of redshifted OH and HI absorption, which is interpreted as the signature of a molecular inflow. The presence of a molecular outflow is also suggested by a U-shaped red optical feature along the northwestern semi-minor axis of the galaxy \citep{sakamoto2013}. 

NGC~4418 was first shown to have a rich molecular chemistry by \citet{nascent} and the high abundance of HC$_3$N \citep[$>$10$^{-8}$,][]{costagliola2010} is not reproduced by models of X-ray-dominated chemistry expected in an AGN scenario \citep{meijerink07}. Together with a low HCO$^+$/HCN $J$=1--0 line ratio, bright HC$_3$N is instead reminiscent of line emission toward Galactic hot-cores, i.e., regions of dense, warm, shielded gas around young stars. This has led some authors to propose that NGC~4418 may be a very young starburst, where the star-forming regions are still embedded in large amounts of dust \citep{roussel03,nascent, costagliola11}. This scenario of a {\it nascent} starburst would be consistent with the galaxy being radio-deficient. However, recent results in chemical modeling by \citet{harada2013} suggest that substantial HC$_3$N abundances may be maintained near an AGN.

The combination of a bright, compact IR source and a high molecular column produce an incredibly rich vibrationally excited spectrum. Vibrationally excited HCN and HNC were first detected in this source by \citet{sakamoto10} and \citet{costagliola2013}, and the galaxy shows very bright emission from vibrationally excited HC$_3$N \citep{costagliola2010,costagliola2013}. 

The velocity dispersion of the molecular emission lines emerging from the core of NGC~4418 is small, of the order of $\sim$100~\kms. This makes the identification of spectral features much more reliable than in other (U)LIRGS \citep[e.g., Arp~220, ][]{martin2011}. The combination of bright molecular emission and narrow lines makes it the ideal target to study molecular chemistry and excitation in compact obscured nuclei.

\section{Observations}

The observations were undertaken during the ALMA Early Science Cycle 0 phase, between April 2012 and January 2013. The target NGC~4418 was observed in 11 tunings out of 38 proposed (see Table \ref{tab:journal}). Each tuning consists in short integrations of $\sim$5 minutes on source, plus observations of the bright radio quasars 3C\,273 or 3C\,279 and Titan for absolute flux calibration (except for 3 dates, for which we used 3C\,273). For each tuning, four different 1.875\,GHz-wide spectral windows were set, each counting 3840 individual channels separated by 0.488\,MHz.  

In order to account for possible edge instabilities and offsets, contiguous spectral windows were overlapped by 250 MHz. As a result, each tuning covers a total span of 7 GHz in sky frequency, including upper and lower sideband. The observed tunings cover 28.7~GHz in band 3, 28~GHz in band 6, and 14~GHz in band 7, corresponding to the 90$\%$, 43$\%$, and 14$\%$ of each band, and a total coverage of 70.7 GHz.

The Cycle~0 array configuration was composed of 19 to 25 antennas, with baseline separations from 20 to 480 meters. The resulting angular resolution is of $\sim$2$''$, $\sim$1$''$, and $\sim$0$''$.8 for bands 3, 6, and 7 respectively. 

The data calibration was done within the CASA\footnote{http://casa.nrao.edu/} package following a standard procedure. We expect an absolute flux accuracy of the order of $\sim$10\% in band 3 and $\sim$20\% in bands 6 and 7. 

\subsection{UV-fit of the spectrum}

Previous interferometric observations by \citet{sakamoto2013} and \citet{costagliola2013} show that 70\% of the molecular emission in NGC~4418 emerges from a region of less than 0$''$.4 in diameter, we therefore expected all the emission to be included in our synthesized beam. 
We verified this assumption by fitting the source size of the emission of the HNC 1-0 and HNC 3-2 lines in band 3 and 6. We performed both a fit of the visibilities and a 2D Gaussian fit of the integrated emission in the image domain. We found the emission to be unresolved, with upper limits to the source size of 0$''$.4 and 0$''$.2 for band 3 and band 6, respectively.

Given the point-like nature of the emission, we chose not to image and clean the whole dataset, but to extract the spectrum from the visibilities. We first obtained the position of the continuum peak by averaging line-free channels in each tuning. Then we fitted the interferometric spectral visibilities of each spectral window assuming a point-source model, with a fixed position but free amplitude. This last step was performed using the CASA {\tt uvmultifit} routine \citep{uvmultifit}.

\begin{figure*}
   \centering
   \includegraphics[width=1\textwidth,keepaspectratio]{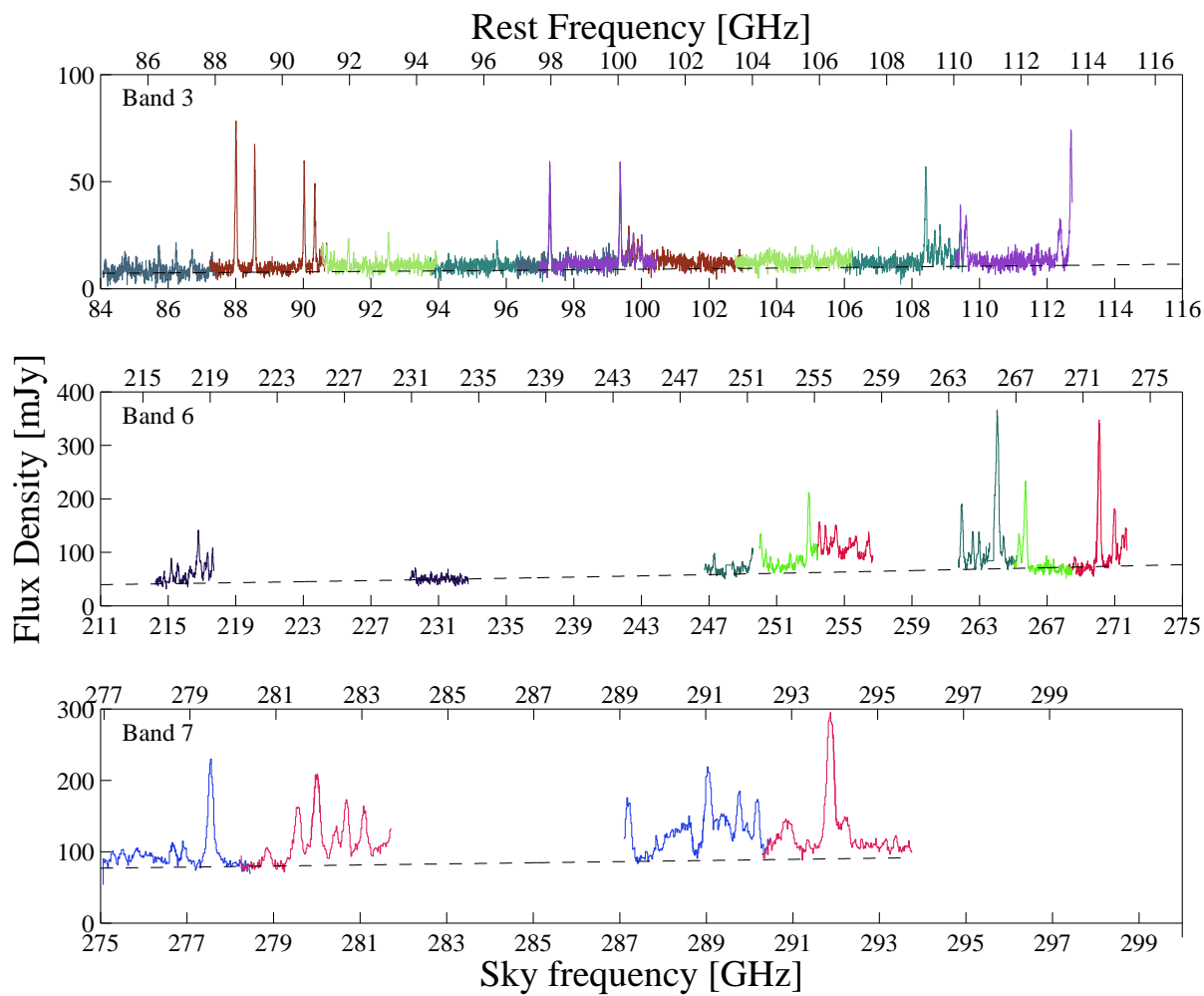}
   \caption{ \label{fig:allbands} Total observed spectrum, extracted by fitting the visibilities with a point source. Each scan is plotted in a different color. The {\it dashed line} represents a 3$\mathrm{rd}$ order fit to the line-free channels, and is used as continuum level in the analysis.}
    \end{figure*}

The extracted channels were then interpolated to a common resolution of 20~\kms across the whole frequency range, and all spectral windows were merged to create a single spectrum. Small offsets between contiguous spectral windows were eliminated by averaging the channels in the 250 MHz overlap region. These offsets were not larger than 10$\%$ in either band, confirming the good quality of the data calibration. The measured {\it rms} of the spectrum is 2~mJy in band~3, 6~mJy in band 6, and 4~mJy in band 7. The total extracted spectrum is shown in Fig. \ref{fig:allbands}.

In Fig. \ref{fig:sdcomp} we compare the extracted ALMA spectrum with IRAM~30~m single-dish observations with beam sizes of 27$''$ and 9$''$ at 3 and 1~mm, respectively. The flux densities measured for the two datasets are within the errors, which is consistent with all the molecular emission to be contained inside the ALMA beam. 

A comparison with the HNC~3-2 and HC$_3$N~30-29 flux densities detected with the SMA in the extended configuration \citep{costagliola2013} is reported in Fig. \ref{fig:smacomp}. The integrated flux density of the different observations is reported in Table \ref{tab:compflux}. We find that between 70\% and 90\% of the total flux is contained by the 0$''$.4 synthesized beam of the SMA.

\begin{table*}[ht]
\caption{\label{tab:journal} Journal of the observations.} 
\begin{center} 
\begin{tabular}{llllllccc}
\hline
Band & L. O. & Date of & Bandpass   & Flux  & Sky frequency & On source  & Number of & rms \\
 & [GHz] & observations & calibrator & calibrator & coverage [GHz] & time [min] & antennas & [mJy]\\
\hline
B3 & 92  & 25-Apr-2012 & 3C\,273 & Titan   & 84-87.5, 96.2-99.7 & 27 & 19 & 2 \\ 
B3 & 95  & 21-Nov-2012 & 3C\,273 & 3C\,273 & 87.2-90.7, 99.5-103 & 17 & 25 & 2 \\ 
B3 & 98  & 21-Nov-2012 & 3C\,273 & 3C\,273 & 90.5-94, 102.7-106.2  & 10 & 25 & 2 \\ 
B3 & 102 & 25-Apr-2012 & 3C\,273 & Titan   & 93.7-97.2, 106-109.5 & 27 & 19 & 2  \\ 
B3 & 105 & 21-Nov-2012 & 3C\,273 & 3C\,273 & 97-100.5, 109.2-112.7  & 18 & 25 & 2 \\ 
B6 & 224 & 31-Aug-2012 & 3C\,279 & Titan   & 214.3-217.8, 229.3-232.8 & 7 & 23  & 6 \\ 
B6 & 253 & 01-Jan-2013 & 3C\,279 & Titan   & 253.2-256.7, 268.2-271.7 & 7 & 24 & 6 \\ 
B6 & 256 & 01-Jan-2013 & 3C\,279 & Titan   & 246.7-250.2, 261.7-265.2  & 7 & 24 & 6 \\ 
B6 & 259 & 01-Jan-2013 & 3C\,279 & Titan   & 250-253.5, 265-268.5 & 7 & 24  & 6\\ 
B7 & 283 & 20-Nov-2012 & 3C\,273 & Titan   & 275-278.5, 287-290.5 & 5 & 25 &  4 \\ 
B7 & 286 & 20-Nov-2012 & 3C\,273 & Titan   & 278.2-281.7, 290.2-293.7 & 5 & 25  & 4 \\ 
\hline
\end{tabular}
\end{center} 
\end{table*}

\section{Molecular emission}
\label{sec:molemission}
The mm-wave spectrum of NGC~4418 shows rich molecular emission in all three ALMA bands. Despite the narrow line width of $\sim$120~\kms, line confusion is reached after short integrations of $\sim$5 minutes in bands 6 and 7. 

\subsection{Removing the continuum}
\label{sec:continuum}
Removing the continuum from the spectrum in band 6 and 7 is complicated by line crowding. In order to fit a continuum level, we assumed the regions in bands 6 and 7 with the lowest emission in the band to be line-free. We then averaged the channels to obtain band-averaged values for the continuum. The derived values are 9$\pm$1~mJy, 56$\pm$6~mJy, and 85$\pm$9~mJy at 98, 244, and 284~GHz, respectively.

\begin{figure}
\includegraphics[width=.5\textwidth,keepaspectratio]{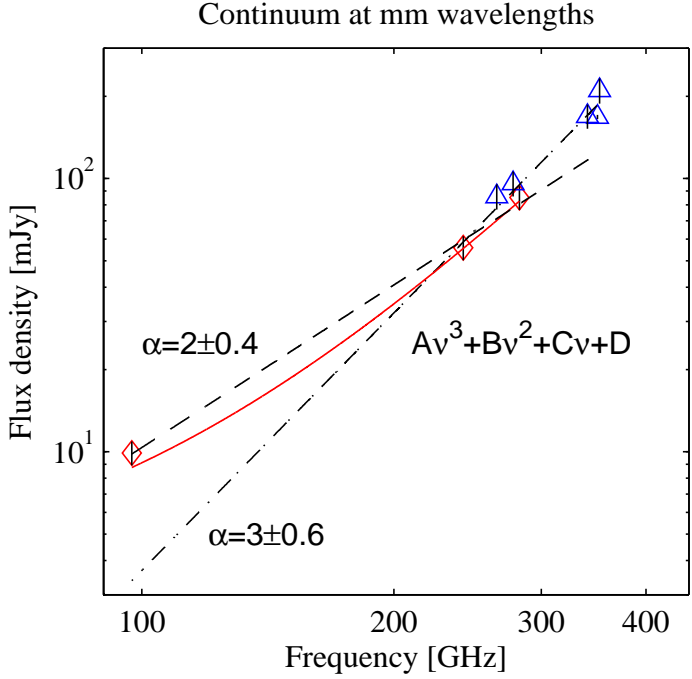}
\caption{\label{fig:cont} Continuum flux density of NGC~4418 at mm wavelengths. The {\it red diamonds} represent the average value of line-free channels in the ALMA bands 3, 6, and 7. Sub-mm continuum levels from \citet{sakamoto10} and \citet{sakamoto2013}   are shown as {\it blue triangles}. Error bars were calculated as the quadrature sum of the spectrum {\it rms} and a calibration uncertainty of 10\%. The {\it dot-dashed}, and {\it dashed lines} represent the power-law fit of data points above 200~GHz ($\alpha=3\pm0.6$) and below 270~GHz ($\alpha=2\pm0.4$), respectively. The {\it red solid line} shows the best fit to the ALMA baseline with a 3$^\mathrm{rd}$-order polynomial. The best fit parameters are A=$2.7\times10^{-9}$~Jy/GHz$^{3}$, B=$6.2\times10^{-8}$~Jy/GHz$^{2}$, C=$3.2\times10^{-5}$~Jy/GHz, D=$2.3\times10^{-3}$~Jy.}
\end{figure}

In Fig. \ref{fig:cont} we compare the derived continuum values with data from SMA observations by \citet{sakamoto10} and \citet{sakamoto2013}. Allowing for a calibration uncertainty of 10\%, the two datasets are consistent. 

We find that the emission is best fitted by two power laws ($S_\nu\propto\nu^\alpha$), one between 97 and 300~GHz with $\alpha=2\pm$0.4, and one at frequencies higher than 270 GHz with $\alpha=3\pm$0.6. If we assume a uniform slab geometry, the dust spectral index in the Rayleigh-Jeans regime can be written as $\alpha=$2+$\beta\tau(e^\tau-1)^{-1}$, where $\beta$ is the the index of power-law frequency dependence of the dust opacity and $\tau$ the opacity of the emission at the wavelength where $\alpha$ is measured. For a typical $\beta$=1.5--2 \citep[e.g., ][]{lisenfeld2000}, an $\alpha=$3 requires opacities of the order of unity. This result is in agreement with previous observations \citep[e.g., ][]{sakamoto2013} which required an optically thick IR source to fit the galaxy's spectral energy distribution. The value of $\alpha=2\pm$0.4 at lower frequencies is difficult to explain as an opacity effect because it would require $\tau>10$ and one would expect opacity to increase with frequency. A better explanation may be that free-free or cold dust emission is contributing significantly to the continuum in band~3. 

The best power-law fit to the ALMA continuum points overestimates the emission at 1~mm by more than 20\%, too much for it to be used for continuum removal in our fitting procedure.  We choose instead to fit the line-free channels in the three bands with a polynomial. We obtain the best fit with a 3$^\mathrm{rd}$ order polynomial, which is shown in Figures \ref{fig:allbands} and \ref{fig:cont}. We will use this estimate in the following analysis of the molecular emission. 

By removing the estimated continuum flux from the spectrum we can derive the contribution of molecular emission to the total observed flux in each band. {\it We find that the molecular emission contributes for 15\% of the total flux in band 3, and for 27\% in bands 6 and 7.} These values are very similar to the 28$\%$ found by \citet{martin2011} for observations at 1~mm in the ULIRG Arp~220. These results clearly show that contamination by molecular emission, if not properly removed, could be a serious issue for studies of continuum emission in the compact cores of (U)LIRGs.

\subsection{Line identification and fit}

The heavy blending of emission features makes the identification of individual molecular species a challenging task, especially in bands 6 and 7. In order to identify the emitting molecules we use a mix of local-thermal-equilibrium (LTE) and non-LTE (NLTE) fitting methods to fit the whole spectrum in the three ALMA bands. In the following paragraphs we describe the fitting routine and its results. 

\subsubsection{Population diagrams}
\label{sec:popdiag}

As a first step to the fit of the whole spectrum, we perform a population diagram analysis for the molecules which have at least two non-blended lines in different ALMA bands. The integrated flux of each line was obtained by fitting a Gaussian profile to the spectrum. In the LTE approximation we assume the lines to be optically thin and the relative intensities to be described by a single excitation temperature T$_\mathrm{ex}$. For each rotational transition, following \citet{popdiag}, we can write the column density of the upper energy state as\\
\begin{equation}
N_u=\frac{8\pi k \nu^2 W}{h c^3 A_{u\ell}},
\end{equation}
where $\nu$ is the transition's frequency, A$_\mathrm{u\ell}$ is the Einstein coefficient, $W$ is the integrated brightness temperature, $h$ is Planck's constant, $k$ is Boltzmann's constant, and $c$ is the speed of light in vacuum. The integrated brightness temperature $W$ is related to the observed integrated flux density $S_\mathrm{line}$ by  
\begin{equation}
W=S_\mathrm{line}\times\frac{\lambda^2}{2k\Omega_\mathrm{s}}, 
\end{equation}
with $\lambda$ the wavelength of the emission, and $\Omega_\mathrm{s}$ the solid angle subtended by the source. At LTE, we have 
\begin{equation}
\label{eq:lte}
N_\mathrm{u}=\frac{N}{Z}g_ue^{-E_u/T_\mathrm{ex}},
\end{equation}
where $N$ is the total molecule's column density, $Z$ is the partition function, and $g_\mathrm{u}$ and $E_\mathrm{u}$ are respectively the degeneracy and the energy of the upper level. By fitting Eq. \ref{eq:lte} to the derived N$_\mathrm{u}$ we can derive $T_\mathrm{ex}$ and $N$ for each molecule. 
Once we obtained $N_\mathrm{u}$ in the optically thin approximation, a correction for opacity effects was applied, following \citet{popdiag}:
\begin{eqnarray}
\label{eq:taucorr}
\tau_\mathrm{J,J-1}=\frac{8\pi^3\mu^2}{3h}\frac{N}{\Delta v}\frac{1}{Z}e^{-\frac{hB_0}{kT_\mathrm{ex}}J(J+1)}\left(e^{\frac{hB_0}{kT_\mathrm{ex}}J}-1\right) \nonumber \\
N^\mathrm{thick}_u=N^\mathrm{thin}_u \times \frac{\tau}{1-e^{-\tau}},
\end{eqnarray}
where $\mu$ is the dipole moment, $\Delta v$ the full width at half maximum (FWHM) of the emission line, and B$_0$ the rotational constant of the molecule.  

The transition parameters ($\nu$, $E_\mathrm{u}$, $A_\mathrm{u\ell}$, $g_\mathrm{u}$)  were taken from the {\it Splatalogue}\footnote{\url{www.splatalogue.net}} database, and were mainly derived from the {\it Cologne Database for Molecular Spectroscopy}\footnote{\url{www.astro.uni-koeln.de/cdms}} \citep{cdms} and {\it Jet Propulsion Laboratory}\footnote{\url{spec.jpl.nasa.gov}} \citep{jpl} catalogs. The partition function was calculated by summation on all the available transitions as $Z=\Sigma_u g_\mathrm{u} e^{-E_\mathrm{u}/kT_\mathrm{ex}}$.  

We take as source size of the emission the upper limit of 0$''$.4 to the HCN and HNC J=3-2 emission observed by \citet{sakamoto2013} and \citet{costagliola2013}. Notice that this value is compatible with the upper limit to the source size of HNC~1-0 in our data. 

\begin{figure*}
\begin{centering}
\includegraphics[width=.2\textwidth,keepaspectratio]{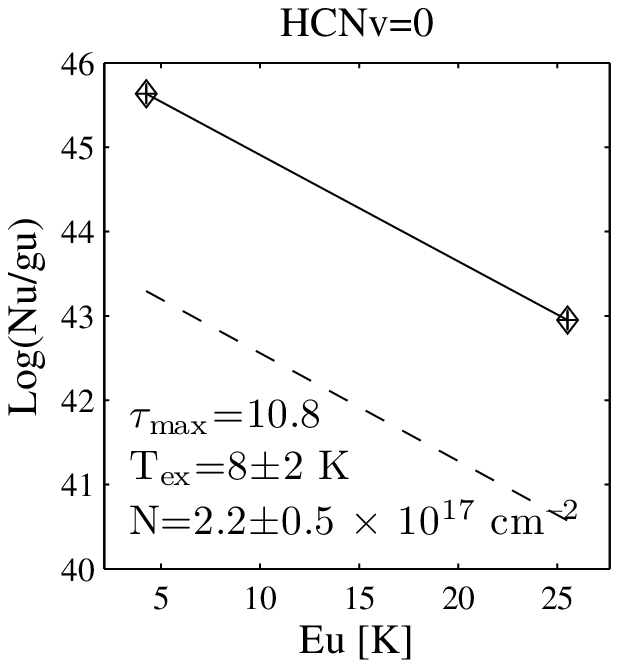}
\includegraphics[width=.2\textwidth,keepaspectratio]{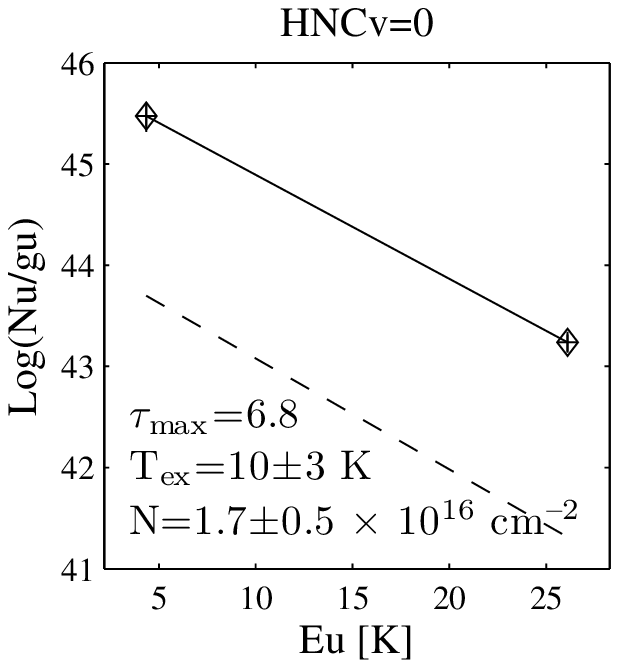}
\includegraphics[width=.2\textwidth,keepaspectratio]{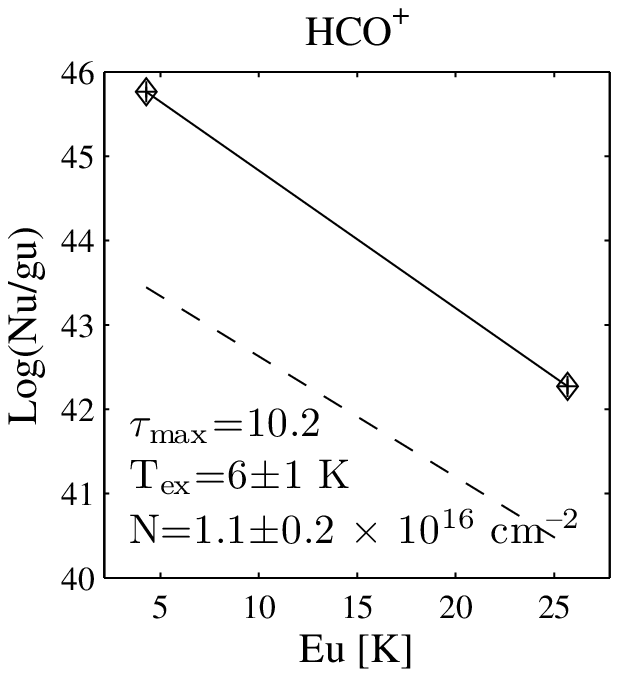}
\includegraphics[width=.2\textwidth,keepaspectratio]{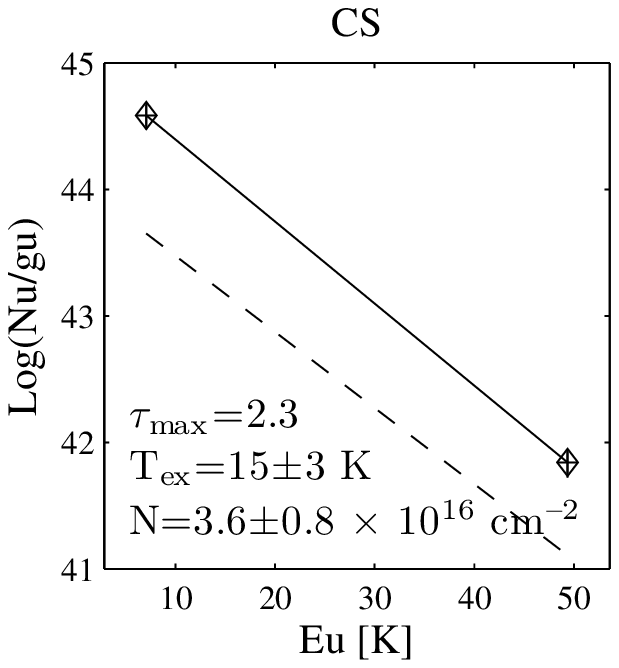}\\
\includegraphics[width=.2\textwidth,keepaspectratio]{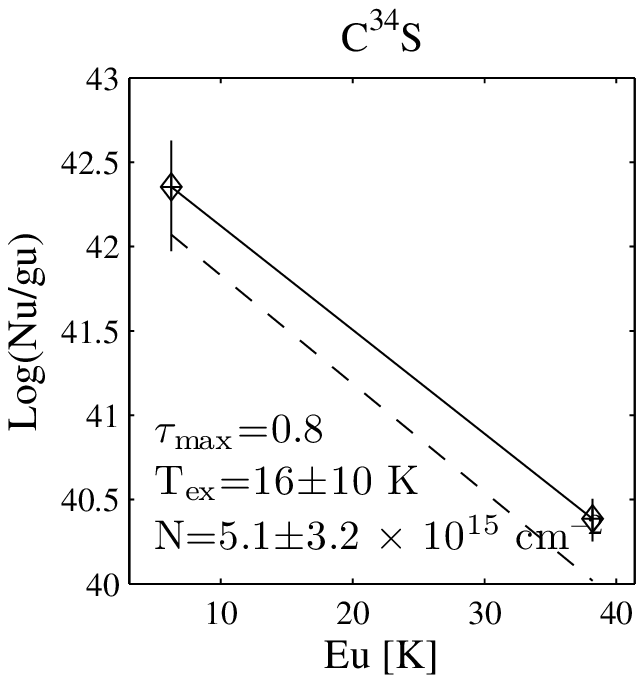}
\includegraphics[width=.2\textwidth,keepaspectratio]{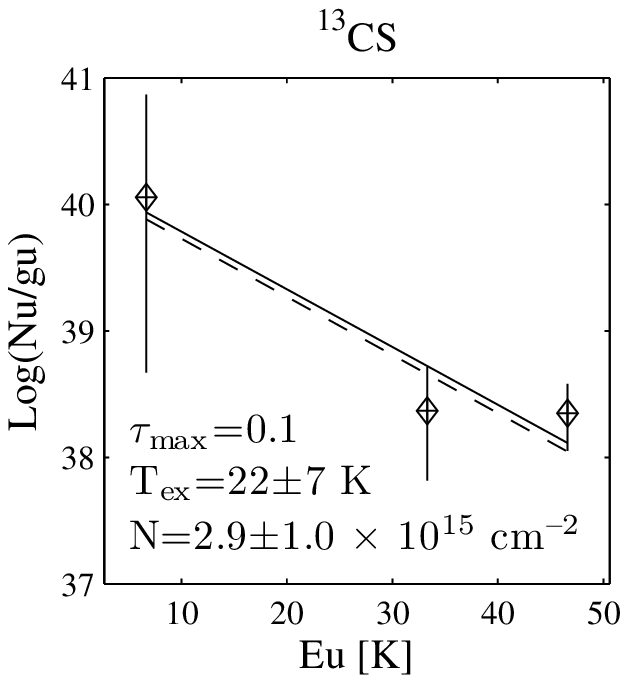}
\includegraphics[width=.2\textwidth,keepaspectratio]{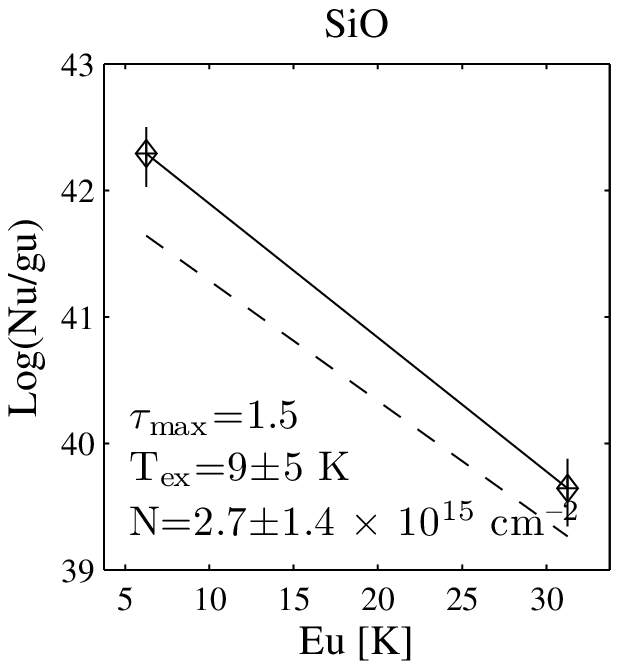}
\includegraphics[width=.2\textwidth,keepaspectratio]{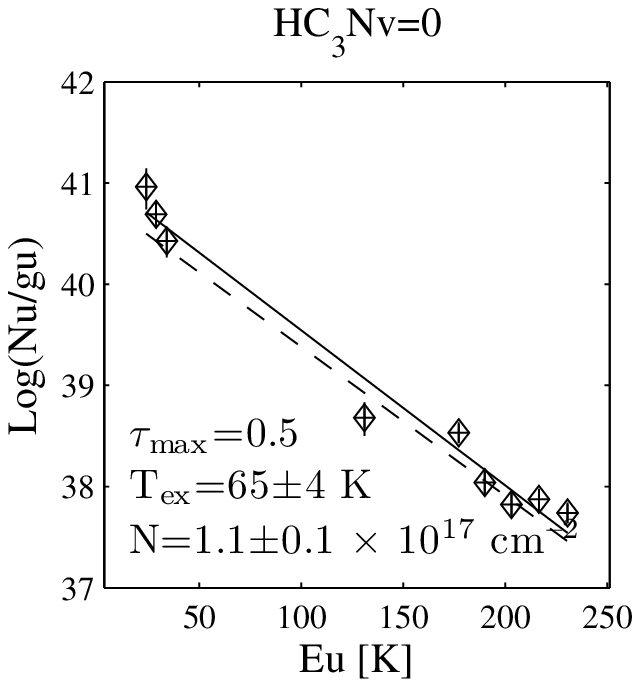}\\
\includegraphics[width=.2\textwidth,keepaspectratio]{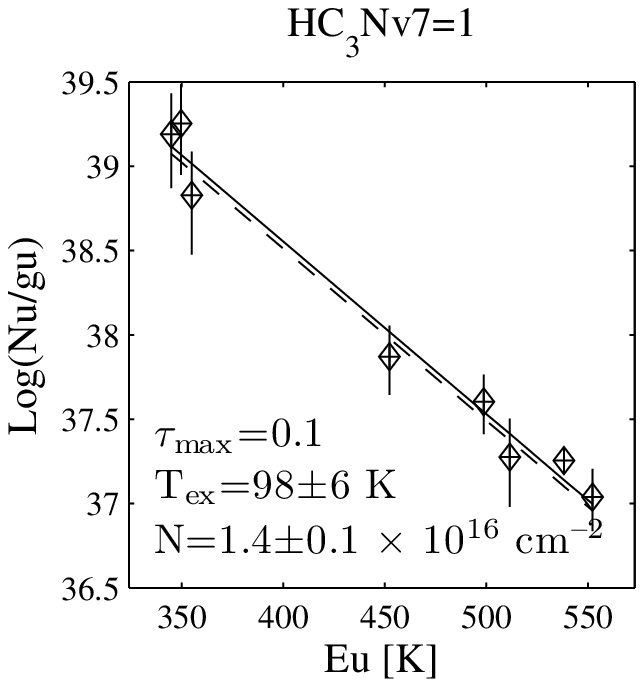}
\includegraphics[width=.2\textwidth,keepaspectratio]{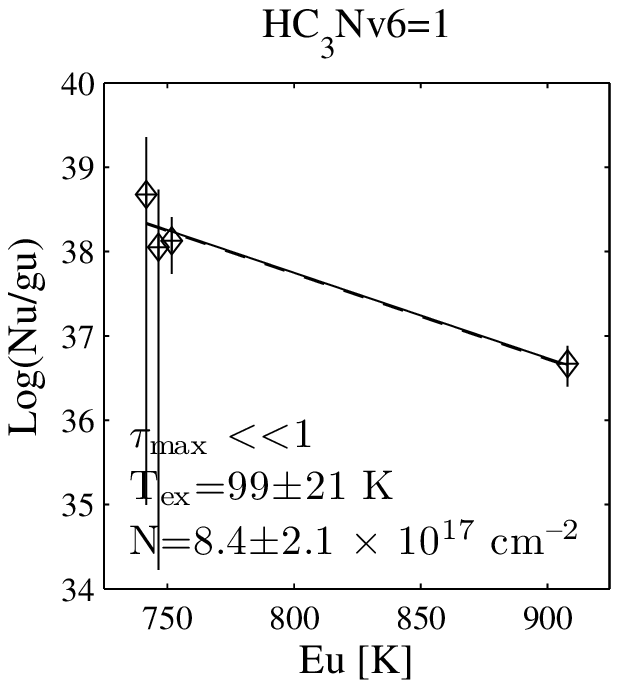}
\includegraphics[width=.2\textwidth,keepaspectratio]{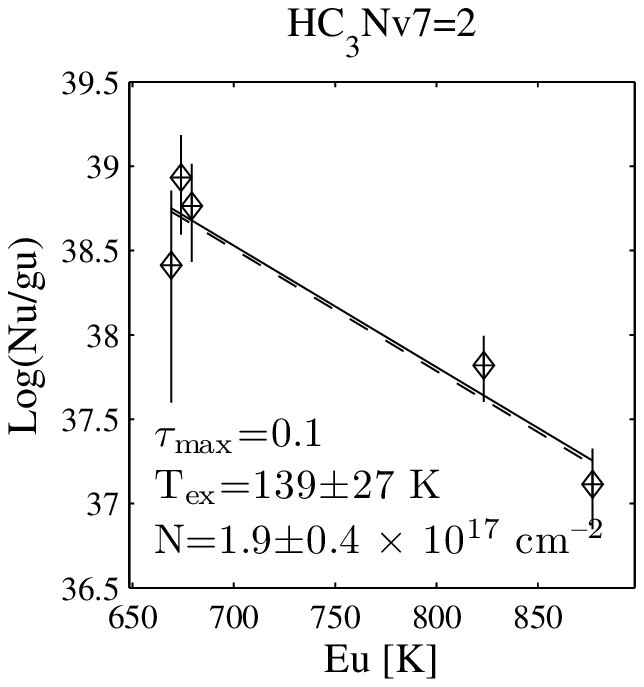}
\includegraphics[width=.2\textwidth,keepaspectratio]{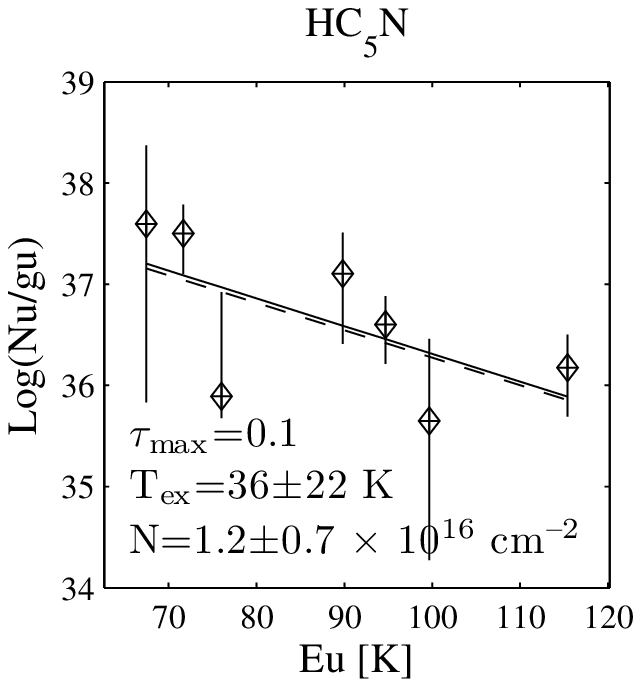}\\
\end{centering}
\caption{\label{fig:popdiag} Population diagrams for molecules with non-blended emission lines. The diamonds represent the values of N$_\mathrm{u}$/g$_\mathrm{u}$ for each transition, derived applying the opacity correction. The solid line shows the fit to the opacity-corrected data, while the dashed line shows the fit assuming optically thin emission. The fitted values for column density (N) and excitation temperature (T$_\mathrm{ex}$) are reported, together with the maximum opacity of the observed transitions. The error bars show 3-$\sigma$ uncertainties.}
\end{figure*}

We fit transitions from twelve different species: HCN, HNC, HCO$^+$, CS, C$^{34}$S, $^{13}$CS, SiO, HC$_{3}$N, vibrationally excited HC$_{3}$N in the v7=1, v6=1, and v7=2 states, and HC$_5$N. The results are summarized in Fig. \ref{fig:popdiag} and Table \ref{tab:popdiag}. We find excitation temperatures ranging from 7 to 140 K, and column densities from 10$^{15}$ to 10$^{17}$ cm$^{-2}$. The scatter in excitation temperatures is larger than the uncertainties and may be due to a steep temperature gradient or NLTE excitation effects.

Standard high-density tracers as CS, HCN, HNC, and HCO$^+$ have low excitation temperatures (<20 K) and high opacities ($>$1), while HC$_3$N transitions have moderate to low opacities and higher excitation temperatures, of the order of 100~K. 
Given the uncertainty on the source size and beam filling factor, the derived column densities have to be considered as lower limits to the real value \citep{popdiag}. This is particularly true for molecules with moderate to high opacities.

\begin{table*}
\caption{\label{tab:popdiag} Gaussian fitting of blend-free lines and results of the population diagram analysis.}
\begin{center}
\medskip
\begin{tabular}{lllllllll}
  \hline
  \hline
  &&&&&&&\\
Molecule & Transition & Frequency& $S_\mathrm{line}^{\scriptscriptstyle (1)}$  & {\it FWHM$^{\scriptscriptstyle (2)}$} & $T_\mathrm{b}^{\scriptscriptstyle (3)}$ & $T_\mathrm{ex}^{\scriptscriptstyle (4)}$& $N^{\scriptscriptstyle (5)}$ & $\tau^{\scriptscriptstyle (6)}$ \\
&& [GHz] & [Jy \kms] &  [\kms] & [K] & [K] & [cm$^{-2}$] &\\
   \hline
\rowcolor{Gray}
HCN	& J=1-0 & 88.63  & 9$\pm$1 & 140$\pm$10 & 65$\pm$7 & 8$\pm$2 & 2.2$\pm$0.5$\times$10$^{17}$ & $>$10\\
\rowcolor{Gray}
	& J=3-2 & 265.89 & 50$\pm$5 & 150$\pm$10 & 35$\pm$4 & & & $>$10\\
HNC	& J=1-0 & 90.66	& 5.2$\pm$0.7 & 110$\pm$15 &  45$\pm$6 & 10$\pm$3 & 1.7$\pm$0.5$\times$10$^{16}$ & 5.9\\
	& J=3-2 & 271.98 & 38.5$\pm$0.3 & 130$\pm$10 & 30$\pm$2 & & & 6.8\\
\rowcolor{Gray}
HCO$^+$	& J=1-0 & 89.19 & 6.2$\pm$0.4 & 110$\pm$7 &  55$\pm$3 & 6$\pm$1 & 1.1$\pm$0.2$\times$10$^{16}$ & $>$10\\
\rowcolor{Gray}
	& J=3-2 & 267.56 & 23.6$\pm$2 & 150$\pm$12 & 17$\pm$2 &  & & 6.6\\
CS	& J=2-1	& 97.98  &  5$\pm$0.5 & 116$\pm$12  &  35$\pm$4   & 15$\pm$3 & 3.6$\pm$0.8$\times$10$^{16}$ & 2.3\\  
	& J=6-5	& 293.91 & 32$\pm$4 & 150$\pm$16 & 19$\pm$2 & & & 1.7 \\
\rowcolor{Gray}
C$^{34}$S & J=2-1 & 96.41  & 1$\pm$0.3 & 100$\pm$30 & 8$\pm$3 &16$\pm$10 & 5$\pm$3$\times$10$^{15}$ & 0.6 \\	  
\rowcolor{Gray}
	  & J=6-5 & 289.21 & 10$\pm$1 & 115$\pm$15 & 8$\pm$1 & & & 0.8\\
$^{13}$CS & J=2-1 & 92.49  & 0.3$\pm$0.2 & 90$\pm$10 & 3$\pm$2 & 22$\pm$7 & 3$\pm$1$\times$10$^{15}$ & 0.1\\	 
	  & J=5-4 & 231.22 & 1.6$\pm$0.7 & 90$\pm$40 & 3$\pm$1 & & & 0.1\\
 	  & J=6-5 & 277.45 & 3.3$\pm$0.8 & 150$\pm$40 & 2$\pm$0.5 & & & 0.1\\
\rowcolor{Gray}
SiO & J=2-1 & 86.85 & 1.2$\pm$0.3 & 90$\pm$20 & 13$\pm$3 & 9$\pm$5 & 3$\pm$1$\times$10$^{15}$ & 1.5\\	 
\rowcolor{Gray}
    & J=5-4 & 217.10 & 4$\pm$1 & 130$\pm$30 & 5$\pm$1 & & & 0.8\\
HC$_3$N & J=10-9 & 90.98   & 4.0$\pm$0.8    &  110$\pm$20 & 34$\pm$7 & 65$\pm$4 & 1$\pm$0.1$\times$10$^{17}$ & 0.5 \\  
	& J=11-10 & 100.08   & 4.6  $\pm$0.6 &  110$\pm$15  & 31$\pm$4 & & & 0.4\\
	& J=12-11 & 109.17   & 5.1 $\pm$0.8     &  120$\pm$20 & 27$\pm$4  & & & 0.4\\
	& J=24-23 & 218.32   & 15.2 $\pm$2.5     &  150$\pm$25  & 16$\pm$3 & & & 0.2\\ 
	& J=28-27 & 254.70   &  23.8 $\pm$2.2   &  150$\pm$15  & 18$\pm$2 & & & 0.3\\
	& J=29-28 & 263.79   & 17.4  $\pm$2.2    &  135$\pm$15 & 14$\pm$2  & & & 0.2\\
	& J=30-29 & 272.88   &  16.2 $\pm$1.9    & 135$\pm$15  & 12$\pm$1   & & & 0.2\\
	& J=31-30 & 281.98  &  19.4 $\pm$2.0   &  150$\pm$15  & 12$\pm$1  & & & 0.2\\
	& J=32-31 & 291.07 & 19.4  $\pm$1.7 & 150$\pm$15 & 12$\pm$1 & & & 0.2\\
\rowcolor{Gray}
HC$_3$N,v7=1 & J=10-9,l=1f & 91.33 & 0.8$\pm$0.2 &  90$\pm$25  & 8$\pm$2  & 98$\pm$6 & 1.4$\pm$0.1$\times$10$^{16}$ & 0.1\\  
\rowcolor{Gray}
	     & J=11-10,l=1f & 100.47 & 1.3$\pm$0.3 & 105$\pm$30  & 9$\pm$2   &  &  & 0.1\\  
\rowcolor{Gray}
	     & J=12-11,l=1f & 109.60 & 1.2$\pm$0.3 &  90$\pm$30  & 8$\pm$2   &  &  & 0.1\\  
\rowcolor{Gray}
	     & J=24-23,l=1f &  219.17 & 7$\pm$ 1.5 & 130$\pm$30  & 9$\pm$2  &  &  & 0.1\\  
\rowcolor{Gray}
	     & J=28-27,l=1f &  255.70 & 10$\pm$2 & 150$\pm$25  & 8$\pm$1   &  &  & 0.1\\  
\rowcolor{Gray}
	     & J=29-28,l=1f & 264.82 & 9$\pm$2 & 130$\pm$30   & 7$\pm$2   &  &  & 0.1\\  
\rowcolor{Gray}
	     & J=31-30,l=1f &  283.07 & 11$\pm$1 & 145$\pm$10  & 7$\pm$1   &  &  & 0.1\\  
\rowcolor{Gray}
	     & J=32-31,l=1f & 292.20 & 10$\pm$2 & 120$\pm$20 & 7$\pm$1  &  &  & 0.1\\  
HC$_3$N,v6=1 & J=10-9,l=1e &  91.13 &  0.50$\pm$0.4 &   140$\pm$100 & 3$\pm$2 & 99$\pm$21 & 8$\pm$2$\times$10$^{17}$ & $\ll$1\\  
	& J=11-10,l=1e &  100.24  & 0.4$\pm$0.3 & 140 $\pm$ 100  & 2$\pm$1 &  & & $\ll$1\\
	& J=12-11,l=1e & 109.35   & 0.6$\pm$0.2   & 120 $\pm$ 40 & 3$\pm$1  & & & $\ll$1\\
	& J=29-28,l=1e & 264.22 &  5$\pm$1 &  140 $\pm$ 30 & 4$\pm$1 &  & & $\ll$1\\  
\rowcolor{Gray}
HC$_3$N,v7=2 & J=10-9,l=2e & 91.56 &  0.4$\pm$0.2  & 70$\pm$40  & 5$\pm$3  &   139$\pm$27 & 1.9$\pm$0.4$\times$10$^{17}$ & $\ll$1\\ 
\rowcolor{Gray}
 & J=11-10,l=2e & 100.71 &  0.9$\pm$ 0.3    & 105$\pm$30   & 7$\pm$2   &  & & $\ll$1\\ 
\rowcolor{Gray}
 &   J=12-11,l=2e & 109.87   & 1.1$\pm$0.3   &  120$\pm$35  & 6$\pm$2   &  & & $\ll$1\\ 
\rowcolor{Gray}
  &  J=28-27,l=2e & 256.31   & 13.1$\pm$2.5     & 170$\pm$30 & 8$\pm$2   & &  & 0.1\\ 
\rowcolor{Gray}
  &  J=32-31,l=2e & 292.91 & 11.2$\pm$2.6 & 170$\pm$40 & 6$\pm$1   &  & & $\ll$1\\ 
HC$_5$N & J=32-31 & 85.20  &  0.6$\pm$0.5  &    150$\pm$100  &  5$\pm$4 &  36$\pm$22 & 1.2$\pm$0.7$\times$10$^{16}$ & 0.1\\ 
 &     J=33-32 &  87.86    & 0.6$\pm$0.2   &   130 $\pm$45  &  4$\pm$1   &  &  & 0.1\\ 
 & 	J=34-33 &   90.52   &    0.2$\pm$0.1    &  100 $\pm$50 &  2$\pm$1  &   &  & $\ll$1\\ 
  &     J=37-36 &  98.51   & 0.6$\pm$0.3     & 130$\pm$65  &  4$\pm$2   &  & & 0.1\\ 
  &     J=38-37 &   101.17  &   0.4$\pm$0.1   &  90$\pm$30 &  4$\pm$1  &  & & 0.1\\ 
  &     J=39-38 &  103.83   & 0.2$\pm$0.1    &   90$\pm$50 &  2$\pm$1   &  & & $\ll$1\\ 
  &     J=42-41 &   111.82  & 0.4$\pm$0.1    & 90$\pm$35  &  3$\pm$1   &  & & 0.1\\ 
\hline 
\end{tabular}
\newline
\end{center}
{\tiny \it (1) Flux density integrated over the fitted Gaussian profile; (2) Gaussian line width; (3) Brightness temperature assuming a source size of 0$''$.4. (4) Excitation temperature from population diagram fit; (5) Column density from the population diagram fit; (6) Opacity of the transition.}

\end{table*}

\subsubsection{LTE/NLTE fit of the whole spectrum}

In order to identify the fainter, heavy blended emission, we fit the whole spectrum with a combined LTE/NLTE model, in a fashion similar to what was done for the ULIRG Arp~220 by \citet{martin2011}. We include in the fit all the molecules identified so far in the extragalactic interstellar medium, as reported in \citet[][Table 5]{martin2011}, for a  total of  46 species and 23 isotopologues.

The fitting procedure is composed of the following steps: {\it 1)} a synthetic LTE spectrum is produced for each molecule, assuming a fiducial column density of 10$^{14}$~\cmt, and an excitation temperature of 80~K \citep[based on observations by][]{costagliola2013};  {\it 2)} the molecules are ordered from the brightest to the faintest based on the flux density from the synthetic spectrum in bands 6 and 7;  {\it 3)} a synthetic spectrum for the first molecule is fitted to the continuum-removed observed spectrum by $\chi^2$ minimization; {\it 4)} The best fit spectrum is removed from the data and the fit parameters stored. Steps {\it 3} and {\it 4} are repeated iteratively for all the molecules in the list.

For the molecules with collisional coefficients available in the LAMBDA database \citep[CS, SO, SiO, HCN, HNC, HCO$^+$, N$_2$H$^+$, H$_2$CO, CH$_3$CN, CH$_3$OH, ][]{lambda}, both an LTE and an NLTE synthetic spectrum was fit to the data, while for the other molecules only an LTE fit was possible. For the NLTE fit, a grid of 100,000 synthetic spectra was produced with the RADEX radiative transfer code \citep{radex}, with molecular hydrogen densities ranging from 10$^{3}$ to 10$^7$\cmth, kinetic temperatures between 5 and 300~K, and molecular column densities between 10$^{15}$ and 10$^{18}$~\cmt. The range in kinetic temperatures was limited by the available collisional coefficients, which in most cases have been computed up to 300~K. The LTE synthetic spectra were produced following the procedure described in \citet{martin2011}. 

We considered a fixed line width of 120~\kms, which is the average of the values found by Gaussian fitting the blend-free lines (see Section \ref{sec:popdiag} and Table \ref{tab:popdiag}). Before calculating the $\chi^2$, the synthetic spectra were red-shifted assuming a single LSR velocity for NGC~4418 of 2100~\kms. We assumed all the emission to be emerging from a region of 0$''$.4 in diameter, which is the upper limit to the size of molecular core of NGC~4418, as observed by \citet{sakamoto2013} and \citet{costagliola2013}. In both the LTE and NLTE fit, we use the 3$^\mathrm{rd}$ order fit of the line-free channels as continuum level, as explained in Section \ref{sec:continuum}.

\begin{table*}
\caption{Summary of species detected in NGC~4418. Tentative detections are shown in italics.}
\begin{center}
\medskip
\renewcommand{\arraystretch}{1.1}
\begin{tabular}{llllll}
\hline
\hline
{\bf 2 atoms} &{\bf 3 atoms} & {\bf4 atoms} & {\bf 5 atoms} & {\bf 6 atoms} & {\bf  7 atoms }\\
\hline
&&&&&\\
CS~~~~\rdelim\}{4}{3mm} & HCN~~~~\rdelim\}{2}{3mm}  &  p-H$_2$CO &  HC$_3$N~~~~~~~\rdelim\}{3}{3mm} & CH$_3$CN  & {\it CH$_3$CCH} \\
$^{13}$CS & H$^{13}$CN  & o-H$_2$CO  & HCC$^{13}$CN & {\it CH$_3$OH} & HC$_5$N \\
{\it C$^{33}$S} & HCN,v2=1 & c-HCCCH & H$^{13}$CCCN &	&	\\
C$^{34}$S& HNC~~~~\rdelim\}{2}{3mm} & H$_2$CS~~\rdelim\}{2}{3mm}  & HC$_3$N,v6=1 & &  \\
$^{13}$CO~\rdelim\}{2}{3mm} & HN$^{13}$C & {\it H$_2^{13}$CS} & HC$_3$N,v7=1 & &  \\
{\it C$^{18}$O} & HNC,v2=1 &  & HC$_3$N,v6=1,v7=1 & &  \\
CN & HCO$^+$~~~~\rdelim\}{3}{3mm} &  & HC$_3$N,v7=2 & &  \\
NS & H$^{13}$CO$^+$ &  & CH$_2$NH & &  \\
SO~~~~\rdelim\}{2}{3mm}   & {\it HC$^{18}$O$^+$} &  & {\it NH$_2$CN} & &  \\
{\it $^{34}$SO} & {\it H$_2$S} &  &  & &  \\
SiO~~~~\rdelim\}{3}{3mm}  & CCH & & & &  \\
~$^{29}$SiO & HCS$^+$ & & & &  \\
~$^{30}$SiO & CCS & & & &  \\
	& N$_2$H$^+$ & & &  &  \\
&&&&&\\
\hline
\end{tabular}
\end{center}

\label{tab:mol}  
\end{table*}

\subsection{Results from the line identification and fit}

A list of the identified molecular species is reported in Table \ref{tab:mol}. We define as {\it detected} those molecules which have more than one transition included in our spectral scan and are well fitted by our LTE or NLTE models. An exception to this rule are bright lines from well-known ISM species or their isotopic variants, such as $^{13}$CO, H$^{13}$CN, HN$^{13}$C, and H$^{13}$CO$^+$, for which one line is enough for a successful identification. We define as {\it tentative} detections those molecules with only one detected transition in the scan (e.g., HCN,v2=1), or those for which no satisfactory fit could be obtained (e.g., CH$_3$OH). Tentative detections are shown in italics in Table \ref{tab:mol}.

{\it We identify 317 emission lines above the three sigma level from a total of 45 molecular species, including 15 isotopic substitutions and six vibrationally excited variants.} The emission lines above three sigma which were fitted by our model are listed in Tables \ref{tab:lines1} to \ref{tab:lines6}. The flux recovered by the fit is 92~\% of the total molecular emission, resulting in a normalized $\chi^2$ of 5 for the whole spectrum. This somewhat high value of the $\chi^2$ is mainly due to unidentified lines and systematic uncertainties, which are discussed in Section \ref{sec:limits}.

The results of the fit are summarized in Table \ref{tab:fit}. The fit details for individual molecules are reported in Section \ref{sec:fitdet}. 

\begin{table*}
\caption{Results from the LTE and NLTE fit of the whole spectrum.}
\begin{center}
\medskip
\begin{tabular}{l|cc|cccc}
\hline
\hline
 & \multicolumn{1}{l}{LTE} & & \multicolumn{1}{l}{NLTE} &&&\\
\hline
&&&&&&\\
Molecule & T$_\mathrm{ex}$ [K]$^{\scriptscriptstyle (1)}$ & N [cm$^{-2}$]$^{\scriptscriptstyle (2)}$ & n(H$_2$) [cm$^{-3}$]$^{\scriptscriptstyle (3)}$ & T$_\mathrm{kin}$ [K]$^{\scriptscriptstyle (4)}$ & \tex$^{3-1\mathrm{mm}}$ [K]$^{\scriptscriptstyle (5)}$ & N [cm$^{-2}$]$^{\scriptscriptstyle (6)}$ \\
CS & 20$\pm$1  & 6$\pm$0.6 $\times$10$^{16}$  &  10$^{5}$ & 55$\pm$2  & 47-30 & 1$\pm$0.2$\times$10$^{17}$  \\
$^{13}$CS & 47  & 1.5$\pm$0.5 $\times$10$^{15}$ & - & - & - & - \\
{\it C$^{33}$S} & 47  & 3$\pm$1 $\times$10$^{15}$ & - & - & - & - \\
C$^{34}$S & 47  & 8$\pm$2 $\times$10$^{15}$  & - & - & - & -\\
$^{13}$CO & 70  & 5$\pm$2 $\times$10$^{18}$ & - & - & - & - \\
{\it C$^{18}$O} & 70  & 6$\pm$1 $\times$10$^{17}$ & - & - & - & -\\
CN & 70  & 5$\pm$0.5 $\times$10$^{17}$ & - & - & - & -\\
NS & 350$\pm$150  & 8$\pm$4 $\times$10$^{16}$ & - & - & - & -\\
SO & $>20$ & 10$^{15}$-10$^{16}$ & 10$^{6}$-10$^{7}$ & 50-200  & 40-100 & 2-20$\times$10$^{15}$ \\
{\it $^{34}$SO} & $>20$ &  10$^{15}$-10$^{16}$  & - & - & - & -\\
SiO & 10$\pm$5  & 3$\pm$1 $\times$10$^{15}$ & $<$10$^{6}$ & $>$20  & 22-10 & $>$10$^{16}$\\
$^{29}$SiO & 22  & 2$\pm$1 $\times$10$^{15}$ & - & - & - & - \\
$^{30}$SiO & 22  & 2$\pm$1 $\times$10$^{15}$ & - & - & - & - \\
HCN & 7$\pm$1  & 1.5$\pm$0.1 $\times$10$^{15}$ & $<$10$^{6}$ & $>$100 & 23-15 & $>$10$^{16}$ \\
H$^{13}$CN & 65  & 1$\pm$0.5 $\times$10$^{16}$  & - & - & - & - \\
HCN,v2=1 & 350  & 3$\pm$1 $\times$10$^{15}$ & - & - & - & -\\
HNC & 8$\pm$1  & 8$\pm$0.5 $\times$10$^{15}$ & $<$10$^{6}$ & $>$100  & 45-40 & $>$10$^{16}$ \\
HN$^{13}$C& 45   & 1$\pm$0.5 $\times$10$^{15}$  & - & - & - & -\\
HNC,v2=1 & 450  & 1.5$\pm$0.5 $\times$10$^{16}$  & - & - & - & -\\
HCO$^+$ & 6$\pm$1  & 6$\pm$0.5 $\times$10$^{15}$ & $<$6$\times$10$^{5}$ & $>100$ & 80-50 & $>$10$^{16}$  \\
H$^{13}$CO$^+$ & 80  & 2.5$\pm$1 $\times$10$^{15}$ & - & - & - & -\\
{\it HC$^{18}$O$^+$} & $<20$ & $>$10$^{14}$ & - & - & - & - \\
{\it H$_2$S} & 70  & 1$\pm$0.5 $\times$10$^{17}$  & - & - & - & -\\
CCH & 70  & 2$\pm$1 $\times$10$^{17}$  & - & - & - & - \\
HCS$^+$ & 20$\pm$10  & 8$\pm$4 $\times$10$^{15}$ & - & - & - & -\\
CCS & 20$\pm$10  & 1.5$\pm$0.5 $\times$10$^{16}$ & - & - & - & -\\
N$_2$H$^+$ & 30$\pm$5  & 5$\pm$1 $\times$10$^{15}$ & 10$^{7}$ & 23$\pm$2  & 23-23 & 5$\pm$2 $\times$10$^{15}$  \\
p-H$_2$CO & 350$\pm$100  & 5$\pm$2 $\times$10$^{16}$ & 10$^{5}$ & $>$300 & 20-100 & 7$\pm$2 $\times$10$^{15}$ \\
o-H$_2$CO & 350$\pm$100  & 8$\pm$4 $\times$10$^{16}$& 10$^{5}$ & $>$300 & 100-500 & 1$\pm$0.2 $\times$10$^{16}$ \\
c-HCCCH & 180$\pm$100  & 1$\pm$0.5 $\times$10$^{17}$ & - & - & - & -\\
H$_2$CS & 35$\pm$10  & 1$\pm$0.5 $\times$10$^{16}$ & - & - & - & -\\
{\it H$_2^{13}$CS} & 35  & 5$\pm$2 $\times$10$^{15}$ & - & - & - & -\\
HC$_3$N & 70$\pm$5  & 1.8$\pm$2 $\times$10$^{16}$ & - & - & - & -\\
H$^{13}$CCCN & 70  & 2$\pm$1 $\times$10$^{15}$ & - & - & - & -\\
HCC$^{13}$CN & 70  & 1$\pm$0.5 $\times$10$^{15}$& - & - & - & -\\
HC$_3$N,v6=1 & 100$\pm$50  & 4$\pm$3 $\times$10$^{15}$ & - & - & - & -\\
HC$_3$N,v7=1 & 100$\pm$25  & 1$\pm$0.2 $\times$10$^{16}$  & - & - & - & -\\
HC$_3$N,v6=1,v7=1 & 70$\pm$30  & 3$\pm$2 $\times$10$^{15}$ &  - & - & - & - \\
HC$_3$N,v7=2 & 100$\pm$33  & 8$\pm$2 $\times$10$^{15}$ & - & - & - & -\\
CH$_2$NH & 45$\pm$30  & 1.5$\pm$0.5 $\times$10$^{17}$& - & - & - & -\\
{\it NH$_2$CN} & 30$\pm$20  & 5$\pm$2 $\times$10$^{15}$ & - & - & - & -\\
CH$_3$CN & $>$1000 & 8$\pm$3 $\times$10$^{15}$ & 10$^{4}$ & $>$300 & 10-20 & 1.5$\pm$0.3 $\times$10$^{16}$ \\
{\it CH$_3$OH} & 60  & $<$1$\times$10$^{16}$ & 10$^{7}$ & 60$\pm$10 & 20-40 & 2$\pm$1$\times$10$^{16}$ \\
{\it CH$_3$CCH} & 350$\pm$100  & 1$\pm$0.5 $\times$10$^{17}$ & - & - & - & -\\
HC$_5$N & 65$\pm$7  & 1$\pm$0.4 $\times$10$^{16}$ & - & - & - & -\\
\end{tabular}

Total normalized $\chi^2$ = 5
  
\end{center}
{\tiny \it (1) Excitation temperature from LTE fit; (2) Column density from LTE fit; (3) Molecular hydrogen density from NLTE fit; (4) Kinetic temperature from NLTE fit; (5) Maximum excitation temperature at 3 and 1~mm from NLTE fit; (6) Column density from NLTE fit.}

\label{tab:fit}  
\end{table*}


\begin{figure}
\begin{centering}
\includegraphics[width=.4\textwidth,keepaspectratio]{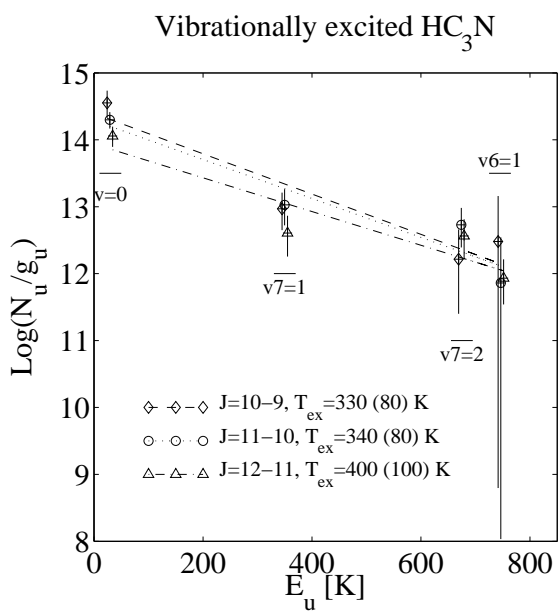}
\end{centering}
\caption{\label{fig:vib} Population diagram for the vibrational excitation of HC$_3$N. For the three rotational levels J=10 ({\it diamonds}), J=11 ({\it circles}), J=12 ({\it triangles}) we compare the relative populations of the v=0, v7=1, v7=2, and v6=1 vibrationally excited states. The vibrational temperature fit is shown as a {\it dashed}, {\it dotted} and {\it dash-dotted} line, for the J=10--9, 11--10 , and 12--11 transition respectively. The three fits result in similar vibrational temperatures in the range 330-400~K.}
\end{figure}

\subsection{Limitations of the fit and main uncertainties}
\label{sec:limits}

The confidence intervals reported in Tables \ref{tab:popdiag} and \ref{tab:fit} represent one-sigma statistical uncertainties derived from the $\chi^2$ minimization. Other systematic uncertainties have to be considered for a correct interpretation of the fit results.

{\it Source size and beam filling factor:} Because the emission is unresolved in the ALMA beam, the main uncertainty in our LTE/NLTE fit is the source size. In our conversion from flux density to brightness temperature we assume a source size of 0.$''$4, which is the upper limit to the diameter of the region containing the bulk of the molecular emission in NGC~4418 \citep{sakamoto2013,costagliola2013}. If we assume that all the molecular emission is coming from this compact region, the derived brightness temperature is a lower limit to the true value.
At LTE this translates into {\it underestimated} column densities or excitation temperatures, respectively for optically thin or optically thick transitions. 

Our fit also assumes that all the molecular species have the same beam filling factor, which is unlikely, given the steep temperature and density gradients expected in this galaxy \citep[e.g., ][]{galfonso2012}. The chemical properties derived for the ISM of NGC~4418 have thus to be intended as an average of the different environments contained in the ALMA beam. The same limitation is true for most of the extragalactic spectral scans observed so far, and will be possible to overcome only with high-resolution observations. Given its extreme compactness, high angular resolution observations (e.g. with the full ALMA array in an extended configuration) will be required to further resolve the compact mm-line emission region.

{\it Line blending:} Severe line confusion in bands 6 and 7 makes the fit non-univocal for the species with no strong blend-free lines detected. This problem is somewhat mitigated by the large frequency coverage of our scan, but its effects are evident in some heavily blended regions as a poor fit to the data. Based on the relative intensity of the blended lines, we estimate the contribution to the  uncertainty on the fitted parameters to be of the order of 20\%.

{\it Opacity:} For optically thick transitions the intensity of the lines depends only marginally on the column density. For the most abundant molecules, such as CS, HCN, HNC, and HCO$^+$, which have opacities greater than one in most of the observed transitions, the column density is thus poorly determined. For this reason we choose not to derive isotopic ratios for optically thick species, and we will not include these molecules in the following abundance analysis.

{\it NLTE effects:} An NLTE analysis of the emission was only possible for the molecules included in the LAMBDA database, while for the others only an LTE fit was performed. The NLTE fit results in a wide range of molecular hydrogen densities, it is thus likely that NLTE effects may affect many of the other species. The LTE assumption implies densities greater than the critical density for the transitions, which in general increases with the rotational quantum number. At lower densities, the de-population of the high-energy levels is interpreted by our LTE code as an underestimated kinetic temperature.  This results evident when comparing  the LTE and NLTE fits for HCN, HNC, and HCO$^+$. A more precise analysis will be possible only when the collisional coefficients for all the relevant species will become available.

{\it Line width: } In order to fit the heavily blended lines in bands 6 and 7, we assume a single line width of 120~\kms, which is an average value derived from the blend-free transitions of Table \ref{tab:popdiag}. However, the measured line widths have a significant scatter, with a trend of an increasing FWHM with critical density (see Fig. \ref{fig:fwhm}). The measured line widths have a standard deviation of 17~\kms, which translates into an error on the fitted integrated intensities of the order of 15\%.
 
Most of the systematic uncertainties in our fitting strategy result in a potential underestimation of molecular column density or kinetic temperature. The effect of other uncertainties (such as those introduced by line blending or variations in line width) on the fit results does not exceed 20\%, comparable to the nominal accuracy of the data calibration.

\section{Discussion}

\subsection{A vibrationally excited spectrum}

The spectrum of NGC~4418 appears to be dominated by bright emission from vibrationally excited HC$_3$N, which was detected in the v7=1, v7=2, v6=1, and (v7=1,v6=1) variants. {\it The detection of emission from the (v7=1,v6=1) state of HC$_3$N is the first obtained in an extragalactic object.} We also detect the J=3-2 transitions of vibrationally excited v2=1 HCN and HNC. Emission from vibrationally excited HC$_3$N, HCN, and HNC amounts to 19$\%$ of the total flux recovered by our fit.

The excitation energies of the HC$_3$N vibrational levels range from 300 to 900~K, corresponding to transition wavelengths in the far infrared (FIR), from 16 to 48~$\mu\mathrm{m}$.  The population analysis of vibrationally excited HC$_3$N shown in Fig. \ref{fig:vib} results in a vibrational temperature of 330-400~K. Even higher values of 350 and 450~K are found respectively for HCN and HNC when comparing the intensities of the J=3-2 transitions of the v=0 and v2=1 vibrational states. 

The  critical densities of the vibrational FIR transitions of HC$_3$N, HCN, and HNC are high, greater than 10$^8$~\cmth. We can thus assume that, for typical ISM densities (10$^3$--10$^7$~\cmth), radiative excitation dominates the population of the vibrational levels of HC$_3$N. Under this assumption, the relative population of the vibrationally excited levels is a direct probe of the FIR radiation field in the core of NGC~4418 \citep[e.g., ][]{costagliola2010}.

In the limit of perfect coupling between the FIR radiation field and the vibrational levels of the three molecules, the derived vibrational temperature is a good approximation to the temperature of the infrared source in the core of NGC~4418. When allowing for a non-uniform coverage of the IR source, inefficient coupling, or contamination from collisional excitation, the derived value has to be interpreted as a {\it lower limit} to the temperature of the exciting IR radiation.

The only extragalactic object where a similar vibrationally excited spectrum has been detected is the ULIRG Arp~220, where \citet{martin2011} find remarkably similar vibrational temperatures (300--400~K) for HC$_3$N. Both objects have strong IR fluxes, very compact molecular cores ($<$20~pc), and high molecular columns (N(H$_2$)$>$10$^{24}$~\cmt); {\it  a vibrationally excited spectrum  may thus be the signature of compact, deeply buried IR sources in extragalactic objects}. 

\subsection{Temperature and density structure}

\begin{figure*}[ht]
\begin{centering}
\includegraphics[width=.9\textwidth,keepaspectratio]{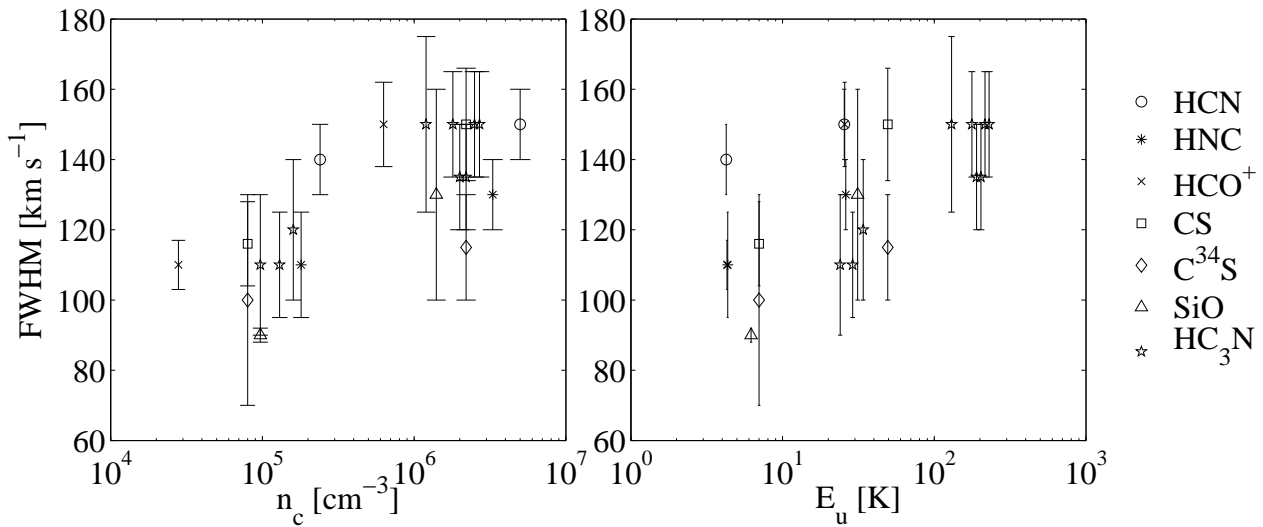}
\caption{\label{fig:fwhm} Full width at half maximum versus critical density and upper state energy. The line width was derived by a Gaussian fit of the blend-free lines (see Table \ref{tab:popdiag}). The critical density (n$_\mathrm{c}$) was calculated from the Einstein's and collisional coefficients in the LAMBDA database. The upper state energies (E$_\mathrm{u}$) of the transitions were taken from Splatalogue.}
\end{centering}
\end{figure*}

Our fit of the ALMA spectral scan finds kinetic temperatures ranging between 20 and 350~K, and molecular hydrogen densities between 10$^5$ and 10$^7$~\cmth.

The vibrational excitation of HC$_3$N, HCN, and HNC, requires the presence of a deeply buried, optically thick source, of brightness temperature higher than 350~K. If we assume all the IR flux of NGC~4418 to be coming from such a source, this must be extremely compact, with a size smaller than 5~pc. A compact IR source of similar temperature and size is also required by models by \citet{galfonso2012} in order to fit the mid-IR continuum. 

The single sphere geometry is a simplifying assumption and our data do not distinguish a single 5~pc sphere from other geometries such as a disk or a group of smaller patches distributed on a larger scale, as long as they have the same total area. We note that a radio source of $\sim$5~pc in diameter has actually been detected by EVN observations \citep{varenius2014} at less than 10 milli arcseconds from the peak of the 860~$\mu$m continuum \citep{sakamoto2013}. This source may be responsible for the compact IR and vibrationally excited emission, however this scenario needs spectroscopic observations at higher spatial resolution in order to be confirmed.

Very high rotational temperatures, exceeding 300~K, are found for molecules such as NS, H$_2$CO, CH$_3$CN, and CH$_3$CCH. The observed transitions have critical densities lower than 10$^6$~\cmth~and we can assume the excitation to be dominated by collisions under normal dense-ISM conditions. This is the first time such high rotational temperatures have been found in NGC~4418, mainly thanks to the wide frequency coverage of our scan. These measurements reveal a warm, $>$300~K gas component, which may be associated with the optically thick IR source producing the vibrationally excited emission. 

The ISM in the core of NCG~4418 appears to be a multi-phase environment, where we can identify three main temperature components: one {\it cold} component at 20-50~K, traced by CS, SiO, HCS$^+$, CCS, N$_2$H$^+$, H$_2$CS, CH$_2$NH, and NH$_2$CN; one {\it warm} component at 60-100~K, traced by CO, CN, HCN, HNC, HCO$^+$, H$_2$S, CCH, HC$_3$N, CH$_3$OH, and HC$_5$N; and one {\it hot} component at $>$300~K, traced by the vibrationally excited HC$_3$N, HNC, and HCN, and by collisionally excited NS, H$_2$CO, CH$_3$CN, and CH$_3$CCH. 

Our analysis is consistent with the results by \citet{galfonso2012} and \citet{costagliola2013}, who modeled Herschel/PACS and SMA observations with a layered density and temperature structure.

Additional evidence of a radial density and temperature gradient comes from the analysis of the velocity dispersion of the blend-free lines of Table \ref{tab:popdiag}. Figure \ref{fig:fwhm}  shows that the line width increases with critical density and upper state energy of the transitions. If we assume the velocity dispersion to be radially increasing towards the center of the galaxy, as it would be the case for a Keplerian rotation, we can interpret these trends as a radial gradient of density and temperature. This result is in agreement with our previous SMA observations \citep{sakamoto2013,costagliola2013}, which found a similar trend for the molecular emission at 1~mm and 850~$\mu$m. These studies also revealed that emission from transitions with high critical densities tend to be more centrally concentrated, supporting the existence of a radial density gradient. 

Most of the transitions considered in Fig. \ref{fig:fwhm} however have opacities greater than one and we cannot exclude the observed line broadening to be caused by radiative trapping. To confirm this scenario, we will need spatially-resolved observations.

\subsection{Extragalactic HC$_5$N}

Emission from HC$_5$N has been observed towards several Galactic hot cores \citep[e.g.,][]{green2014}, but until recently the molecule had not been detected outside the Galaxy. Aladro et al. ({\it in prep.}) report the detection of eight HC$_5$N transitions in the starburst galaxy NGC~253, to our knowledge NGC~4418 is the only other extragalactic object where the molecule has been detected.

From the eleven detected transitions we derive an excitation temperature of 60-70~K, and a column density of  0.5-2$\times$10$^{16}$~\cmt. Assuming a column density of H$_2$ towards the nucleus of NGC~4418 of 10$^{24}$~\cmt \citep{costagliola2013}, this corresponds to an abundance of 0.5-2$\times$10$^{-8}$. This value is about one order of magnitude higher than what found in NGC~253. 

Chemical models by \citet{chapman09} show that cyanopolyynes can reach high abundances in hot cores, thanks to gas-phase neutra-neutral reactions. The main formation path of HC$_5$N for gas temperatures of the order of 100~K is
\begin{eqnarray}
C_2H+C_2H_2 \rightarrow H_2CCCC+H \nonumber \\
CN+H_2CCCC \rightarrow HC_5N+H,
\end{eqnarray}
which for molecular hydrogen densities close to 10$^5$~\cmth~can lead to detectable column densities (N>10$^{14}$~\cmt). 

The abundance of HC$_5$N  is directly linked to acetylene, which is highly abundant in the earlier stages of the core evolution when it is evaporated from grain mantles. The enhancement of HC$_5$N in the gas phase is extremely short lived, it abundance increasing and decreasing over a period of the order of 10$^2$-10$^3$~yr. For this reason, \citet{chapman09} suggest that the molecule could be used as a chemical clock in hot-cores. The fact that HC$_5$N has been detected so far only in two external galaxies may be due to its short lifetime in the gas phase.
The detection of HC$_5$N in NGC~4418 may thus be coming from an extremely young, dust-embedded starburst. Observations by \citet{sakamoto2013} and \citet{costagliola2013} indeed suggest that if a starburst is powering the central 100~pc of the galaxy, this should be younger than 10 Myr. The fact that the molecule has been detected in NGC~253 but not in M~82 supports the interpretation that its abundance may be a good indicator of the starburst evolution.

\subsection{Standard line ratios: HCN, HNC, and HCO$^+$}

Line ratios of the dense gas tracers HCN, HNC, and HCO$^+$ have often been used in the literature to investigate the nature of the dominant energy source in LIRGs, either a compact starburst or an AGN \citep[e.g., ][]{imanishi07,nascent,krips08,costagliola11}.

{\it HCO$^+$/HCN ratios:} The HCO$^+$/HCN J=1--0 line ratio has been proposed as a diagnostic tool to distinguish between AGN- and starburst-powered galaxies \citep[e.g., ][]{kohno2001,imanishi04,imanishi07}. Interferometric observations by \citet{kohno2001} reveal that Seyfert galaxies have lower HCO$^+$/HCN ratios compared with starbursts. Seyfert galaxies with starburst-like HCO$^+$/HCN ratios are generally interpreted as mixed AGN-starburst objects \citep{imanishi07}. Single dish surveys \citet[][]{krips08,costagliola2013} find that starbursts and LIRGs have, on average, higher HCO$^+$/HCN ratios than AGNs, which in this scenario would suggest a dominant UV-dominated (PDR) chemistry. The theoretical explanation of the observed low HCO$^+$/HCN in AGNs is however still being debated. Early work by \citet{lepp96} suggest that the observed trend would be due to an enhancement of HCN abundance in the X-ray-dominated region surrounding an AGN. However, more recent chemical models find that the HCO$^+$/HCN abundance ratio is a strong function of density and ionizing radiation, \citep[e.g., ][]{meijerink07}, and that a low line ratio could be either caused by low-density($<$10$^5$~\cmth) XDRs or dense ($>$10$^5$~\cmth) PDRs. Also, given the different critical densities of the J=1--0 transition of HCN and HCO$^+$, localized variations in the ratios could be explained by variations in density \citep[e.g., ][]{meier12}. 

In our spectral scan of NGC~4418 the HCO$^+$/HCN line ratio varies between 0.8 to 0.5 for the J=1--0 and J=3--2 transitions respectively. The abundance ratio can be estimated by H$^{13}$CN and H$^{13}$CO$^+$ J=1--0 transitions. Assuming optically thin emission, and an excitation temperature of 7 and 6~K, we find LTE column densities of 2$\pm$1\tenfift~\cmt~ and 7$\pm$4\tenfort~\cmt~ for H$^{13}$CN and H$^{13}$CO$^+$, respectively. If we assume an extragalactic $^{12}$C/$^{13}$C ratio between 40 and 100 \citep{henkel2014}, these correspond to column densities of (0.8--2)$\times$10$^{17}$~\cmt~for HCN and (2.8--7)$\times$10$^{16}$~\cmt~ for HCO$^+$. The derived HCO$^+$/HCN abundance ratio in NGC~4418 is thus 0.35$\pm$0.2.

The molecular gas density cannot be higher than 10$^5$~\cmth~because at higher densities the HCO$^+$~3--2 transition would be thermalized, while we observe a J=3--2/1--0 brightness temperature ratio of 0.3. For such densities, the observed HCO$^+$/HCN line and abundance ratios are consistent with PDR models by \citet{meijerink07}, while XDR models would result in much higher ratios (between 3 and 5).

However, high-resolution ALMA observations of NGC~1068 by \citet{gburillo2014} find a HCO$^+$(4--3)/HCN(4--3) brightness temperature ratio of $\sim$0.4 in the region within 140~pc from the AGN, and a ratio of $\sim$0.8 at the AGN position. These values are very close to the ones observed in NGC~4418 for the J=1--0 and J=3--2 transitions and provide a strong observational evidence for the low HCO$^+$/HCN ratio to be associated with AGN activity. The higher ratio measured directly towards the AGN in NGC~1068 suggests that the relative intensity for the HCO$^+$ and HCN emission may be regulated by mechanical heating or shock chemistry from the AGN feedback rather than by X-ray dominated processes.

{\it HNC/HCN ratios:}  In the Galaxy, the HNC/HCN observed line ratio ranges from 1/100 in hot cores \citep{schilke92} to values as high as 4 in dark clouds \citep{hirota98}.The abundance of the HNC molecule decreases with increasing gas temperature. \citet{hirota98} suggest that this may be due to the temperature dependence of neutral reactions, which, for temperatures exceeding 24 K, selectively destroy HNC in favour of HCN. Bright HNC emission is commonly observed in extragalactic objects \citep[e.g., ][]{aalto02,wang04,meier05,perez07}. In particular, over-luminous HNC J=3--2 is found in LIRGs \citep[e.g., ][]{aalto07,costagliola11}, where gas temperatures, derived by the IR dust continuum and mid-IR molecular absorption, are usually higher than 50~K and can reach values as high as a few 100 K. Here, ion-molecule chemistry in PDRs may be responsible for the observed ratios. \citet{meijerink07} find that the HNC/HCN J=1–0 line ratio is enhanced in PDRs and can reach a maximum value of one for H$_2$ column densities exceeding 10$^{22}$~\cmt.  

In NGC~4418 we find HNC/HCN line ratios of 0.7 and unity for the J=1--0 and J=3--2 transitions, respectively. These values are in agreement with a previous single-dish extragalactic survey by \citet{costagliola11}, which finds and average HNC/HCN J=1--0 of 0.5, with a maximum value of one, for a sample of 23 galaxies, including LIRGs. The observed ratios can be easily explained by  PDR chemistry  \citep[][]{meijerink05,meijerink07}.

Although the line ratios of HCN, HNC and HCO$^+$ observed in NGC~4418 are compatible with PDR models, we advise extreme caution when interpreting these results. Extragalactic surveys have shown that these line ratios show only little variations, of the order of a few, across different galaxy types and their interpretation is still controversial \citep[e.g.,][]{kohno2001,aalto07,meijerink07,costagliola11}. Recent studies \citep[e.g., ][]{meijerink2011,kazandjian2012,kazandjian2015} show that including mechanical heating (e.g., from star-formation or AGN feedback) and cosmic rays (e.g., from supernovae) in a PDR network can change the line ratios of HCN, HNC and HCO$^+$ by more than one order of magnitude. The observed values may be thus due to a combination of factors which are very difficult to disentangle in spatially unresolved observations.

\subsection{Chemical composition of NGC~4418}

\begin{figure*}[ht]
\centering
\includegraphics[width=.9\textwidth,keepaspectratio]{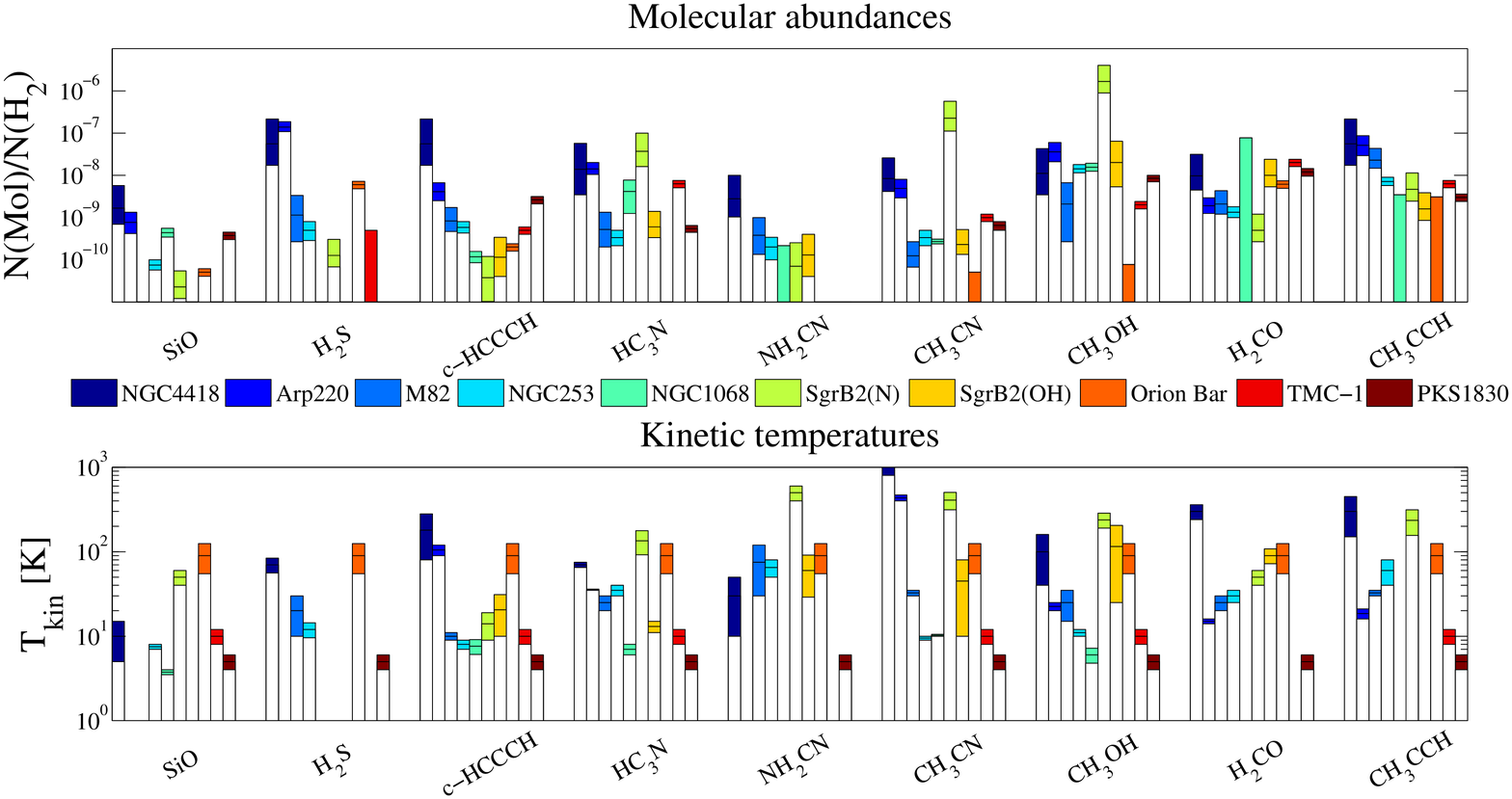}
   \caption{ \label{fig:abgal} Molecular abundances and kinetic temperatures for Galactic and extragalactic objects. Data for Sgr~B2(N), Sgr~B2(OH), the Orion Bar, TMC-1, M~82, and NGC~253 were taken from \citet{martin06}, and references therein. Data for Arp~220, NGC~1068, and PKS~1830-211 come from spectral scans, respectively by \citet{martin2011}, \citet{aladro_1068}, and \citet{muller_pks}. The color bars represent 1-$\sigma$ confidence intervals. In case no confidence interval was reported in the literature, a 20~\% uncertainty was assumed. Filled bars represent upper limits. Abundances are reported in Table \ref{tab:abtab}}
\end{figure*}

 \begin{figure*}[ht]
\centering
\includegraphics[height=.42\textwidth,keepaspectratio]{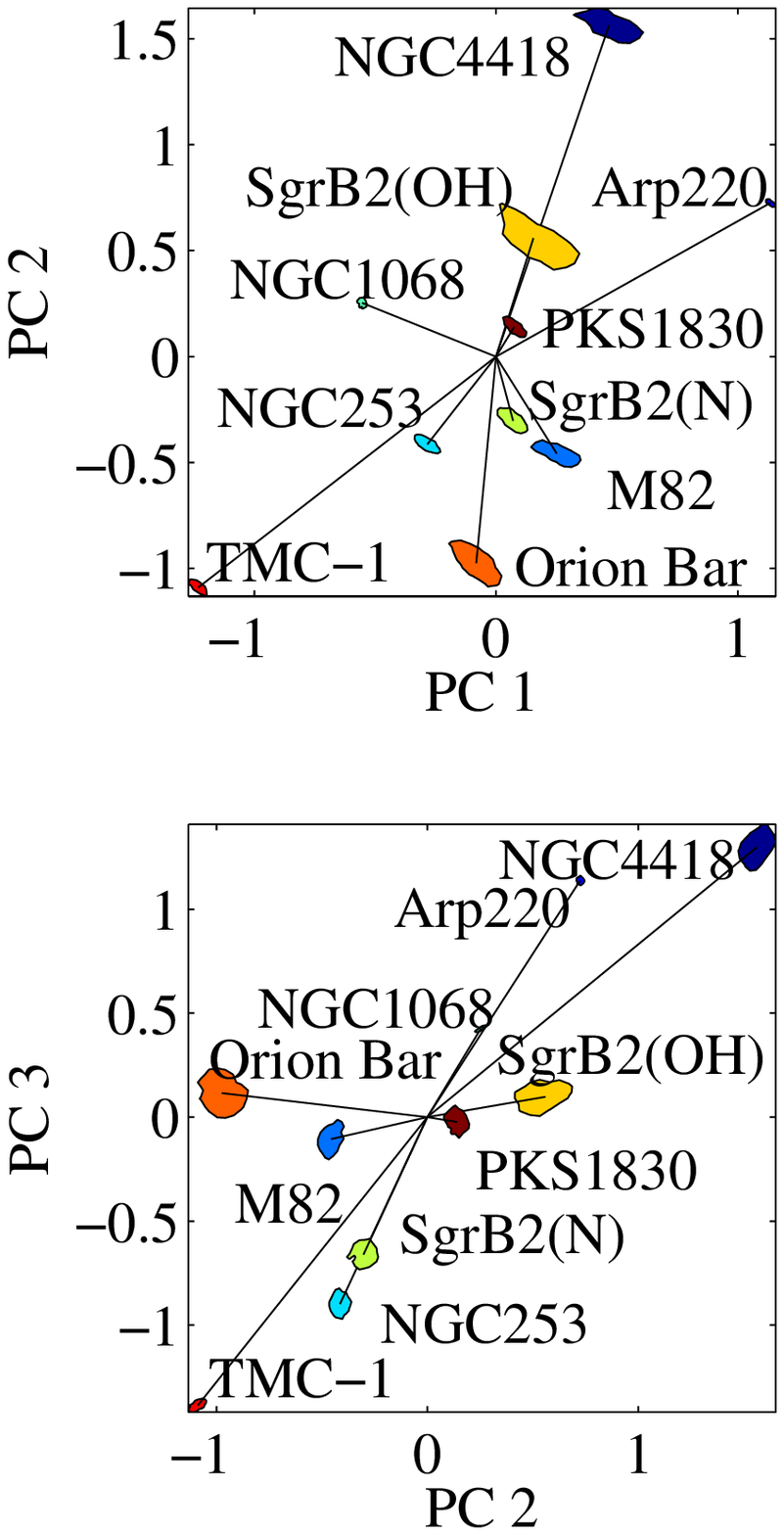}
\includegraphics[height=.42\textwidth,keepaspectratio]{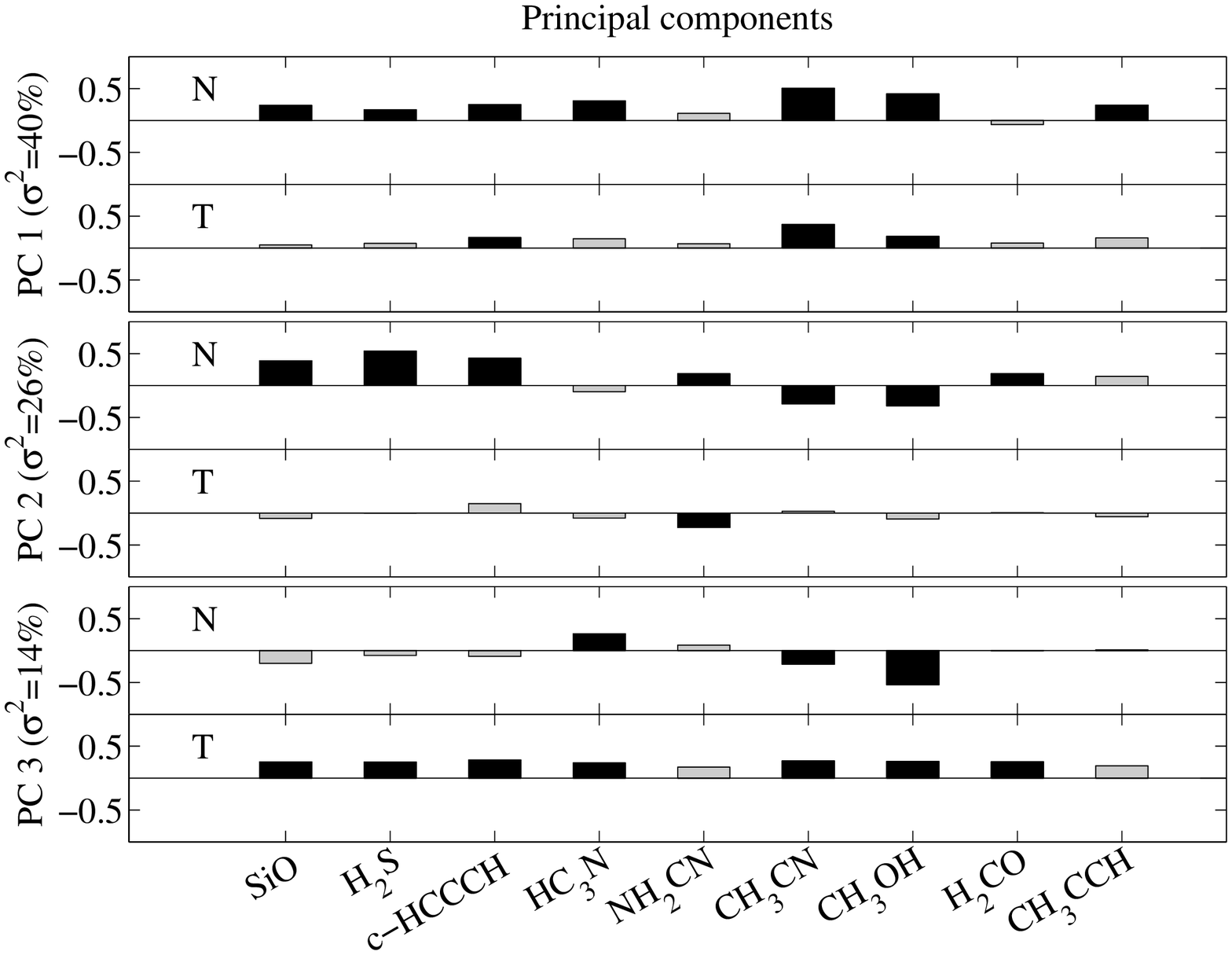}
   \caption{ \label{fig:pc} Principal component analysis of molecular abundances. {\it Left:} Projection of the molecular abundances of Fig. \ref{fig:abgal} onto the first three principal components (PC). For each source, the 1-$\sigma$ confidence interval is shown as a shaded area. The color legend is the same as in Fig. \ref{fig:abgal}. {\it Right: }Relative contribution of molecular abundances and kinetic temperatures to the first three principal components. For each PC, the upper and lower panels show contributions from abundance and temperature, respectively. The variables contributing for up to 75~\% in total to each PC are highlighted in black. The fraction of the total variance ($\sigma^2$) represented by each PC is shown in parentheses.}
\end{figure*}

In Fig \ref{fig:abgal} we compare the molecular abundances found for NGC~4418 with those observed in other Galactic and extragalactic objects. We choose to include in the analysis only molecules with low to moderate opacities, and for which reliable abundances have been reported in the literature for most of the sources.  In total we consider nine molecular species: SiO, H$_2$S, c-HCCCH, HC$_3$N, NH$_2$CN, CH$_3$CN, CH$_3$OH, H$_2$CO, and CH$_3$CCH. Their abundances, together with the relevant references, are reported in Table \ref{tab:abtab}. For NGC~4418 the abundances were calculated assuming a molecular hydrogen column of 0.7--2.9$\times$10$^{24}$~\cmt, as derived by \citet{costagliola2013} from CO 2--1 observations.

Our analysis includes well known Galactic objects, for which extensive literature is available, and six extragalactic sources for which mm/submm spectral scans have been performed. The Galactic objects span a wide range of physical conditions, including  a {\it hot-core} (Sgr~B2(N)), a typical {\it nuclear bulge cloud} (Sgr B2(OH)), the Orion Bar {\it photo-dissociation region}, and the {\it dark cloud} TMC-1 \citep[][ and references therein]{martin06}. The extragalactic sources also span a wide range of galaxy types, including the {\it ULIRG} Arp~220 \citep{martin2011}, the {\it evolved starburst} M~82, the {\it young starburst} NGC~253 \citep{martin06}, and the {\it Seyfert-2} NGC~1068 \citep{aladro_1068}. We also include the abundances derived in absorption by \citet{muller_pks} in the diffuse ISM of a z=0.89 galaxy located in the foreground of the blazar PKS~1830-211. 

The abundances and kinetic temperatures derived for each source are shown in Fig. \ref{fig:abgal}. A significant scatter in molecular properties is evident, even when considering the large uncertainties in the fit (color bar in the graph). Differences as high as four orders of magnitude are found, much higher than the typical differences in line ratios found for standard dense gas tracers, such as HCN, HNC and HCO$^+$, by extragalactic surveys \citep[factors of a few, e.g., ][]{krips08,baan08,costagliola11}. 

It is evident that molecular abundances in NGC~4418 are very similar to those found in the much more massive and luminous ULIRG Arp~220. The kinetic temperatures of most molecular species are of the order of 100~K, with complex molecules as CH$_3$CN and CH$_3$CCH having T$_\mathrm{kin}>$300~K. These temperatures are of the same order of magnitude of the ones found in hot cores and PDRs in the Galaxy, but in general higher than those found for the molecular gas of starburst galaxies.

\subsubsection{Principal component analysis}
In order to obtain a statistical description of the molecular properties of the sample, we performed a principal component analysis \citep[e.g., ][]{pca86}. This is a standard procedure used to reduce the dimensionality of a problem and to find trends in multivariate data, and has been applied to extragalactic molecular data by various authors \citep[e. g., ][]{meier05,costagliola11,meier12}. The uncertainty in the column density (N) and kinetic temperature (T) fit was taken into account by producing 1000 virtual samples for each source, with N and T values uniformly distributed inside their confidence intervals. Assuming the abundance and kinetic temperature of all the analyzed molecular species as starting coordinates, an orthogonal set of principal components (PC) was derived following the procedure described in \citet{costagliola11}. As a result, we obtained the contribution of each observable to the nine PCs, and the projection of the sources on the PC basis. The uncertainties in the N and T fit result in a spread in the projected values, which is represented by the shaded areas in  Fig. \ref{fig:pc}({\it left}).

The first three PCs contain 80\% of the variance in the sample, the contribution of each of the molecular properties being summarized in Fig.~\ref{fig:pc}({\it right}). The first principal component (PC1) is dominated by temperature and abundance of dense gas tracers, efficiently separating the cold gas of TMC-1 and the dense, warm environment of Arp~220 (see Fig.~\ref{fig:pc}, {\it upper left}). The second component (PC2) is dominated by SiO, H$_2$S, and c-HCCCH abundances, which appear to be anti-correlated with CH$_3$CN and CH$_3$OH abundances. Along this component, NGC~4418 has the highest projected value, while TMC-1 and the Orion Bar lie at the opposite end of the scale. The third component (PC3) is characterized by high HC$_3$N and low CH$_3$OH abundances, and high kinetic temperatures for all the considered tracers. This component efficiently  separates NGC~4418 and Arp~220 from the rest of the Galactic and extragalactic objects.

\subsection{A new CON chemistry?}

Our PC analysis suggests that NGC~4418 and Arp~220 have molecular abundances and excitation which set them apart from all well-studied Galactic and extragalactic environments. Both objects are characterized by high kinetic temperatures (T$_\mathrm{k}>$100~K) and high abundances of HC$_3$N, SiO, H$_2$S, and c-HCCCH. Compared to Galactic hot cores, the two sources show CH$_3$OH abundances lower by more than one order of magnitude. 

Both galaxies have very compact ($<$100 pc) and warm ($>$100~K) molecular cores, and have been suggested to harbor an obscured AGN and/or an extreme compact starburst \citep[e.g., ][]{spoon2001,spoon2004,imanishi04,iwasawa2005,downes07,aalto09,batejat11,costagliola2013,sakamoto2013, bmunoz2015}. It is thus possible that the combination of strong radiation fields, extreme obscuration, and compactness of the molecular core are producing a peculiar chemistry which has not yet being investigated in detail. Indeed, current chemical models of PDRs, XDRs, and hot--cores struggle to reproduce the molecular abundances observed in the compact cores of LIRGs and ULIRGs \citep[e.g., ][]{krips08,costagliola11,viti2014}.

{\it HC$_3$N and c-HCCCH: Hot gas chemistry?} In general, high HC$_3$N abundances are associated with warm gas, shielded from UV radiation, typical of Galactic hot cores. In standard models, the molecule forms in the gas phase from acetylene and it is easily destroyed by ions and UV photons \citep[C$_2$H$_2$+CN$\rightarrow$HC$_3$N+H, e.g., ][]{turner98}. This is in agreement with observations of IC~342, where \citet{meier05} find that HC$_3$N  anti-correlates with PDRs. However, the molecule shows bright emission in dust-obscured LIRGs \citep{lindberg2011,costagliola11}, which are thought to be powered either by a compact starburst or an AGN. Furthermore, recent observations in the Seyfert galaxies  NGC~1097 \citep{martin2015} and NGC~1068 \citep{takano2014} reveal bright HC$_3$N emission concentrated in the central 100~pc around the AGN (circumnuclear disk, CND) and not in the more extended starburst ring. How the molecule can survive in such energetic environments is somewhat puzzling. A possible explanation may come from recent chemical models by \citet{harada10} and \citet{harada2013}, which show that an enhancement of HC$_3$N is possible in an AGN torus, thanks to hot ($>$300~K) gas phase chemistry. Abundances up to 10$^{-6}$ can be reached in the plane of the disk, where the molecule is shielded from the ionizing radiation by large gas columns (i.e., N$>$10$^{24}$~\cmt). Such a hot gas phase could also explain the high c-HCCCH abundance observed in NGC~4418 \citep{harada10}. 

{\it CH$_3$OH, SiO, and H$_2$S: Shocks and XDRs ?} Methanol forms on dust grains and is easily destroyed by UV for Av$<$5~mag and by ions such as C$^+$, He$^+$ \citep[e.g.,][]{turner98}. It is observed in the gas phase after evaporation from dust mantles, and it is usually associated with young star formation (e.g., Galactic hot-cores). The hot-core Sgr~B2(N) is a clear example, having a CH$_3$OH abundance which is more than ten times higher than that found in all other sources in our sample (see Fig. \ref{fig:abgal}). If the central 100~pc of NGC~4418 were dominated by a young compact starburst \citep[$<$10~Myr, ][]{sakamoto2013}, we should observe an enhancement of CH$_3$OH when compared to more evolved starbursts. The observed low abundance seems to be at odds with the young starburst scenario. 

However, the nearby starburst galaxy Maffei~2, \citet{meier12} find that CH$_3$OH strongly correlates with SiO emission, and anti-correlates with sites of massive star formation.  The SiO molecule is liberated in the gas phase by the sputtering of dust particles due to strong shocks. The authors thus suggest that the two molecules may be tracing shocks rather than star formation.  

The presence of strong shocks in the nuclei of NGC~4418 and Arp~220 is supported by the high H$_2$S abundance. In their spectral scan with the SMA, \citet{martin2011} find it to be higher in Arp~220 than in the starburtst NGC~253 by about two orders of magnitude. They interpret this overabundance to be the result of grain disruption by shocks, and subsequent injection into the gas phase. This scenario is supported by the detection of a SiO outflow in the western nucleus of Arp~220 by \citet{tunnard2015}. 

Since liberation of methanol from the grain mantles does not require sputtering, the relative abundances of SiO and CH$_3$OH observed in NGC~4418 and Arp~220 may depend on the strength of the shocks. The two molecules have very different dissociation rates when exposed to ionizing radiation, the rate of CH$_3$OH being more than ten times higher than for SiO \citep[e.g., ][]{sternberg:1995aa}. Therefore, an alternative explanation of the observed abundances may be the dissociation of CH$_3$OH by X- or cosmic rays. 

In this scenario,  HC$_3$N should be also dissociated by the hard radiation and one would not expect such a high HC$_3$N/CH$_3$OH abundance ratio. However, methanol forms mainly on dust grains and has no efficient gas-phase formation route, while HC$_3$N can be formed efficiently by hot gas phase reactions. The observed ratio may thus be a result of the different gas formation efficiencies when the dust grains have been destroyed by shocks.

In summary, the common chemistry of NGC~4418 and Arp~220 resembles that observed in Galactic hot-cores, but with a much lower methanol abundance and higher SiO. {\it Young, dust-embedded star formation alone cannot explain the observed abundances and some kind of additional ionizing radiation is needed, together with mechanical feedback either from star formation or from an AGN.} Dissociation of CH$_3$OH may be due either to X-rays from an AGN or cosmic rays from supernova remnants. If a starburst is powering the IR emission of NGC~4418, it must be in a young pre-supernova stage \citep[$<$10~Myr][]{sakamoto2013}, and it is unclear whether the low supernova rate could sustain the cosmic ray flux needed to dissociate methanol. As an alternative, our observations are compatible with a composite starburst/AGN system in the inner 20~pc of the galaxy.

\section{Summary and Conclusions}

We report the results of a 70.7~GHz-wide spectral scan of the LIRG NGC~4418 with ALMA Cycle~0 observations in bands 3, 6 and 7. Our spectral scan confirms that the chemical complexity in the nucleus of NGC~4418 is one of the highest ever observed outside our Galaxy. We identify 317 emission lines above the three sigma level from a total of 45 molecular species, including 15 isotopic substitutions and six vibrationally excited variants. 

We find that the molecular emission accounts for 15$\%$ of the total flux in band 3, and for 27$\%$ in bands 6 and 7. Our observations clearly show that line contamination can have a serious impact on studies continuum emission in (U)LIRGs. This is especially true for galaxies with broad lines, where the contamination from blended molecular emission could amount to up to 30$\%$ of the flux.

A combined LTE/NLTE fit of the spectrum reveals a multi-phase ISM, with temperatures ranging between 20 and more than 500~K, and densities between 10$^4$ and 10$^7$~\cmth. The spectrum of NCG~4418 is dominated by bright vibrationally excited HC$_3$N, HCN , and HNC, with vibrational temperatures exceeding 350~K. We interpret this emission as the signature of an optically-thick, compact IR source ($<$5~pc of effective area) with brightness temperature $>$350~K in the core of the galaxy. We suggest that the presence of bright emission from vibrationally excited states of HC$_3$N, HCN, and HNC may be a good tracer of dust-embedded compact sources in extragalactic objects.

We compare the molecular abundances derived by our fit with those found in other Galactic and extragalactic environments by means of a principal component analysis. We find that NGC~4418 and the ULIRG Arp~220 have similar molecular abundances and excitation, which set them apart from the other sources in the sample. {\it We interpret this as the signature of a common, peculiar chemistry, which may be typical of the compact obscured nuclei of (U)LIRGs}. 

While the line ratios of standard dense gas tracers such as HCN, HNC, and HCO$^+$ can be reproduced by PDR models, other more complex molecules such as HC$_3$N and c-HCCCH require either a hot-core chemistry or hot gas-phase reactions to be explained. The relative abundances of SiO and CH$_3$OH seem to require a mix of strong shocks and dissociation by X- or cosmic-rays. The inferred chemistry, together with the strong vibrationally excited emission and compactness of the IR core, is consistent with a composite starburst/AGN system.

Compared to Arp~220, NGC~4418 has a similarly bright and complex molecular spectrum, but much narrower line widths, which greatly reduce blending and simplify the identification of spectral features. The galaxy is thus the ideal target for future studies of molecular chemistry and excitation in IR-bright, obscured sources.

\begin{acknowledgements}
This paper makes use of the following ALMA data: ADS/JAO.ALMA\#2011.0.00820.S . ALMA is a partnership of ESO (representing its member states), NSF (USA) and NINS (Japan), together with NRC (Canada) and NSC and ASIAA (Taiwan), in cooperation with the Republic of Chile. The Joint ALMA Observatory is operated by ESO, AUI/NRAO and NAOJ.
\end{acknowledgements}


\bibliographystyle{aa} 
\bibliography{bibliototale}

\begin{appendix}

\section{Fit details for individual molecules}

\label{sec:fitdet}
{\bf $^{13}$CO, \textit{C$^{18}$O}:} Carbon monoxide was detected in the J=1-0 transitions of the two isotopic variants $^{13}$CO and C$^{18}$O. The latter is only a tentative detection, the peak flux being only at the 2-sigma level. Since we only detected one line per species, we could not fit an excitation temperature to the emission. By fixing the excitation temperature to 70~K, which is the lower limit to the CO J=2-1 brightness temperature detected by \citet{costagliola2013}, a column density of the order of 10$^{18}$ and 10$^{17}$~\cmt~is found for  $^{13}$CO and C$^{18}$O, respectively.

{\bf CS, $^{13}$CS, \textit{C$^{33}$S}, C$^{34}$S: }
We detect the J=2-1 and J=6-5 rotational transitions of carbon monosulfide and its $^{13}$C, $^{33}$S, and $^{34}$S isotopic variants. For CS, both the population diagram analysis and the LTE fit find an excitation temperature of $\sim$20 K, and a column density of 6$\pm$0.6$\times$10$^{16}$ \cmt. In order to calculate isotopic ratios, we fix the excitation temperature of the isotopic variants at 20~K. This implies the emission to be co-spatial and the excitation to be similar for all isotopic variants. Under this assumption, the LTE fit gives column densities of the order of 10$^{15}$~\cmt for $^{13}$CS and C$^{33}$S, and of the order of 10$^{16}$~\cmt for C$^{34}$S. An NLTE fit of CS emission was also performed. We find that the emission is best fit by a model with hydrogen density \hdens=10$^5$~\cmth, kinetic temperature \tkin=50 K, and column density $N$=4$\times$10$^{17}$~\cmt. The NLTE values for \tkin~and $N$ are significantly higher than the LTE results. This may be due to the high critical density of the J=6-5 line, which would require \hdens$>$10$^{7}$~\cmth~to be efficiently excited by collisions.  

{\bf CN:} We detect the CN multiplet at 3~mm, including nine merged lines spanning from 113.12 to 113.52 GHz in rest frequency. The lines have very similar upper-state energies ($\sim$5~K) and do not allow for a reliable fit. In order to obtain a lower limit to the molecule's column density, we performed an LTE fit with a fixed temperature of \tex=70~K. The resulting column is $N\sim$5\tensevt~\cmt. As discussed for $^{13}$CO and C$^{18}$O, the \tex value was chosen as the lower limit to the observed CO brightness temperature. 

{\bf NS:} We detect the strongest lines of the J=11/2-9/2 multiplet of nitric sulfide at rest frequencies 253.57-255.6 GHz. Even if so close in frequency, the detected lines span a wide interval in upper state energies ($E_\mathrm{u}$=40-360~K). We derive an excitation temperature of \tex=350$\pm$150 and a column density of $N$=8$\pm$4\tensixt \cmt. The large uncertainties on the derived quantities are due to the line blending in band~6.

{\bf SO, \textit{$^{34}$SO}:} Emission from sulfur monoxide is heavily blended, the only sure detections being the lines 3$_2$-2$_1$ (99.3 GHz), 2$_3$-1$_2$ (109.25 GHz), and 6$_6$-5$_5$ (258.255). Because of the strong blending, the $\chi^2$ surface of our LTE fit is not well defined, giving only a lower limit for the excitation temperature of \tex$>$20~K. The fitted value for the column density is relatively well defined, $N$=4$\pm$3\tenfift~\cmt. We also performed a NLTE fit of the isolated lines, resulting in hydrogen densities between 10$^6$ and 10$^7$~\cmth, kinetic temperatures of 50-200~K, and column densities from 2\tenfift~ to 2\tensixt~\cmt. The isotopic variant  $^{34}$SO is only tentatively detected as two highly blended lines at 251.4 and 255.1~GHz. The LTE fit with a fixed excitation temperature of 20~K results in a poorly constrained column density of 10$^{15}$-10$^{16}$~\cmt.

{\it\bf SiO, $^{29}$SiO, $^{30}$SiO :} Strong silicon monoxide emission is detected in band 3 (J=2-1) and band 6 (J=5-4). The two rotational lines are free of blending and well identified. Our LTE fit gives an excitation temperature of 10$\pm$5~K and a column density of 3$\pm$1\tenfift~\cmt, comparable with what found with the population diagram analysis. The NLTE fit results in an hydrogen density lower than 10$^6$~\cmth, a kinetic temperature greater than 20~K, and a column density greater than 10$^{16}$~\cmt. We also detect the isotopic variants $^{29}$SiO and $^{30}$SiO. The emission lines from these two species are strongly blended, but still clearly identified. In our LTE fit we fix their excitation temperature to that of SiO, i.e. \tex=10~K. The resulting column density estimation is of the order of 10$^{15}$~\cmt, with large uncertainties ($>$50\%) due to the line blending.

{\it\bf HCN, H$^{13}$CN:} Hydrogen cyanide is detected in its J=1-0 and J=3-2 transitions. The J=3-2 line is the strongest molecular line detected in our scan, with an integrated flux density of 50$\pm$5 Jy~\kms, and a peak brightness temperature of 35~K. Both emission lines have $FWHM>$140~\kms, broader than than the average 120~\kms assumed in our fit.  The derived properties are thus representative only of the line core. The LTE fit gives \tex=7$\pm$1~K and $N$=2$\pm$1\tensixt~\cmt, while the NLTE analysis results in a hydrogen density lower than 10$^6$~\cmth, \tkin$>100$, and $N>$10$^{16}$~\cmt. The large discrepancy in excitation temperature between the two fitting methods is most likely due to a limitation of the LTE fit, which gives unreliable results for high opacities. An excitation temperature of 8~K seems too low when compared with the CO brightness temperature of $>$70~K reported by \citet{sakamoto2013} and \citet{costagliola2013}, and may be an indication of sub-thermal excitation of HCN. The brightness temperature for a source size of 0$''$.4 is $\sim$35~K and $\sim$65~K for the two lines respectively (see Table \ref{tab:popdiag}). This sets a lower limit for the excitation temperature of the molecule, and thus to the kinetic temperature of the gas, of at least 65~K. For these reasons, we will assume the NLTE fit to best represent the HCN emission in NGC~4418. 

The  H$^{13}$CN J=1-0 line was also detected. A column density estimate of 2$\pm$1\tenfift~\cmt was obtained by fixing the excitation temperature at 7~K. If we assume an extragalactic $^{12}$C/$^{13}$C ratio between 40 and 100 \citep{henkel2014}, this correspond to a column densities of HCN of (0.8--2)$\times$10$^{17}$~\cmt.

{\it\bf HNC, HN$^{13}$C:} We detect the J=1-0 and J=3-2 rotational transitions of hydrogen isocyanide. Both the population diagram analysis and the LTE fit of the total spectrum result in an excitation temperature of 8-9~K, and column densities in the range 3-8\tenfift~\cmt. An NLTE fit of the emission was also performed, resulting in an hydrogen density  $<$10$^{6}$~\cmth, a kinetic temperature  greater than 100~K, and a column density $>$10$^{16}$~\cmt. The results of the LTE and NLTE fitting of HCN and HNC are very similar, including the discrepancy of the fitted excitation temperatures for the two methods. The LTE fit results in very low \tex, lower than the brightness temperature of the two detected lines of HNC (\tbr=45~K and 30~K for the J=1-0 and J=3-2 transitions, respectively). The NLTE fit results in \tex$\simeq$40~K, which is consistent with the brightness temperature of the lines for the assumed source size of 0$''$.4.
The J=1-0 transition of HN$^{13}$C is also detected. The LTE fit with a fixed excitation temperature of 8~K results in a column density of 1.5$\pm$0.5\tenfift~\cmt.

{\it\textbf{HCN,v2=1}, \textbf{HNC,v2=1}:} Vibrationally excited HCN and HNC are tentatively detected in our scan. For each molecule we only detect the J=3-2,v2=1f transitions, therefore we cannot produce a reliable fit to the rotational temperature. However, we can derive the vibrational temperature by comparing the intensity of the J=3-2 transitions of both the v2=1 and v=0 states. {\it We find that the relative population of the v2=1 and v=0 states is well described by a vibrational temperature of $\sim$350~K for HCN and 450~K for HNC.} When assuming these temperatures to be also the rotational excitation temperatures of the v2=1 emission, we obtain line intensities which are consistent with the observed spectrum.

{\it\bf HCO$^+$, H$^{13}$CO$^+$, \textit{HC$^{18}$O$^+$}:} The J=1-0 and J=3-2 transitions of HCO$^+$ (formylium) are detected in band~3 and band~6, respectively. The population diagram analysis and the LTE model fit result in an excitation temperature of 6-8~K and a column density of 1-6\tenfift~\cmt. The RADEX models that best fits the emission have \hdens$<$6$\times$10$^5$~\cmth, \tkin$>$100~K, and $N>$10$^{16}$~\cmt. For HCO$^+$ we find the same discrepancy between the LTE and NLTE fitted excitation temperatures that we discuss in the HCN and HNC case. The NLTE fit provides more physically reasonable results, consistent with an observed brightness temperature of the lines of $>$20~K (see Table \ref{tab:popdiag}).

We also detect the J=1-0 emission line of H$^{13}$CO$^+$. By fixing the excitation temperature of the molecule at 6~K, we obtain an LTE estimate of the column density of $N$(H$^{13}$CO$^+$)=7$\pm$4\tenfort~\cmt. If we assume an extragalactic $^{12}$C/$^{13}$C ratio between 40 and 100 \citep{henkel2014}, this corresponds to a column density of HCO$^+$ of  (0.8--2)$\times$10$^{17}$~\cmt. 

The HC$^{18}$O$^+$, J=3-2 transition is tentatively detected in band 6, with heavy blending. The LTE fit is highly degenerate, and only results in an upper limit of 20~K for the excitation temperature and a lower limit to the column density of 10$^{14}$\cmt.

{\it \textbf{H$_2$S}:} At least six transitions of hydrogen sulfide are included in the scan, but only one transition (2$_{(2,0)}$-2$_{(1,1)}$ at 216.71 GHz) is free of blending and clearly identified. By fixing the excitation temperature of H$_2$S at 70~K, the LTE fit results in a column density of 1$\pm$0.5\tensevt~\cmt. 

{\bf CCH:} We detect the N=1-0 multiplet of ethynyl at rest frequencies between 87.28 and 87.45 GHz (sky frequency $\sim$86.76 GHz). The six lines have very similar upper state energies ($\sim$4~K) and thus do not allow for an accurate fit of the excitation temperature. By fixing \tex=70~K we obtain an LTE estimate of the column density of 2$\pm$1\tensevt~\cmt.

{\bf HCS$^+$:} The J=2-1 and J=6-5 rotational transitions of HCS$^+$ (thioformylium) are detected in band~3 and band~6, respectively. The 1~mm line is heavily blended with emission from vibrationally excited HC$_3$N, while the 3~mm line is free of blending. The emission can be reproduced by our LTE model with \tex=20$\pm$10~K and $N$=8$\pm$4\tenfift.   

{\bf CCS:} We detect six emission lines of thioxoethenylidene, with upper energy levels ranging from 20 to 110 K. The emission is free of blending and is well fit by our LTE model. The best fit values are \tex=20$\pm$10~K and $N$=1.5$\pm$0.5\tensixt~\cmt.

{\bf N$_2$H$^+$ :} We detect the J=1-0 triplet at 93.17 GHz and the J=3-2 line of N$_2$H$^+$ (diazenylium). The triplet emission is totally blended, but free of contamination from other species. The LTE and NLTE fit procedures give very similar results (see Table \ref{tab:fit}) with \tex$\sim$30~K and $N\sim$5\tenfift~\cmt. The best fit hydrogen density in our NLTE analysis is 10$^7$~\cmth, which is higher than the critical density of the detected transitions ($n_\mathrm{c}\sim$10$^5$-10$^6$~\cmth), and the \tkin~and \tex~coincide for both lines. We conclude that the RADEX best fit represents an LTE excitation of the molecule.

{\bf H$_2$CO:} We detect eight transitions of formaldehyde, four from para-H$_2$CO and four from ortho-H$_2$CO. The LTE fit for both species results in an excitation temperature of 350$\pm$100~K and column a column density of 5-8\tensixt~\cmt. The NLTE analysis results in an hydrogen density of 10$^5$~\cmth, a kinetic temperature greater than 300~K and a column density of 7\tenfift-10$^{16}$~\cmt. In our NLTE fit we only obtain a lower limit to the kinetic temperature because the LAMBDA database does not contain collision coefficients for temperatures greater than 300 K. 

{\bf c-HCCCH:} We detect 15 emission lines of cyclopropenylidene at the 3-$\sigma$ level across ALMA bands 3 and 6. The detected lines have upper state energies ranging from 40 to 500~K.  The LTE model well fits the emission, resulting in  an excitation temperature of 180$\pm$100~K and a column density of 1$\pm$0.5\tensevt~\cmt. The large uncertainties are mostly due to the blending in band 6.

{\bf H$_2$CS, \textit{H$_2^{13}$CS}:} Three isolated emission lines of thioformaldehyde  are detected at sky frequencies 103.89, 268.66, and 276.95 GHz. Three more lines from H$_2^{13}$CS are tentatively detected ($\sim 2\sigma$) in band 3. The detected transitions of H$_2$CS have upper state energies ranging from 20 to 70~K. The LTE fit results in an excitation temperature of 35$\pm$10~K and a column density of 1$\pm$0.5\tensixt~\cmt. By fixing the excitation temperature at 35~K we derive an upper limit for the column density of H$_2^{13}$CS of 5$\pm$2\tenfift~\cmt.

{\bf HC$_3$N, H$^{13}$CCCN, HCC$^{13}$CN:} We detect bright cyanoacetylene emission in all three ALMA bands. Nine isolated lines are detected,  with upper state angular momentum ranging from 10 to 32, and upper state energies from 24 to 230~K  (see Table \ref{tab:popdiag}). The population diagram analysis and the LTE fit of the whole spectrum give an excitation temperature of $\sim$70~k and column densities in the range 2-9\tensixt~\cmt   (see Tables \ref{tab:popdiag}, \ref{tab:fit}). 
The H$^{13}$CCCN and HCC$^{13}$CN isotopomers are also detected in all three bands, for a total of nine isolated emission lines with upper state energies ranging from  20 to 240~K. We performed an LTE fit of both species, assuming the same excitation temperature of 70~K as HC$_3$N. The fitted column densities are very similar, in the range of 1-2\tenfift~\cmt, with uncertainties of the order of 50\%. 

{\bf Vibrationally excited HC$_3$N:} Emission from vibrationally excited HC$_3$N dominates the mm-wave spectrum of NGC~4418. We detect rotational transitions from the four vibrationally excited states HC$_3$N,v6=1, HC$_3$N,v7=1, HC$_3$N,v6=1,v7=1, and HC$_3$N,v7=2, for a total of 25 transitions. The upper state energies of such vibrationally excited lines are very high, ranging from 300 to 900~K. 
The emission from all four species is well fitted by our LTE model and by the rotational diagram analysis. We find similar excitation temperatures, in the range 70-100~K, and column densities ranging from 10$^{15}$-10$^{16}$~\cmt.
From the relative population of the vibrational levels it is possible to derive the vibrational temperature of the molecule via a population diagram analysis. The results are reported in Fig. \ref{fig:vib}. We are able to derive the vibrational temperature of three different angular momentum states (J=10,11,12) for which emission of more than two vibrational states was detected. The three fits give remarkably similar results, with vibrational temperatures ranging from 330 to 400~K.

{\bf CH$_2$NH:} We identify three spectral features as emission from methanimine at rest frequencies 105.79, 257.11, and 266.27 GHz, with upper state energies ranging from 30 to 60~K. The LTE fit of the spectrum results in an excitation temperature of 45$\pm$30~K and a column density of 1.5$\pm$0.5\tensevt~\cmt.

{\it \textbf{NH$_2$CN}:} Emission from cyanamide is tentatively detected in NGC~4418. Only one isolated line at sky frequency  98.62 GHz is unambiguously identified as the NH$_2$CN J=5-4 transition, while most emission across the observed ALMA bands is heavily blended. Our LTE fit results in an excitation temperature of 30$\pm$20~K and a column density of 5$\pm$2\tenfift~\cmt. However, the derived parameters are highly uncertain because of the blending of the lines at 1~mm. 

{\bf CH$_3$CN:} Bright emission from methyl cyanide is detected as four line forests at sky frequencies 91.3-91.35, 109.56-109.62, 254.87-255.78, 292.11-292.28 GHz. The detected lines have upper state energies ranging from 13 to 500~K, with bright emission from lines with $E_\mathrm{up}>$200~K. The LTE fit results in a very high excitation temperature of 1000~K, {\it but does not reproduce the intensity of the lines at 3~mm}. The NLTE procedure provides a better fit of the emission across all three ALMA bands, resulting in a low hydrogen density of 10$^4$~\cmth, and a high kinetic temperature $>$500~K, required in order to fit the high energy lines. The critical density of the observed transitions varies between 10$^5$~\cmth~for the lines at 3~mm, and 10$^7$~\cmth~for the lines at 1~mm. It appears that the only way to fit both the 3~mm and the high-energy 1~mm lines is to lower the hydrogen density until the 3~mm lines are not efficiently excited by collisions ($n$(H$_2$)$<n_\mathrm{c}$, i.e. they are not at LTE). This could imply that  CH$_3$CN emission is originating from a gas phase that is significantly more diffuse than for the other molecular tracers, or that the low- and high-excitation lines are emerging from different environments.

{\it \textbf{CH$_3$CCH}:} Methyl acetylene is tentatively detected in NGC~4418. The emission is heavily bended and no isolated line can be identified. Emission from the molecule is required in order to fit the line forests at sky frequencies 254-254.6, 271.2-271.6, and 288-288.6 GHz. The LTE fit results in an excitation temperature of 350$\pm$100~K, and a column density of 1$\pm$0.5\tensixt. We advise extreme caution when interpreting these results, since the strong blending makes the fit highly degenerate. 

{\bf HC$_5$N:} Emission from cyanobutadiyne is required in order to explain 11 spectral features across all three ALMA bands, which could not be associated to any other molecule. Nine of the detected lines are isolated lines free of blending with other species. The upper state angular momentum of the detected transitions ranges from 32 to 42, corresponding to energies above ground of 65 to 85~K. The population analysis and the LTE fit of the whole spectrum result in an excitation temperature of 60-70~K, and column densities in the range 0.5-2$\times$10$^{16}$~\cmt. 

{\bf \textit{CH$_3$OH}:} Methanol emission is needed in order to fit the highly blended region between sky frequencies 287.8 and  288.3 GHz. No other transition from the molecule is detected above the 3--$\sigma$ level across the whole observed frequency range. The NLTE fit results in a kinetic temperature of 100$\pm$60~K and a column density of 2$\pm$1\tensixt~\cmt, for an hydrogen density of 10$^7$~\cmth. The fit is poorly constrained because of line confusion. Non-LTE excitation of methyl acetylene could give a significant contribution to the emission in this region, thus the derived column density of CH$_3$OH should be considered as an upper limit.

\begin{figure*}[ph]
   \centering
   \includegraphics[width=.8\textwidth,keepaspectratio]{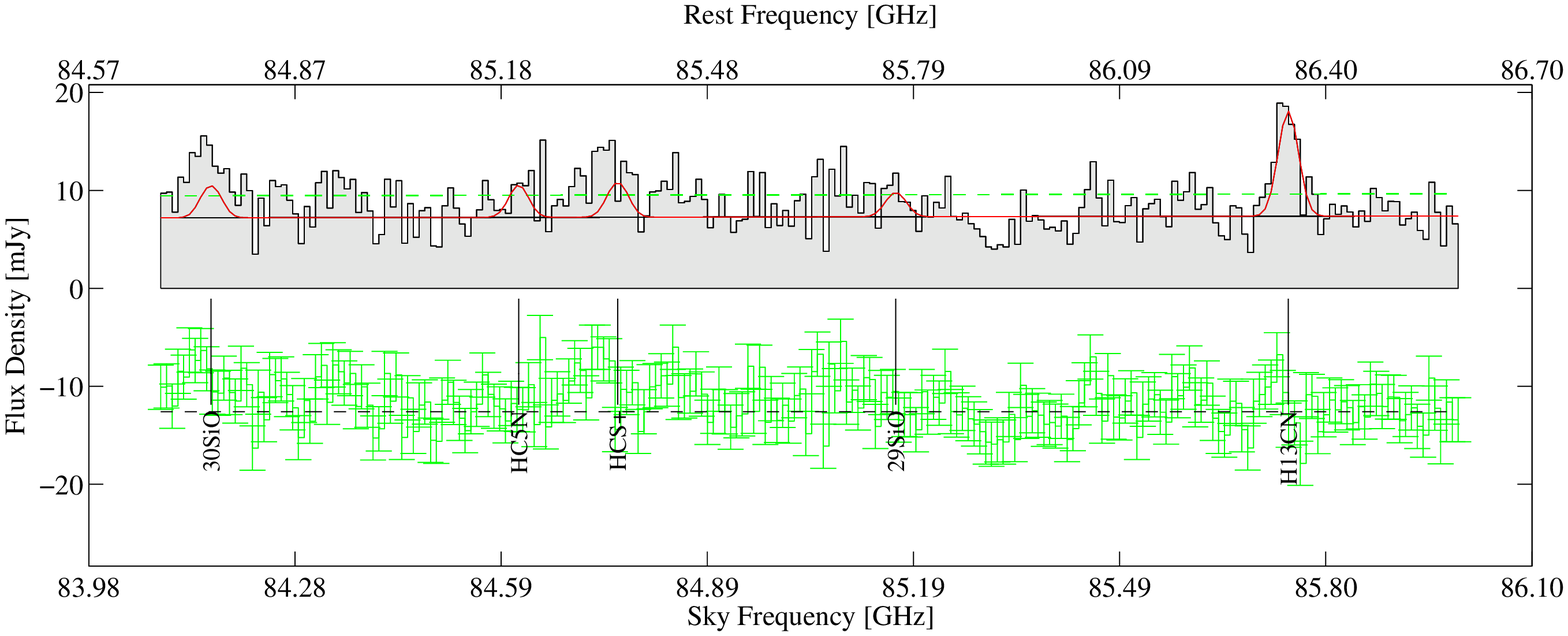}\\
   \includegraphics[width=.8\textwidth,keepaspectratio]{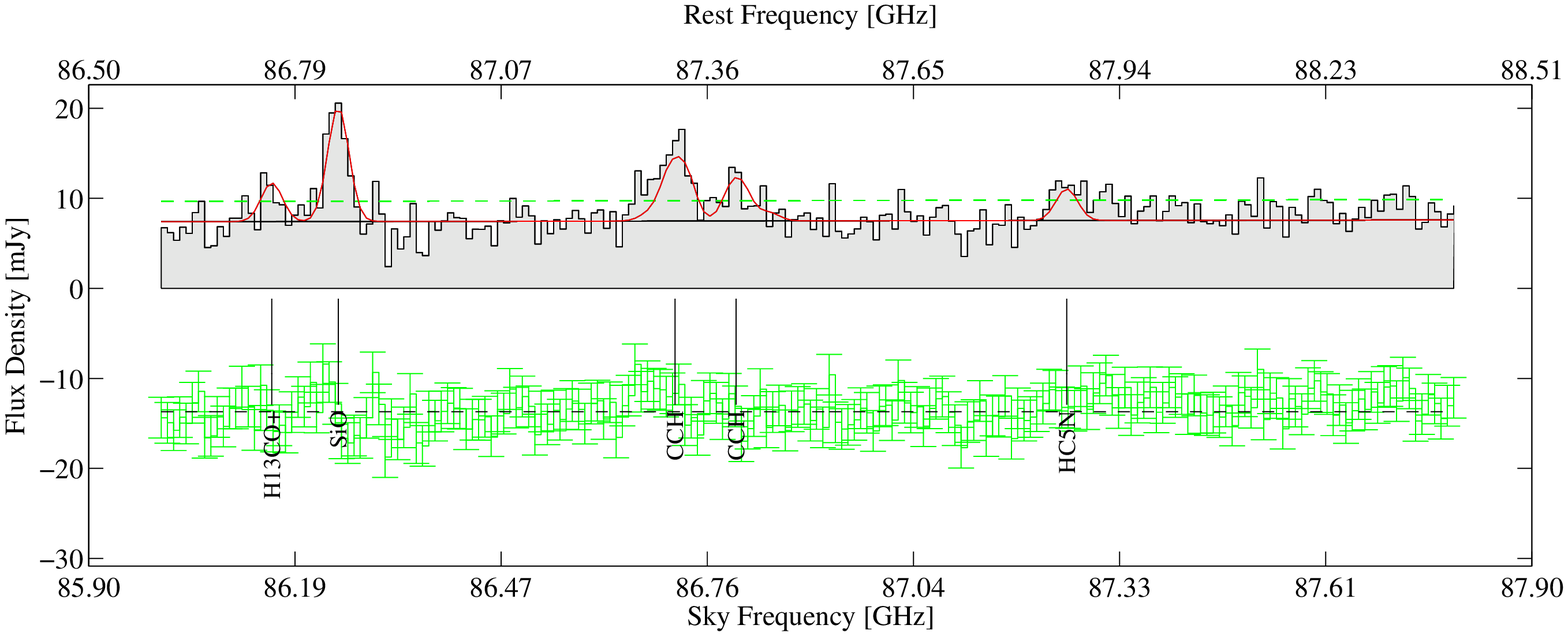}\\
   \includegraphics[width=.8\textwidth,keepaspectratio]{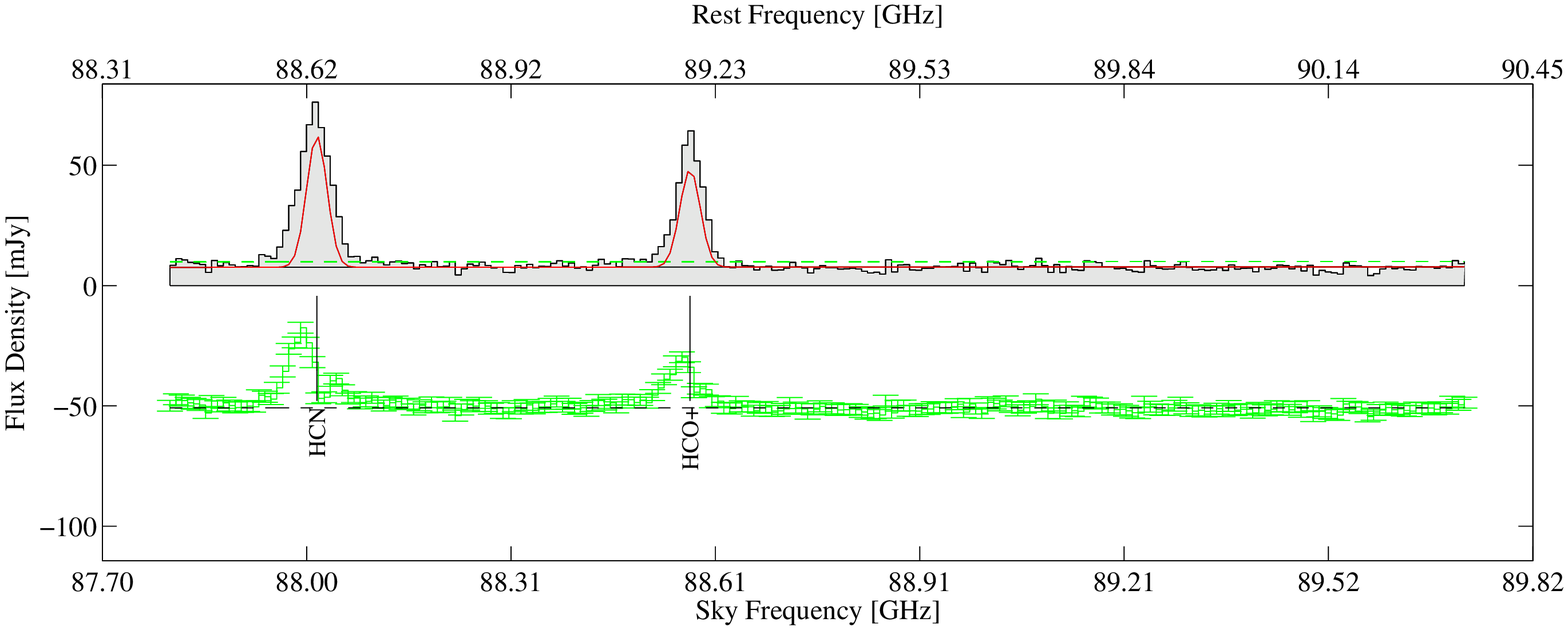}\\
   \includegraphics[width=.8\textwidth,keepaspectratio]{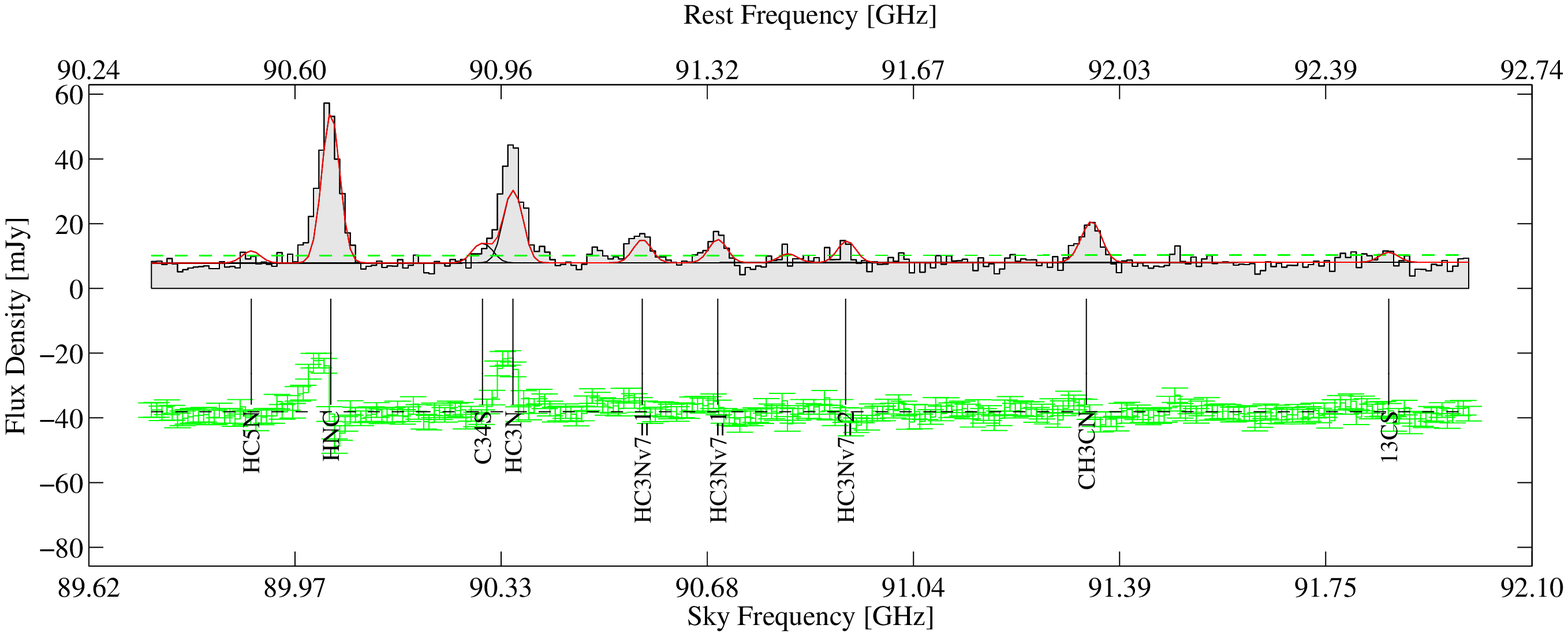}\\
      \caption{\label{fig:fit1} The ALMA spectral scan of NGC~4418. The data are shown as a shaded {\it histogram plot}, the best fit for each molecule is shown as {\it solid black lines}. The total fit to the data is shown as a {\it red solid line}. The 1-$\sigma$ level is shown as a dashed green line. The residuals, together with the data rms, are shown below the spectrum as a {\it green histogram plot} with error bars. Emission lines above 3-$\sigma$ are labeled.}   
              
    \end{figure*}
\clearpage

\begin{figure*}[ph]
   \centering
   \includegraphics[width=.8\textwidth,keepaspectratio]{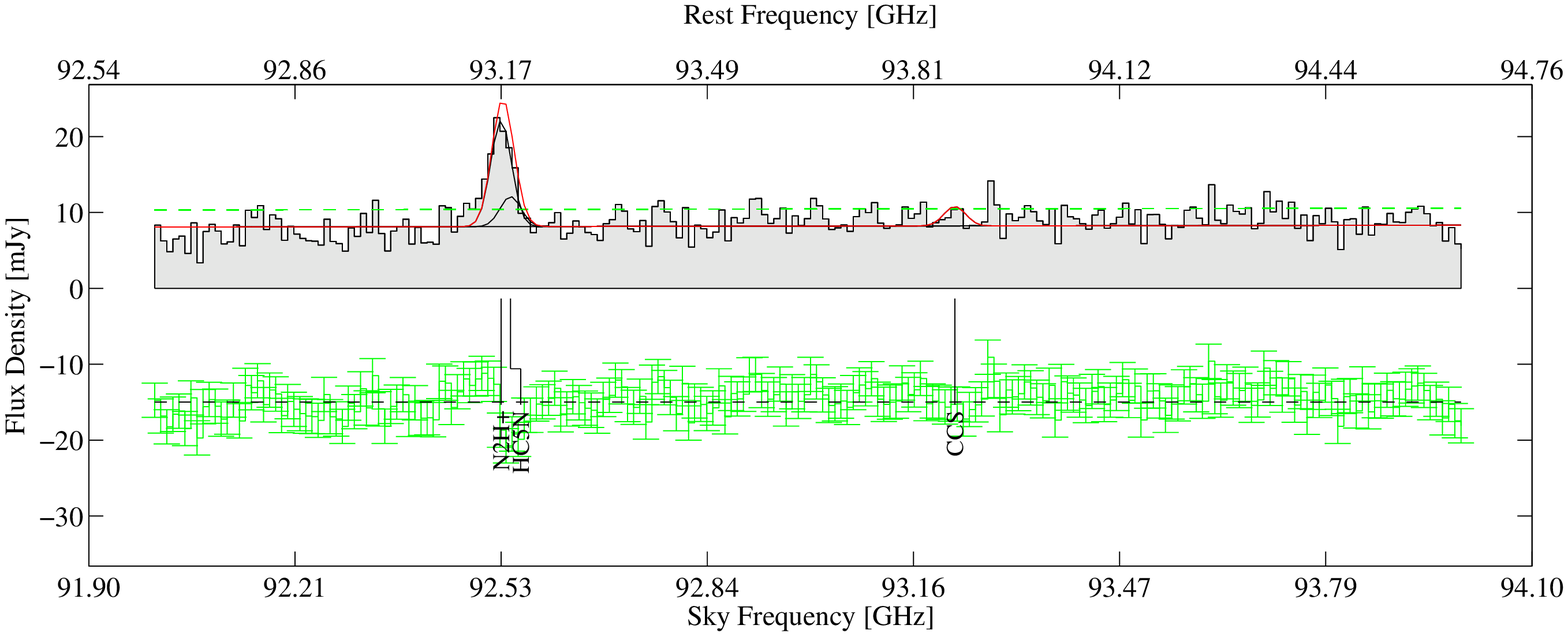}\\
   \includegraphics[width=.8\textwidth,keepaspectratio]{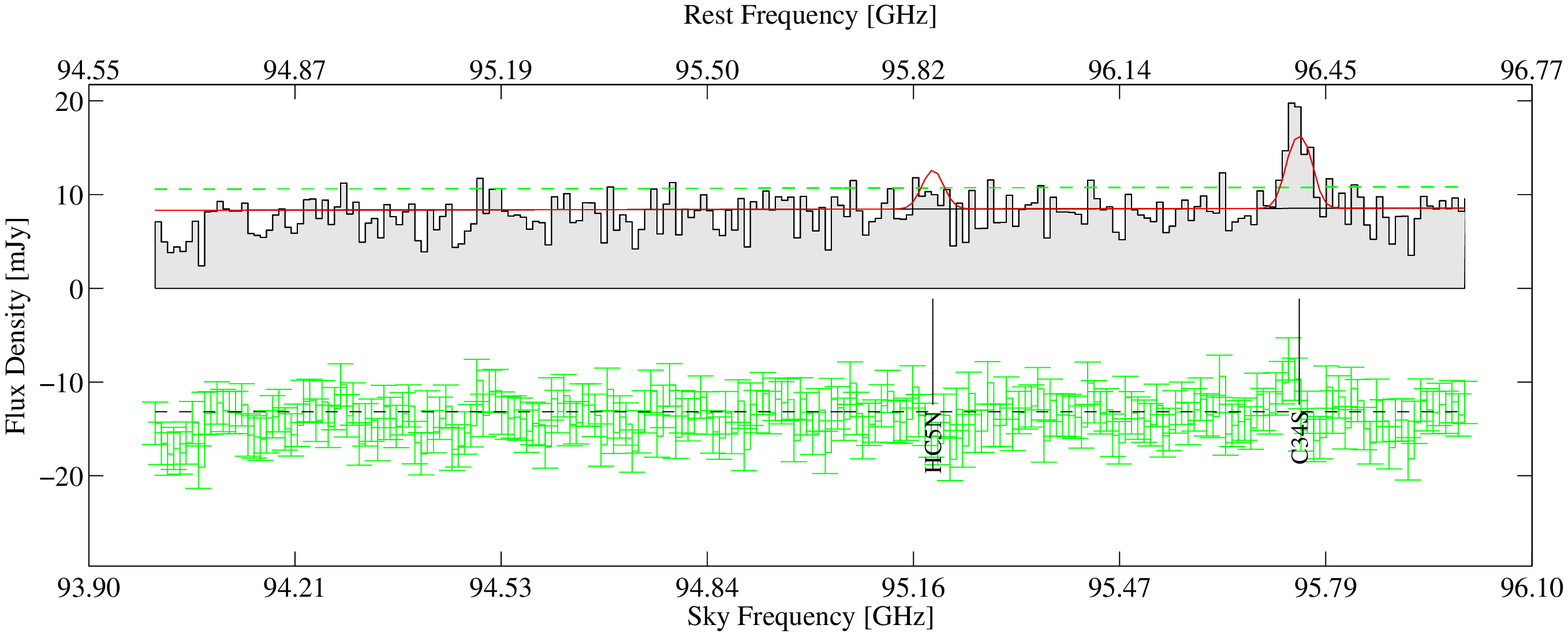}\\
   \includegraphics[width=.8\textwidth,keepaspectratio]{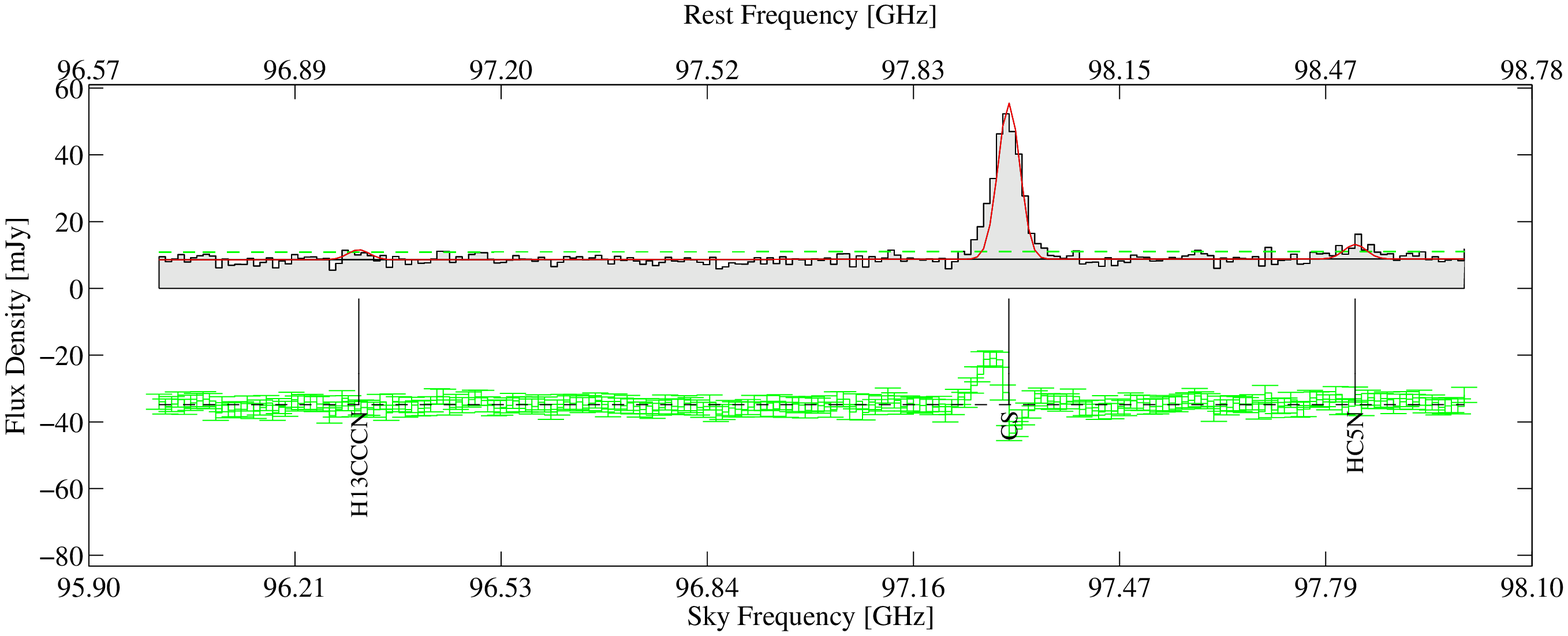}\\
   \includegraphics[width=.8\textwidth,keepaspectratio]{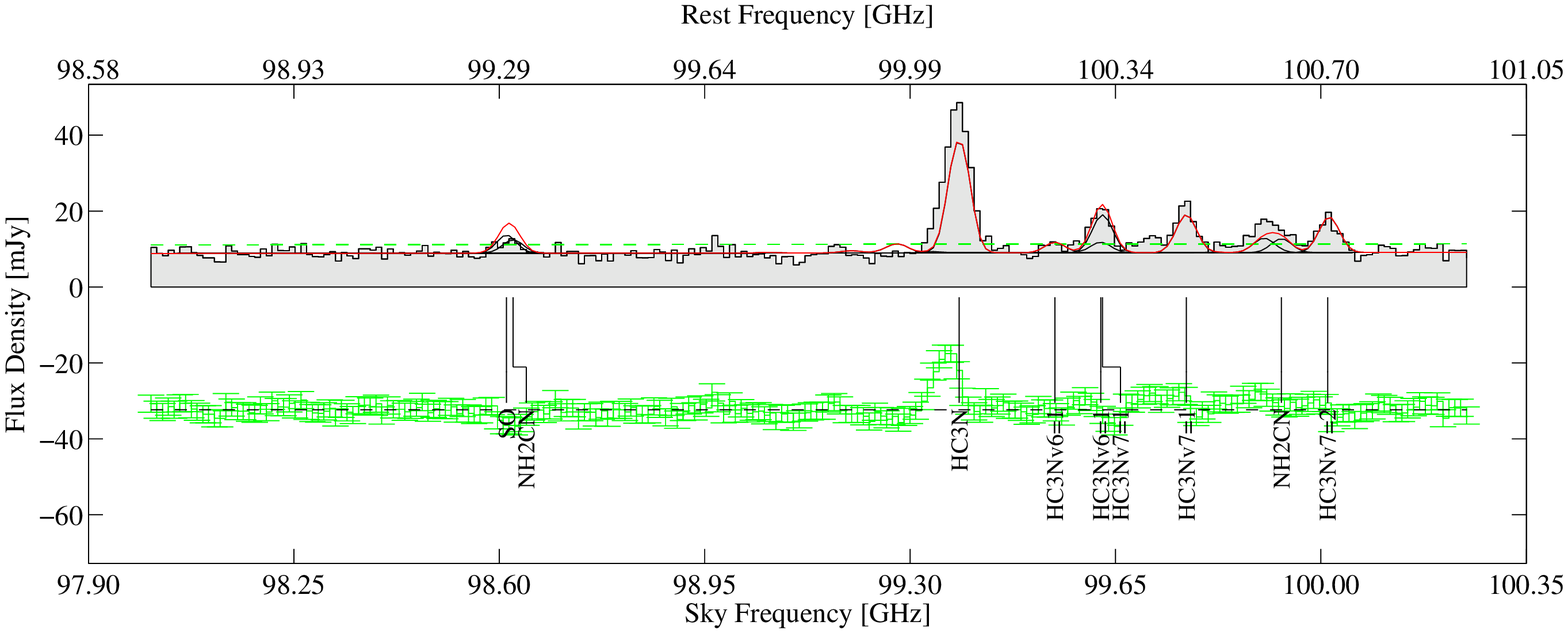}\\
      \caption{\label{fig:fit2} Continues from Fig. \ref{fig:fit1}.}
              
    \end{figure*}
\clearpage

\begin{figure*}[ph]
   \centering
   \includegraphics[width=.8\textwidth,keepaspectratio]{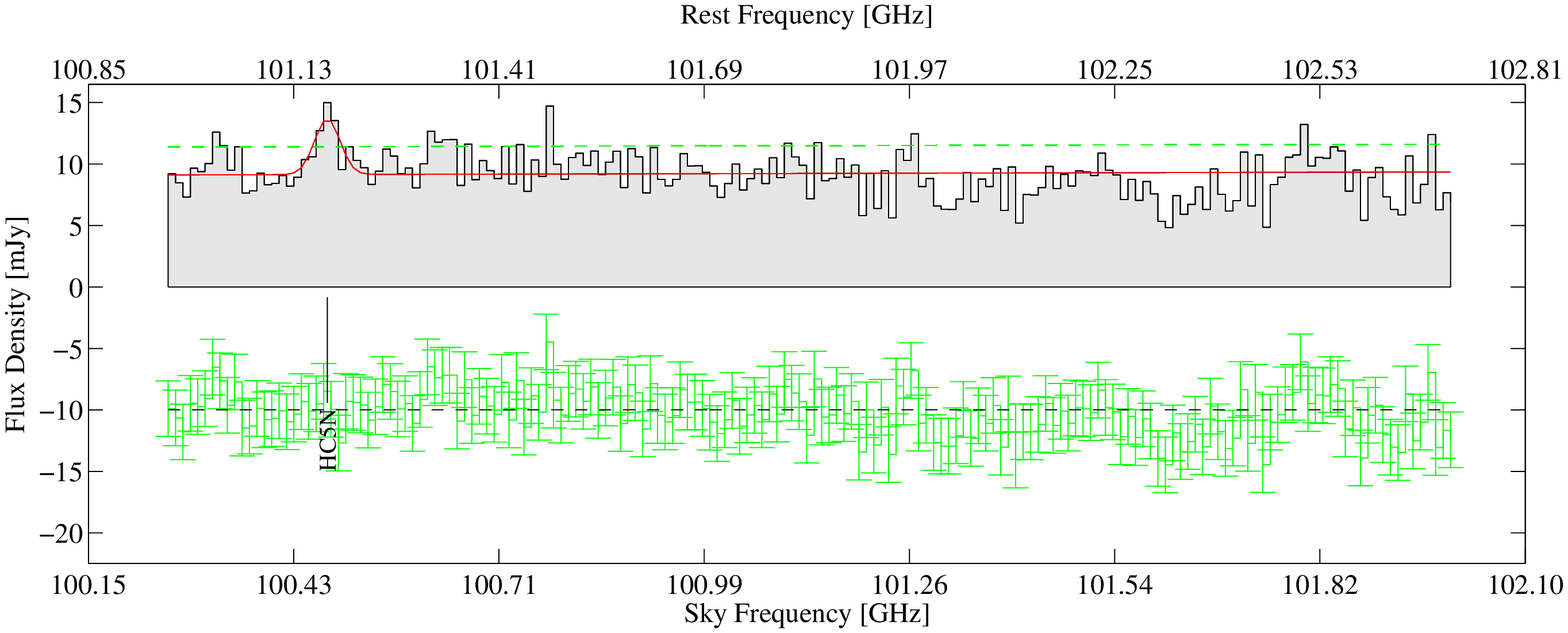}\\
   \includegraphics[width=.8\textwidth,keepaspectratio]{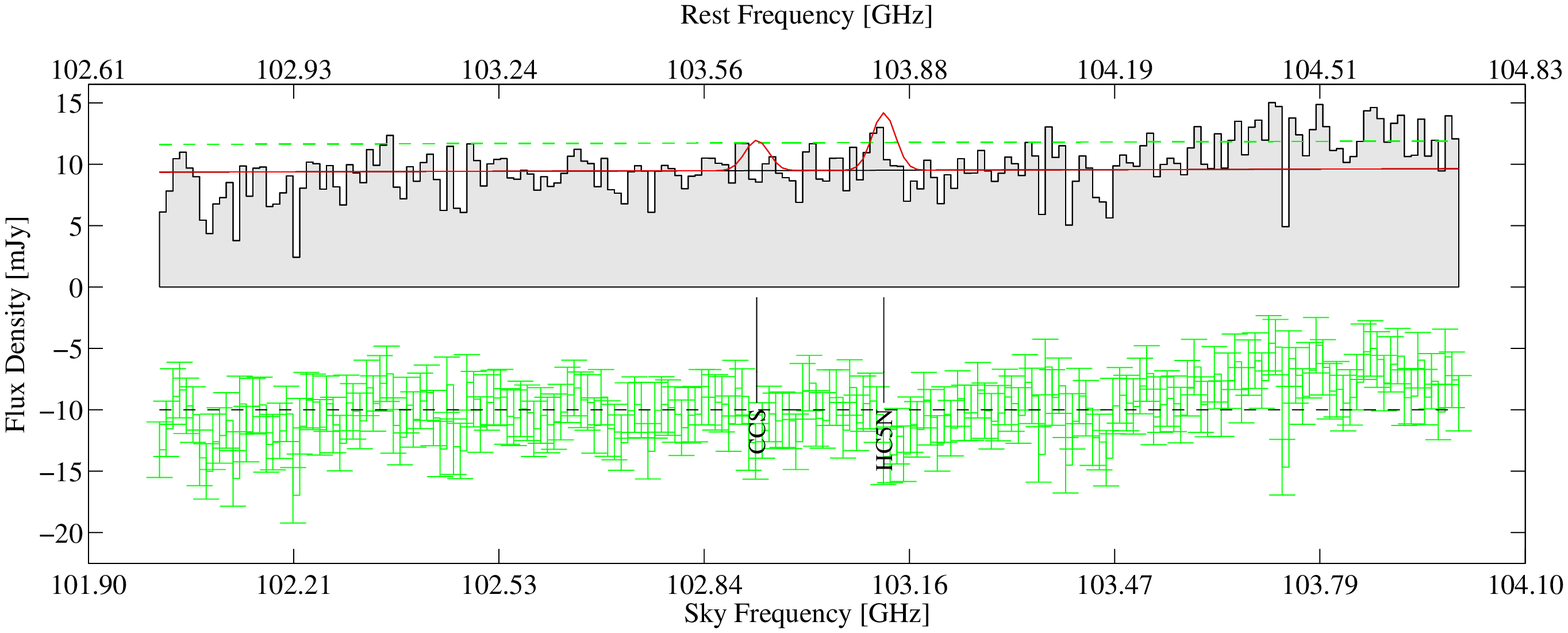}\\
   \includegraphics[width=.8\textwidth,keepaspectratio]{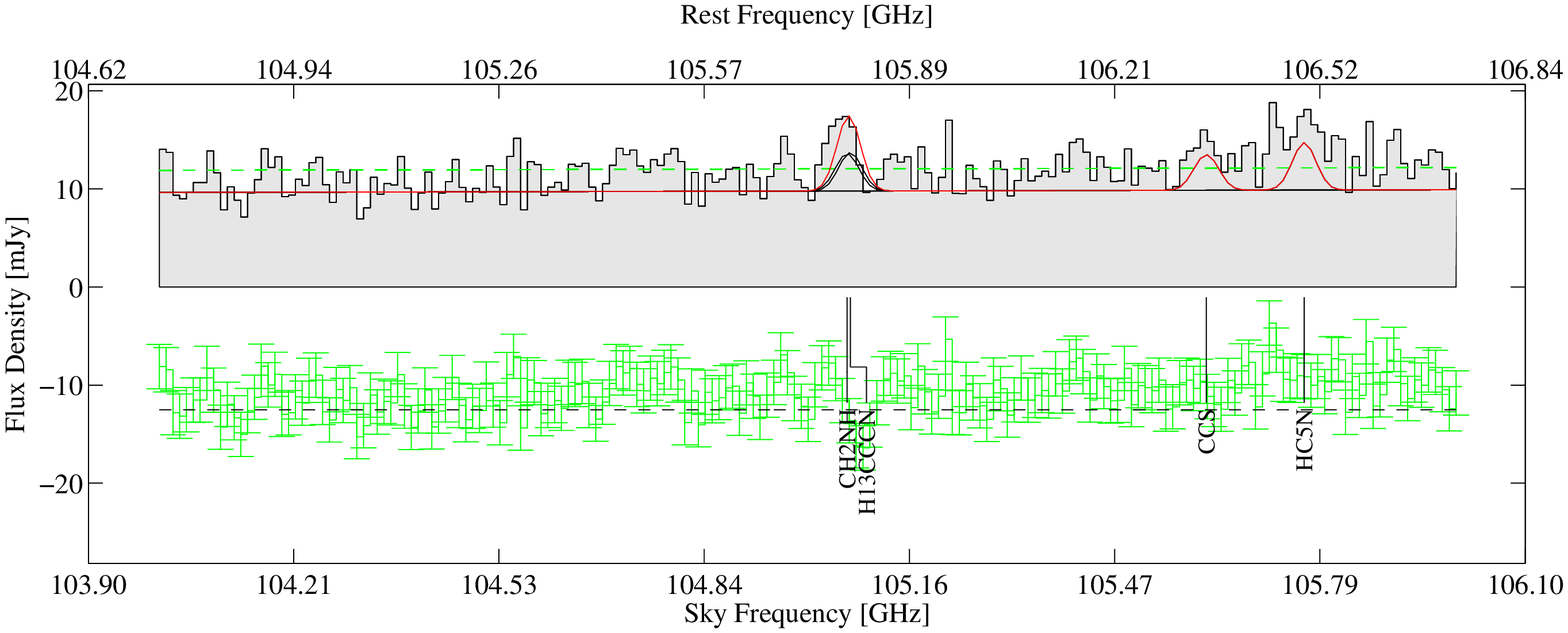}\\
   \includegraphics[width=.8\textwidth,keepaspectratio]{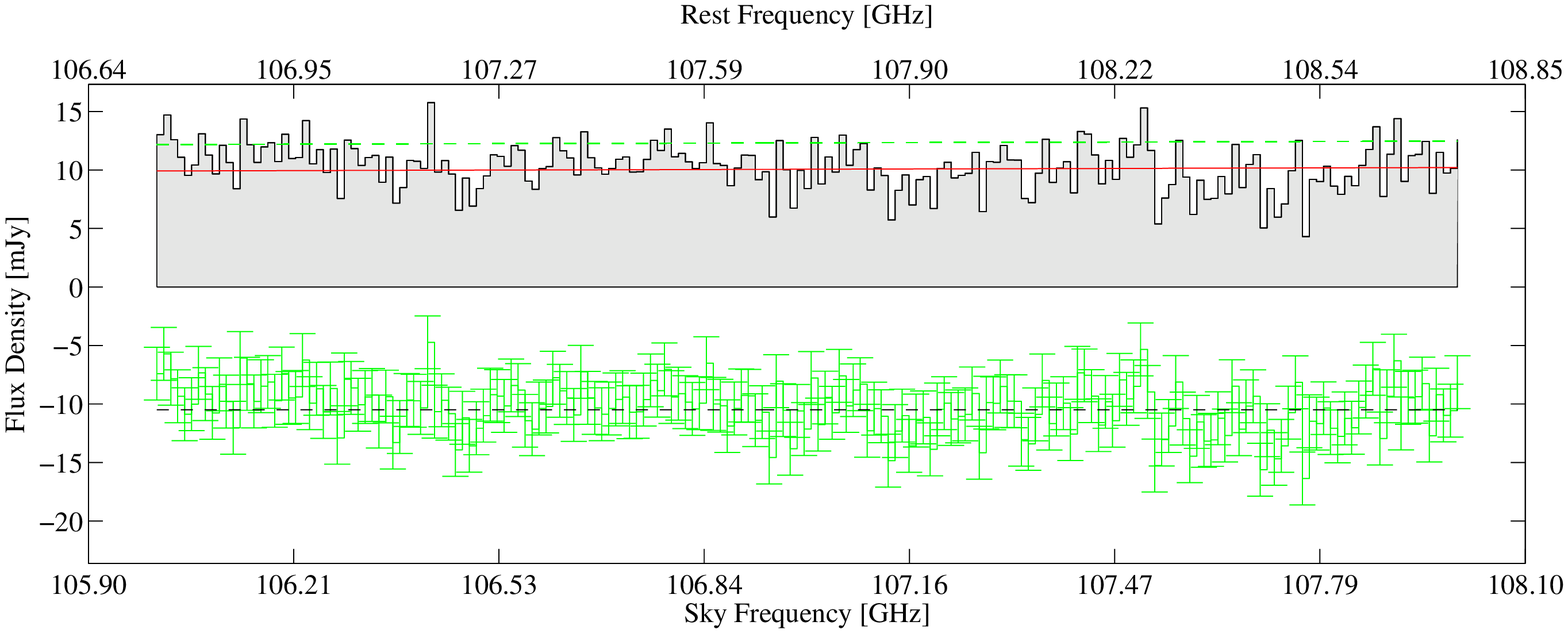}\\
      \caption{\label{fig:fit3} Continues from Fig. \ref{fig:fit1}.}
    \end{figure*}
\clearpage

\begin{figure*}[ph]
   \centering
   \includegraphics[width=.8\textwidth,keepaspectratio]{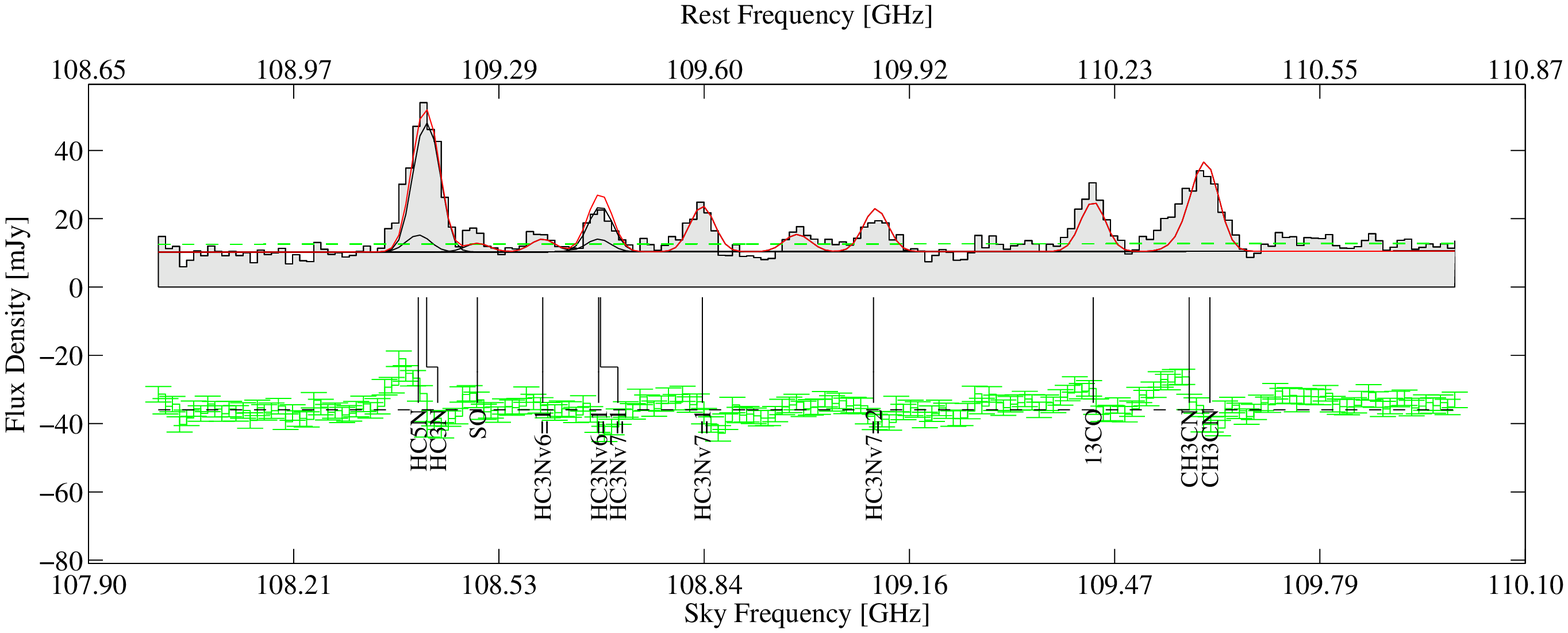}\\
   \includegraphics[width=.8\textwidth,keepaspectratio]{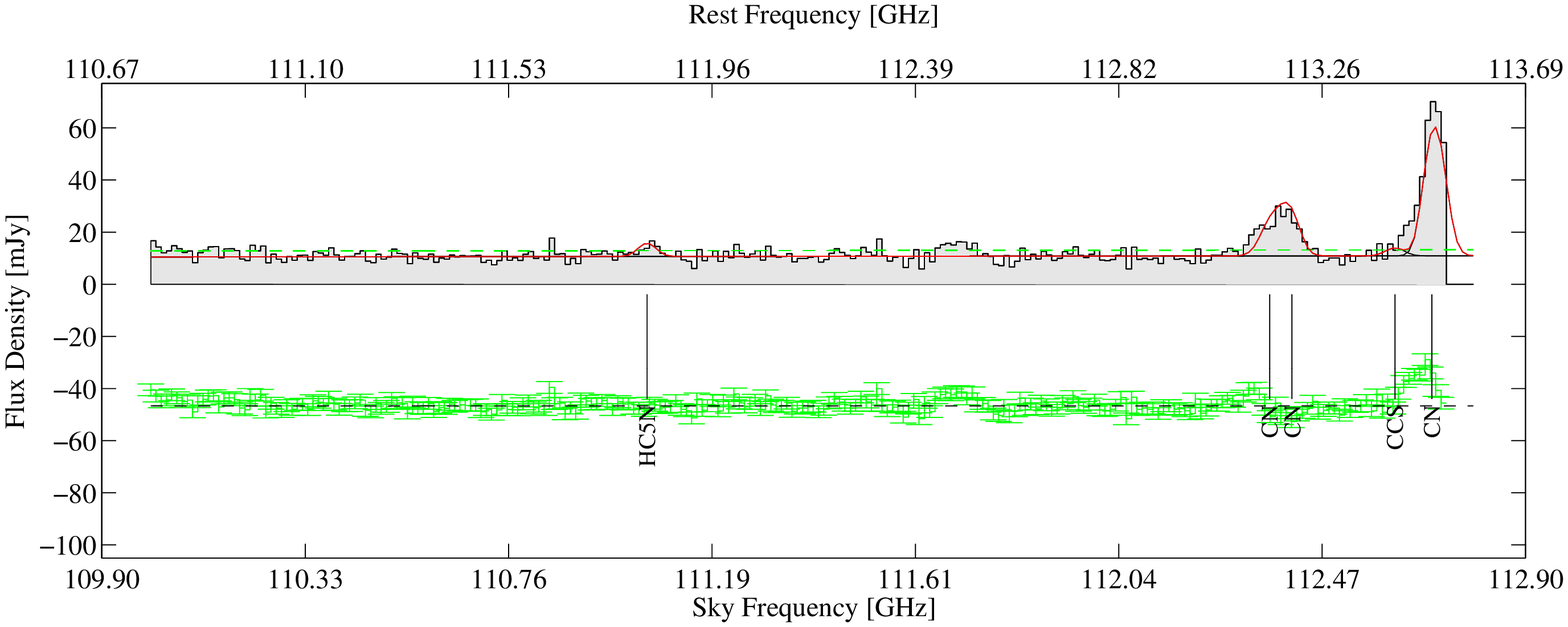}\\
   \includegraphics[width=.8\textwidth,keepaspectratio]{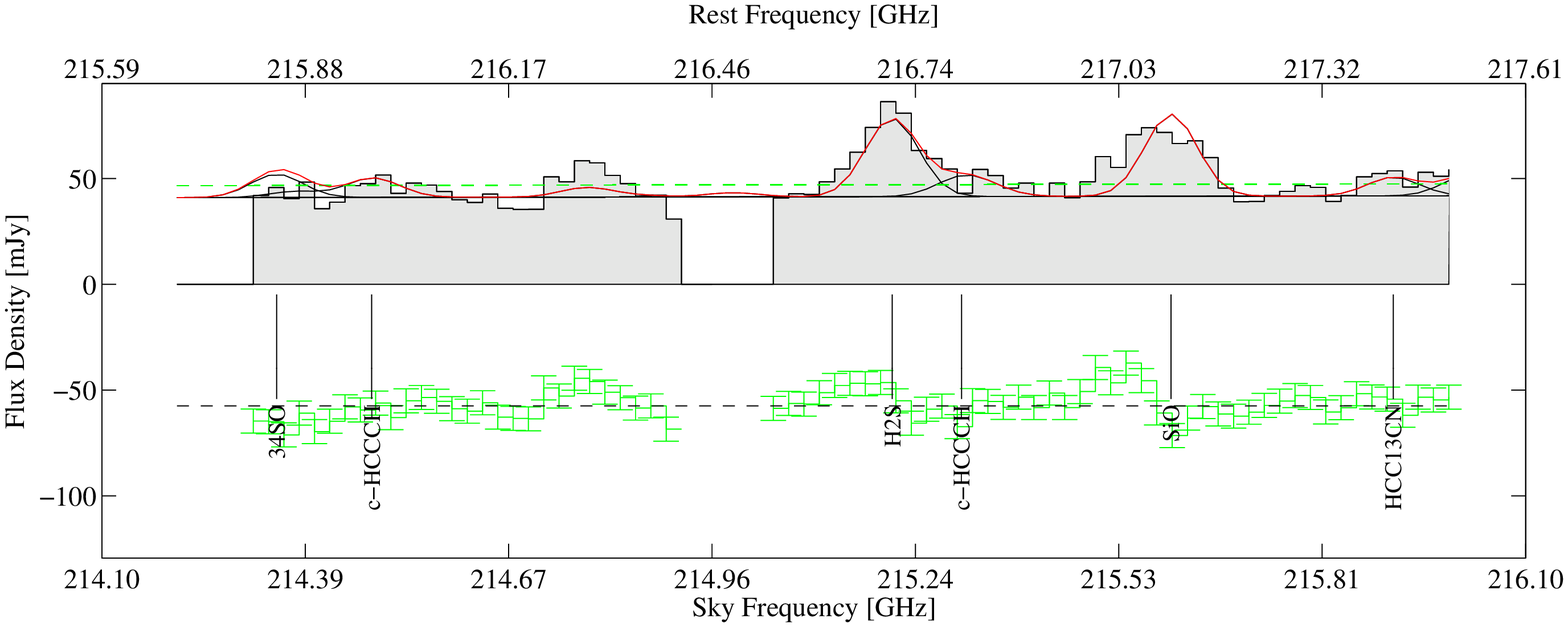}\\
   \includegraphics[width=.8\textwidth,keepaspectratio]{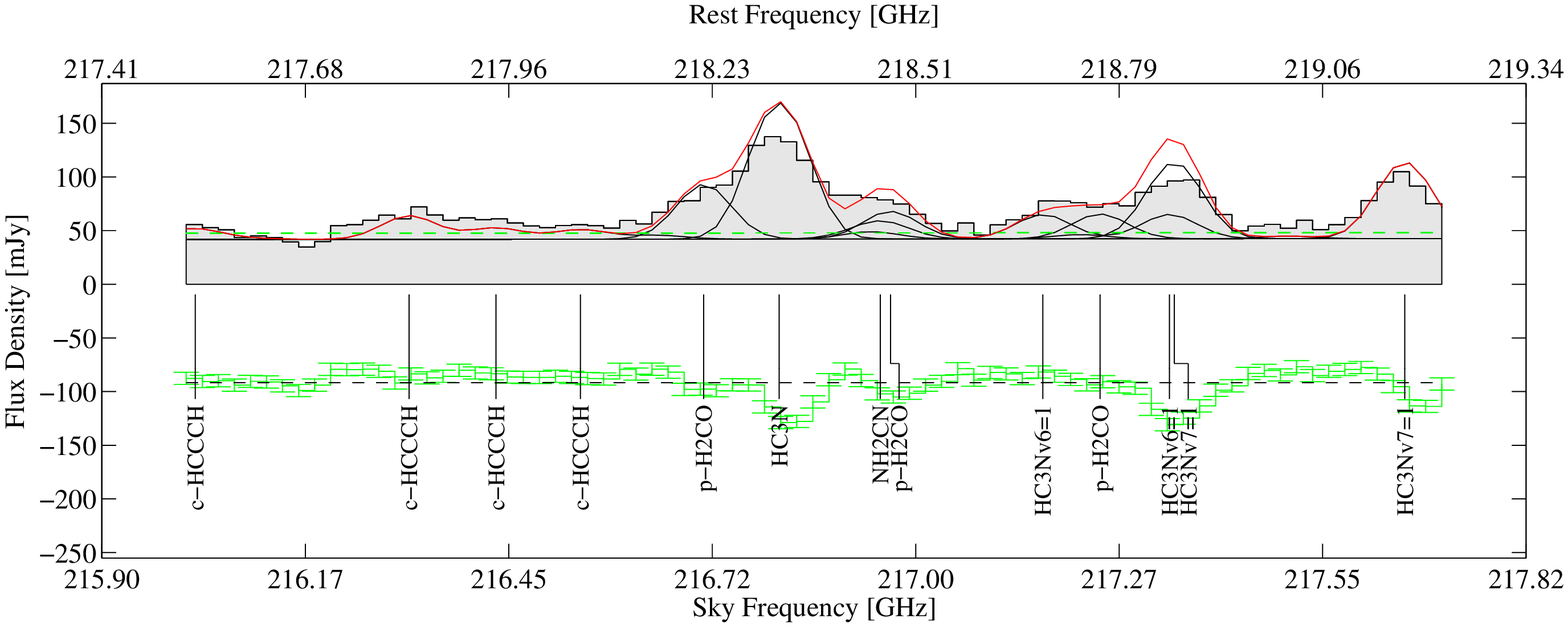}\\
       \caption{\label{fig:fit4} Continues from Fig. \ref{fig:fit1}.}
    \end{figure*}
\clearpage

\begin{figure*}[ph]
   \centering
   \includegraphics[width=.8\textwidth,keepaspectratio]{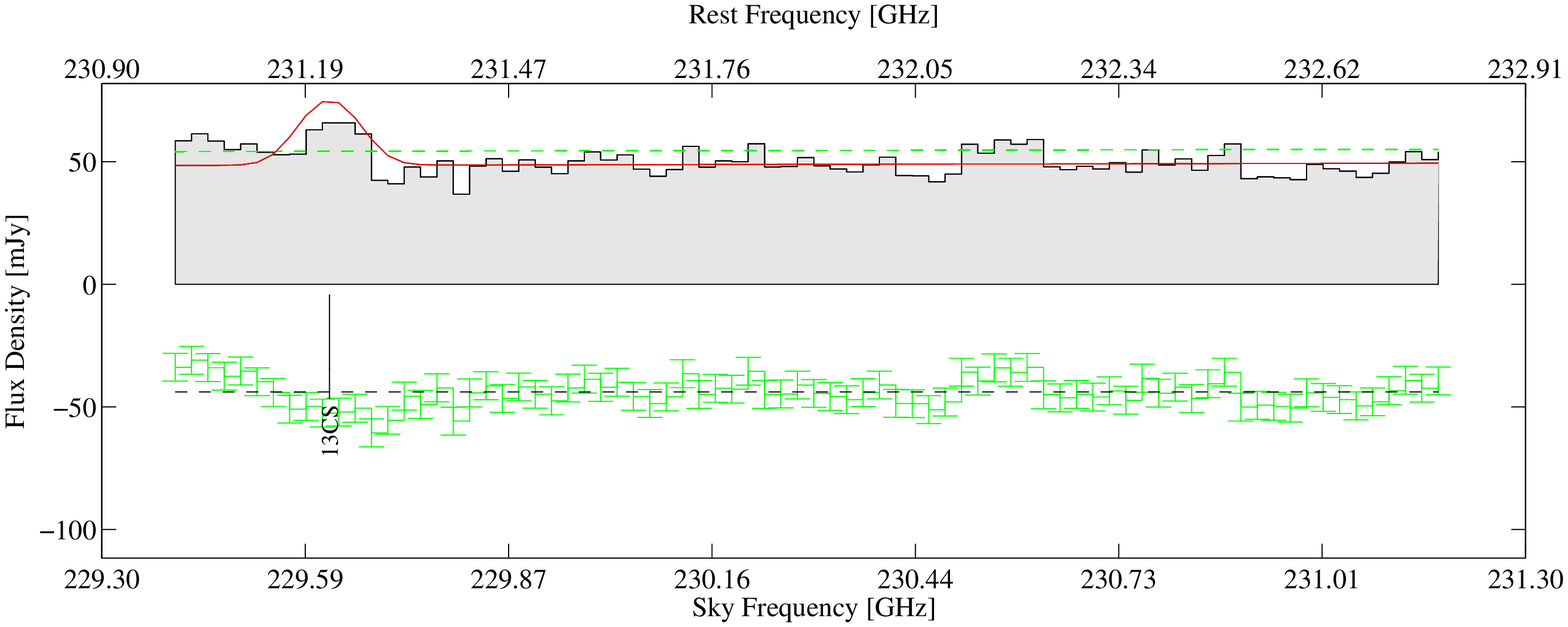}\\
   \includegraphics[width=.8\textwidth,keepaspectratio]{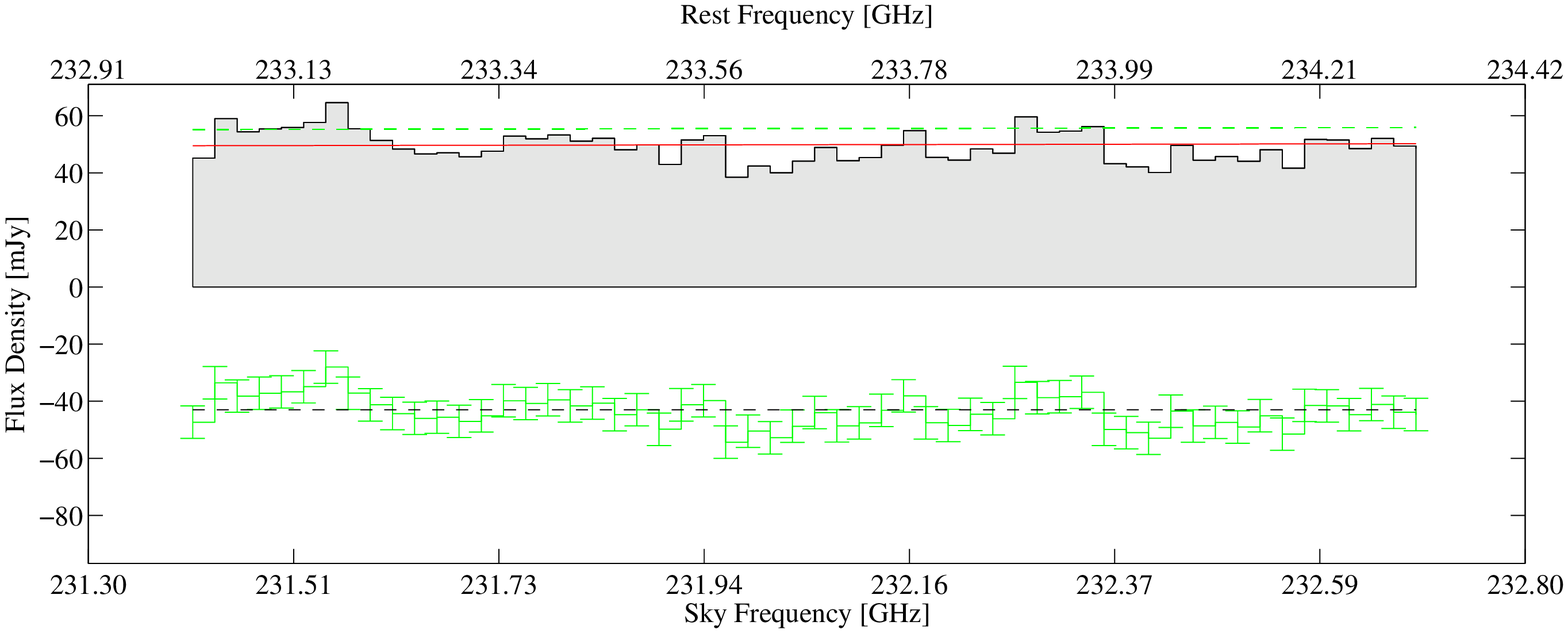}\\
   \includegraphics[width=.8\textwidth,keepaspectratio]{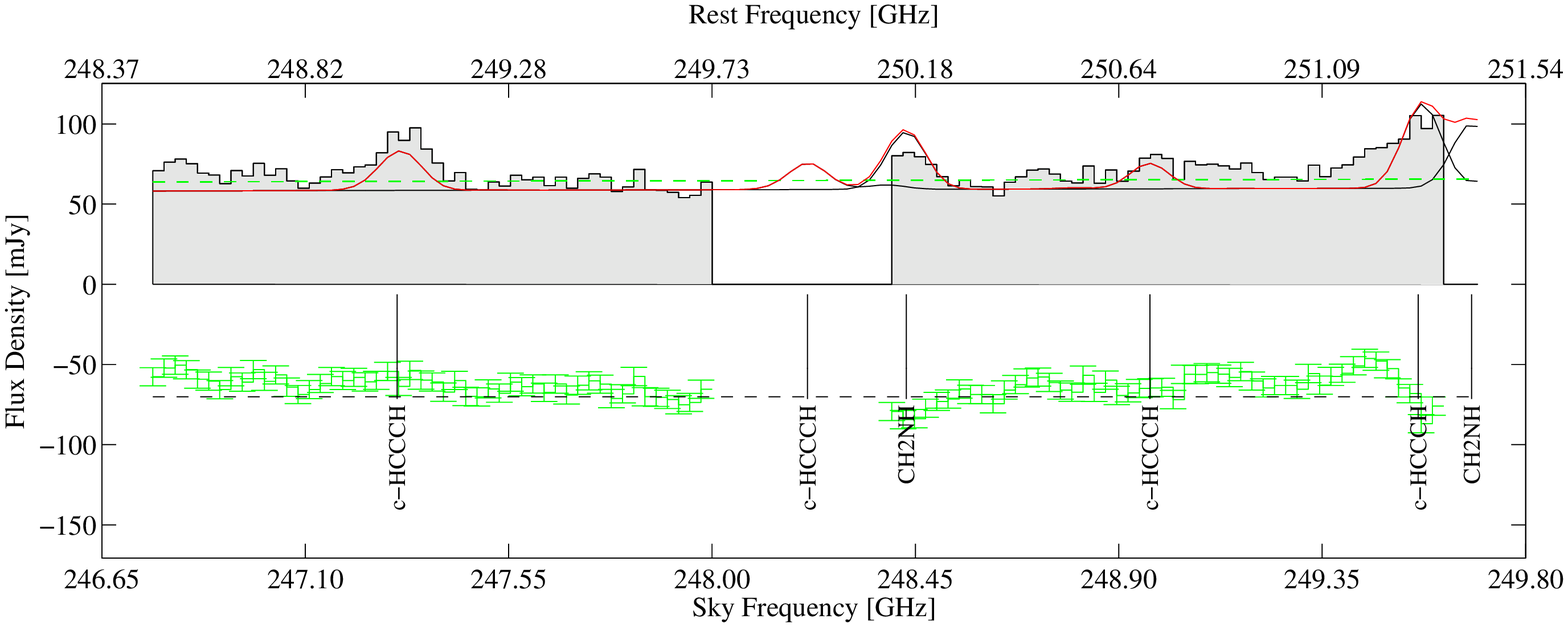}\\
   \includegraphics[width=.8\textwidth,keepaspectratio]{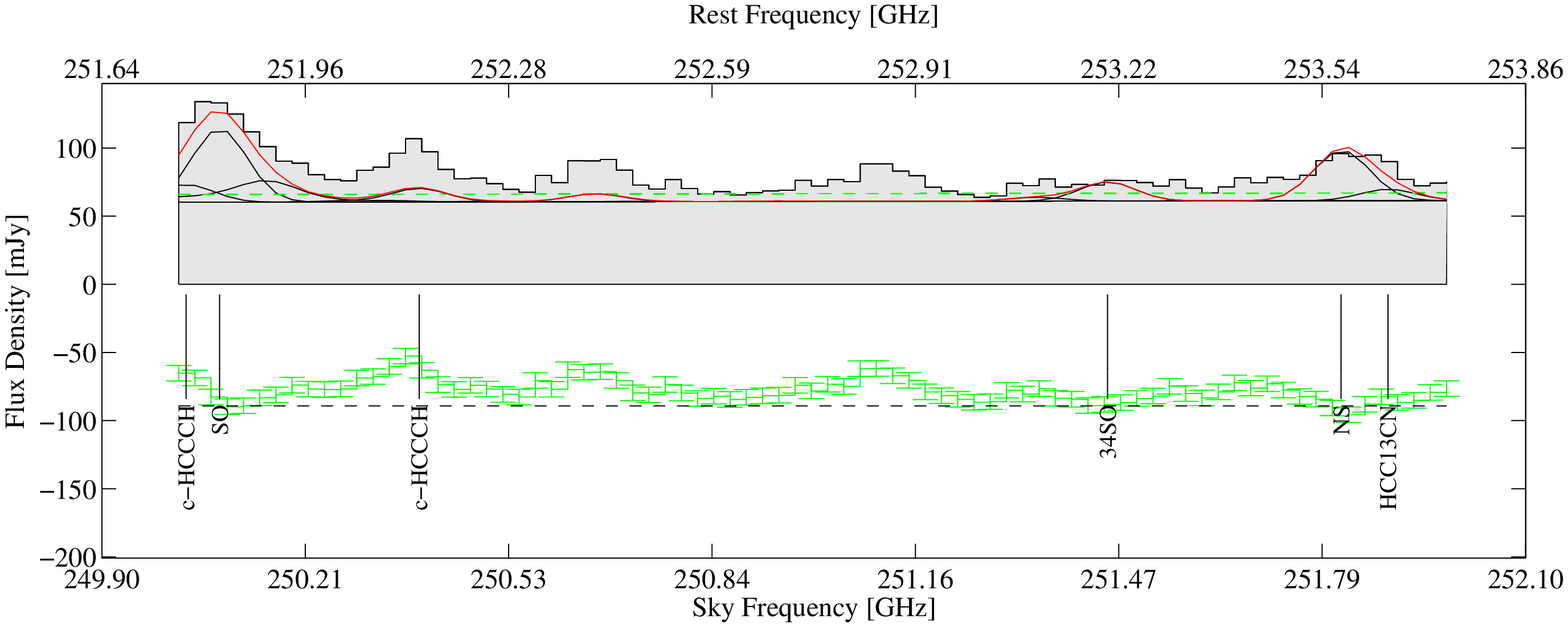}\\
      \caption{\label{fig:fit5} Continues from Fig. \ref{fig:fit1}.}
    \end{figure*}
\clearpage

\begin{figure*}[ph]
   \centering
   \includegraphics[width=.8\textwidth,keepaspectratio]{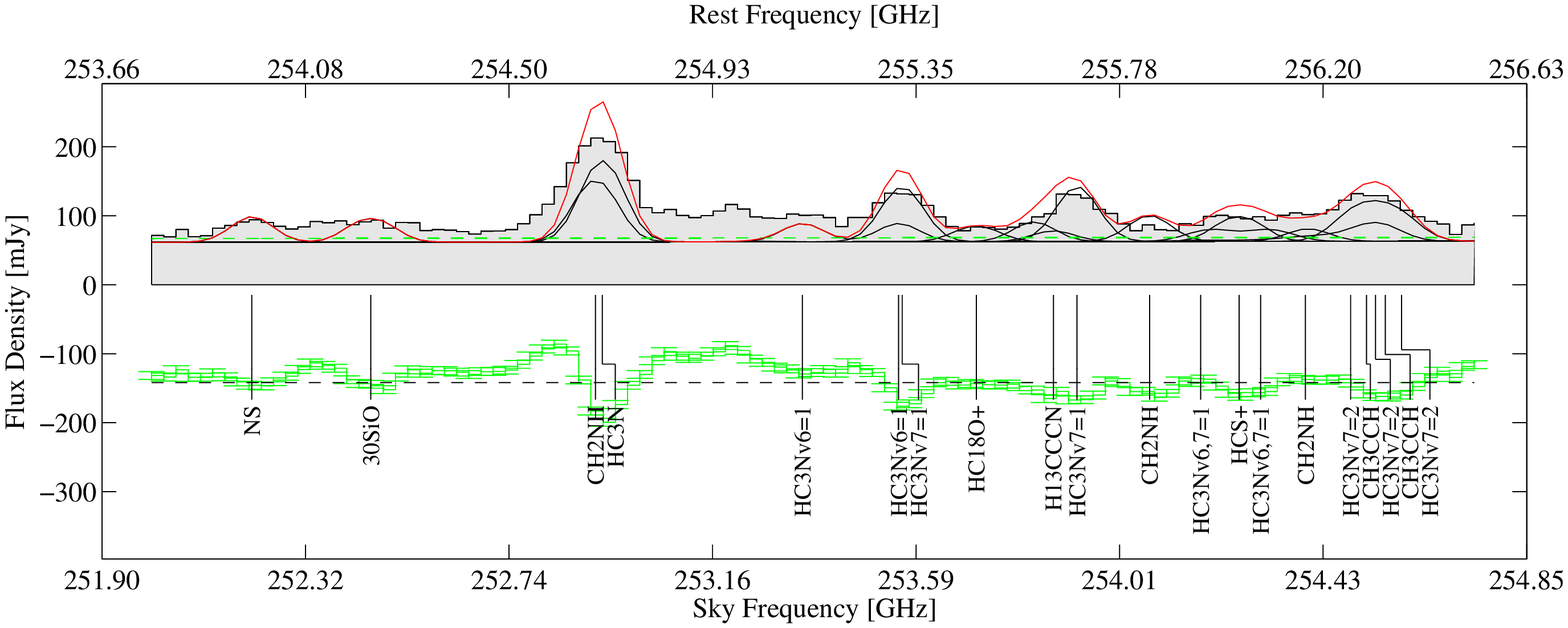}\\
   \includegraphics[width=.8\textwidth,keepaspectratio]{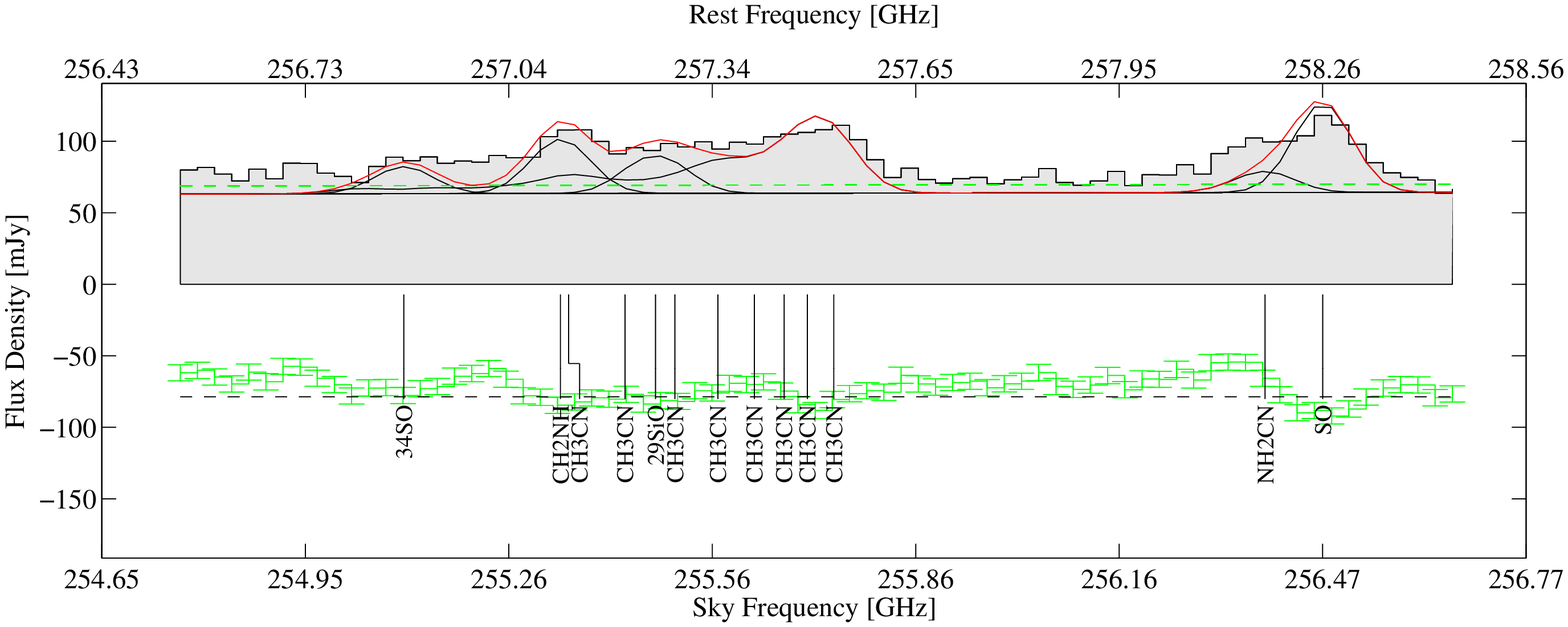}\\
   \includegraphics[width=.8\textwidth,keepaspectratio]{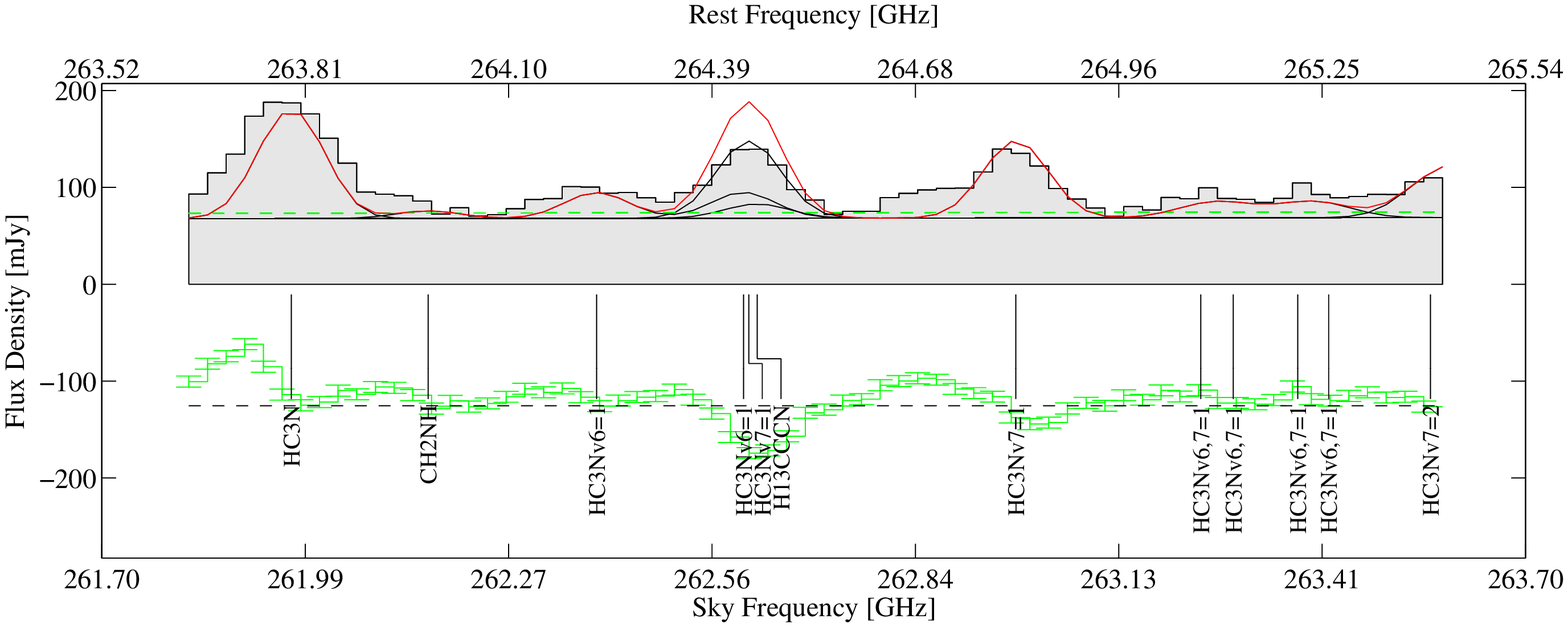}\\
   \includegraphics[width=.8\textwidth,keepaspectratio]{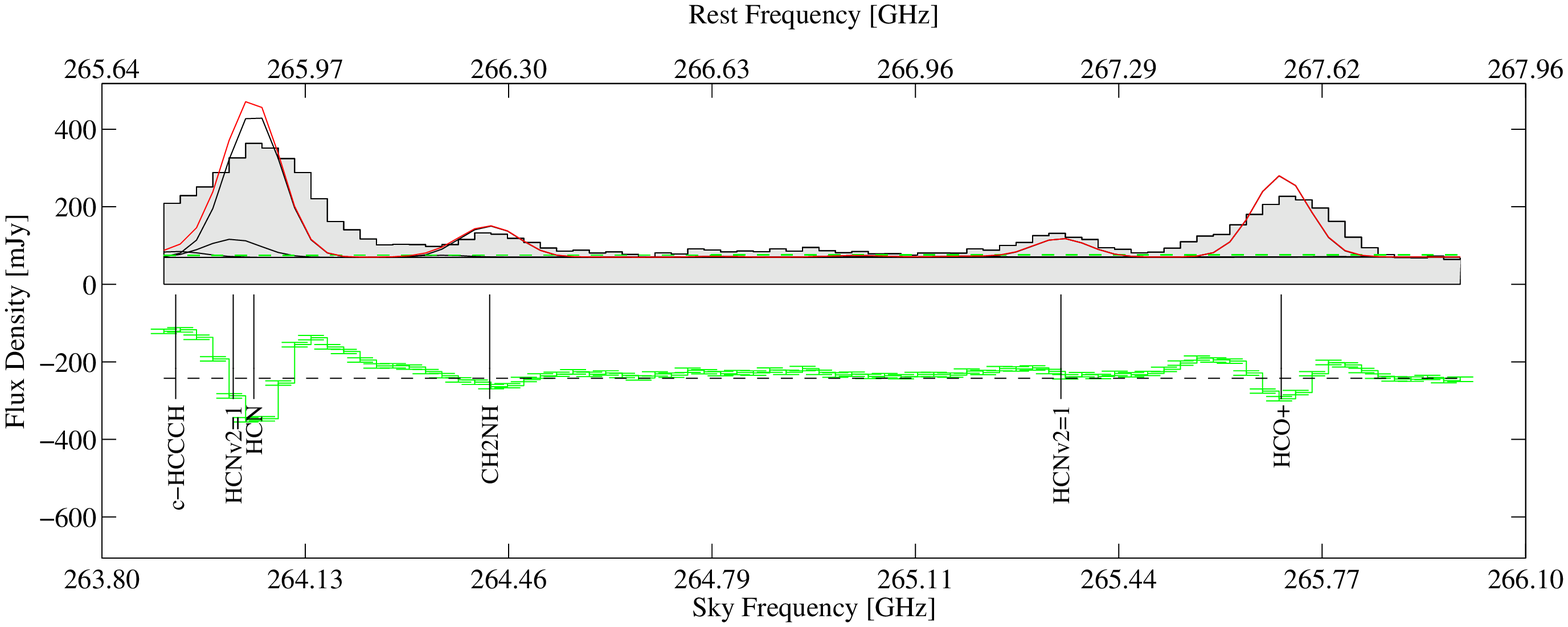}\\
       \caption{\label{fig:fit6} Continues from Fig. \ref{fig:fit1}.}
    \end{figure*}
\clearpage

\begin{figure*}[ph]
   \centering
   \includegraphics[width=.8\textwidth,keepaspectratio]{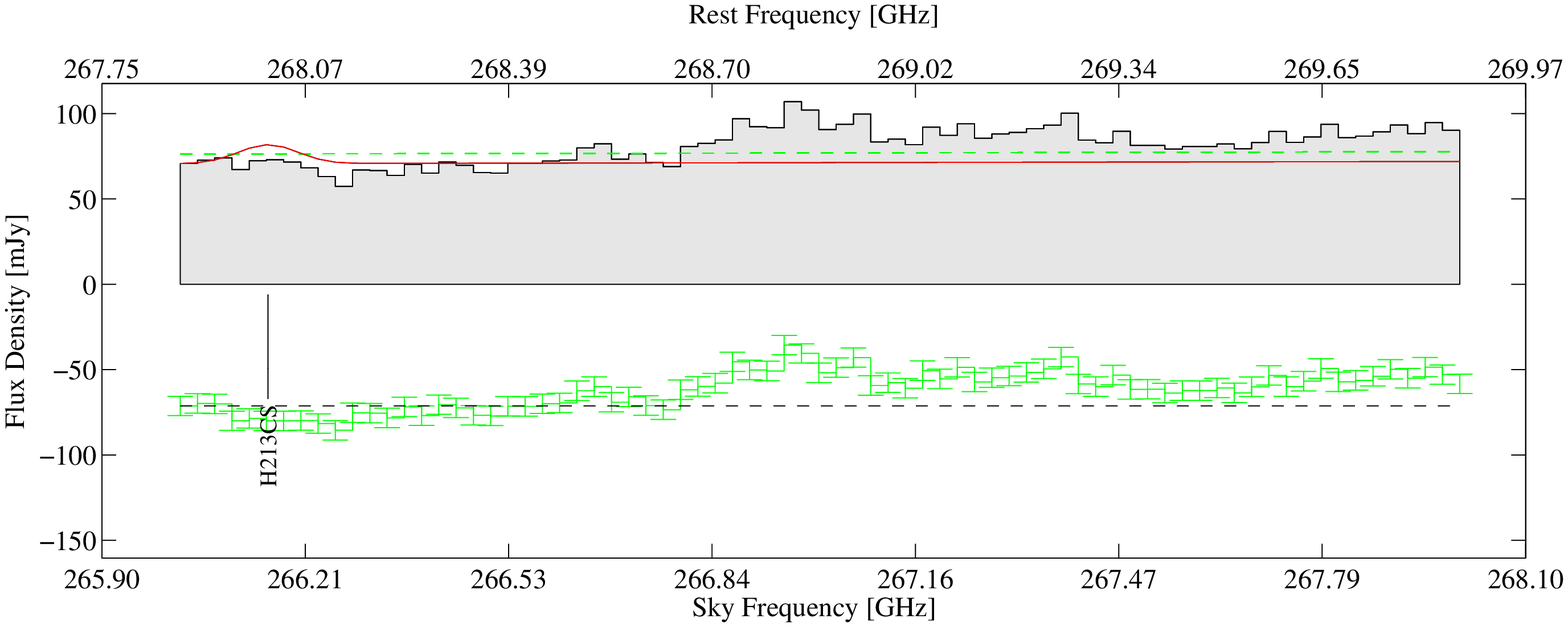}\\
   \includegraphics[width=.8\textwidth,keepaspectratio]{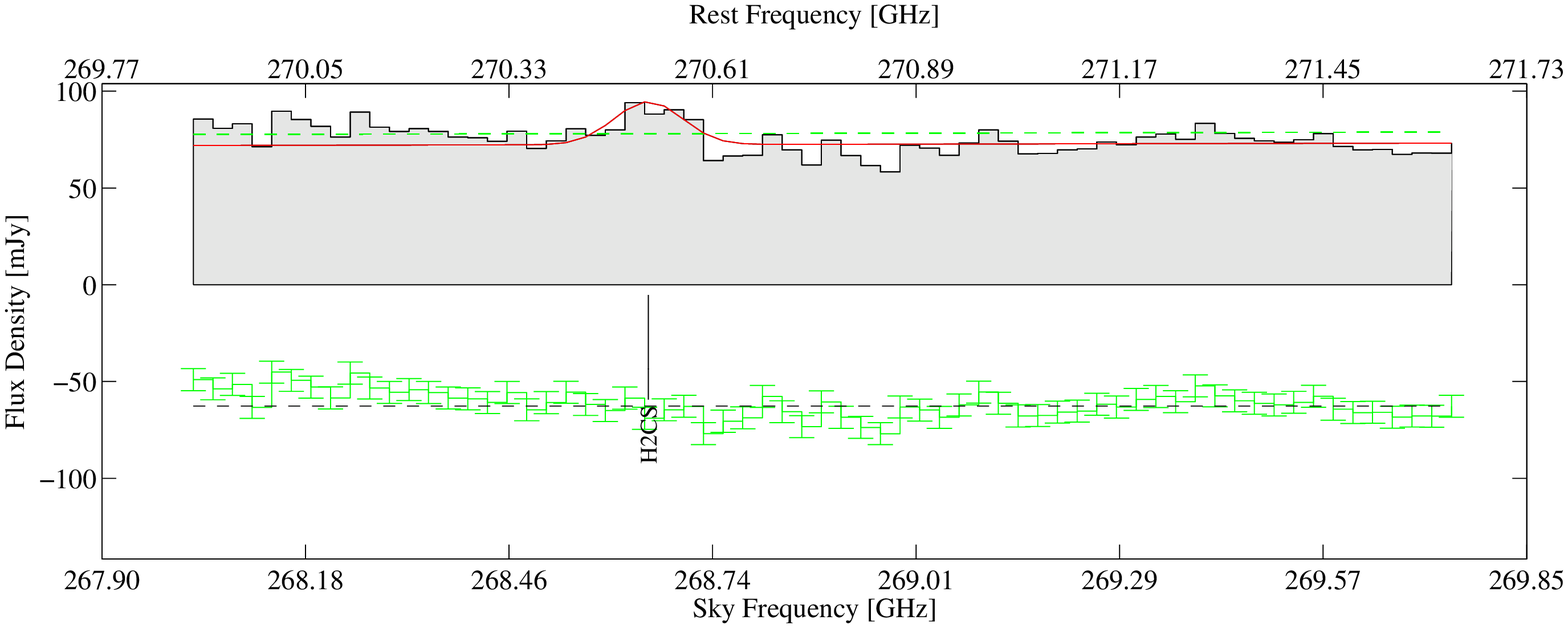}\\
   \includegraphics[width=.8\textwidth,keepaspectratio]{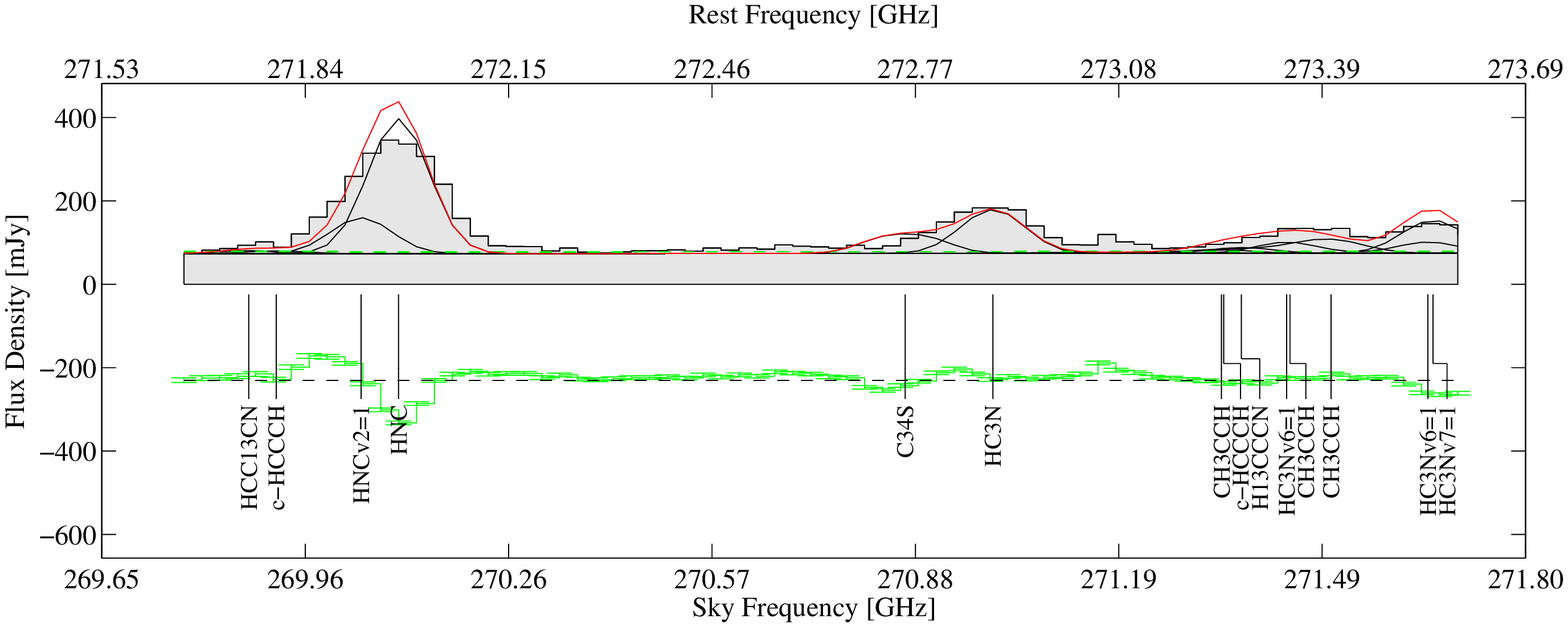}\\
   \includegraphics[width=.8\textwidth,keepaspectratio]{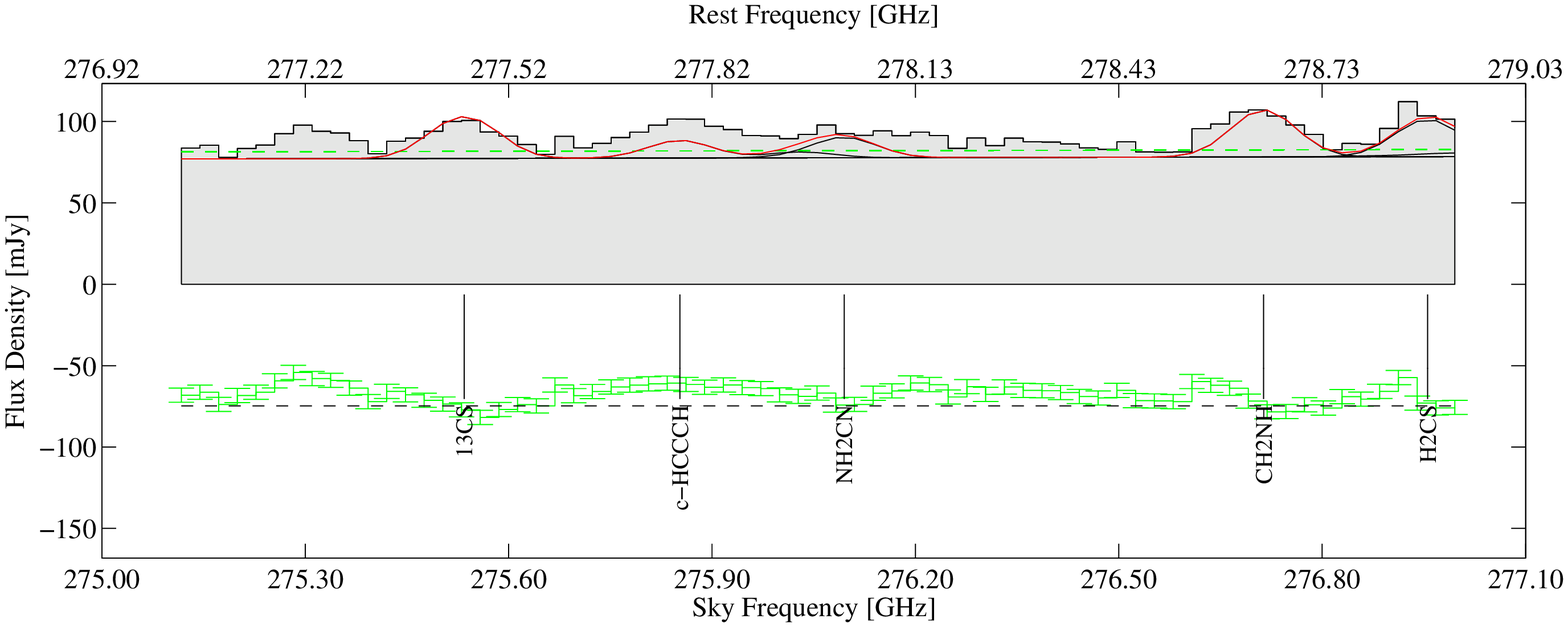}\\
       \caption{\label{fig:fit7} Continues from Fig. \ref{fig:fit1}.}
    \end{figure*}
\clearpage

\begin{figure*}[ph]
   \centering
   \includegraphics[width=.8\textwidth,keepaspectratio]{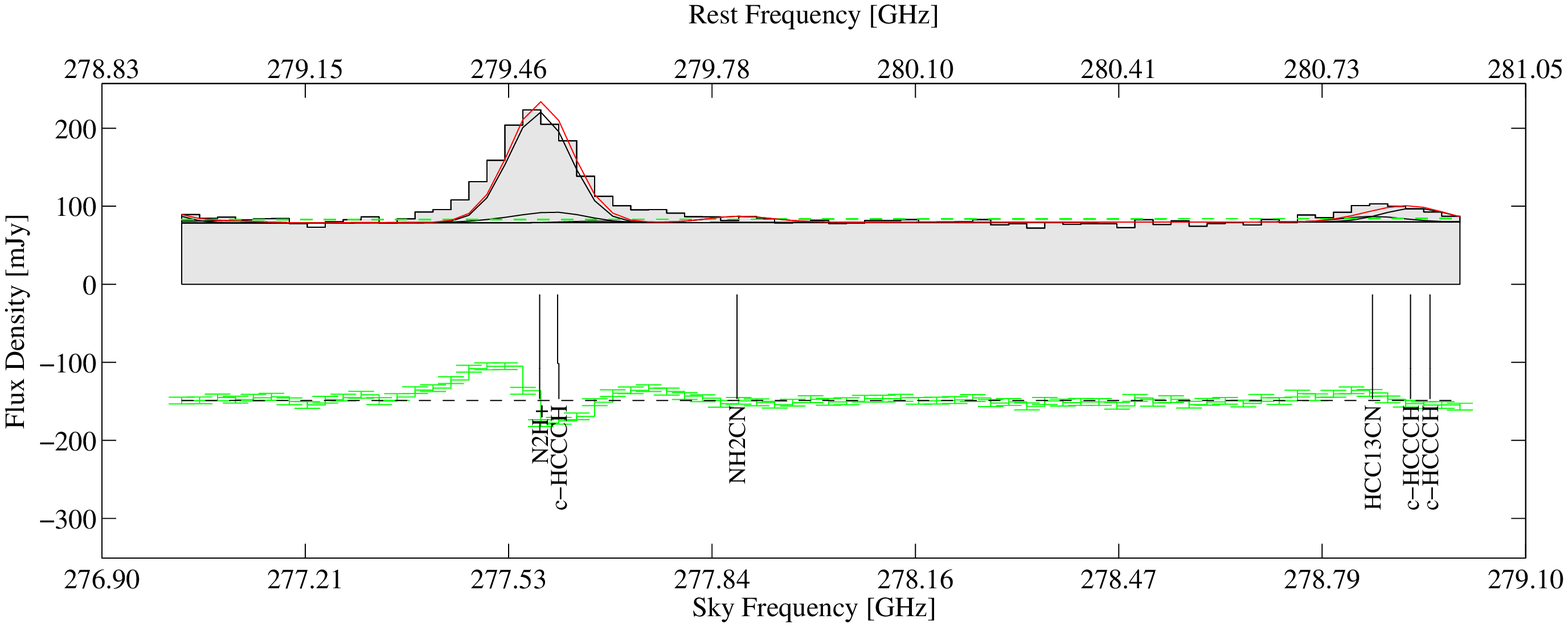}\\
   \includegraphics[width=.8\textwidth,keepaspectratio]{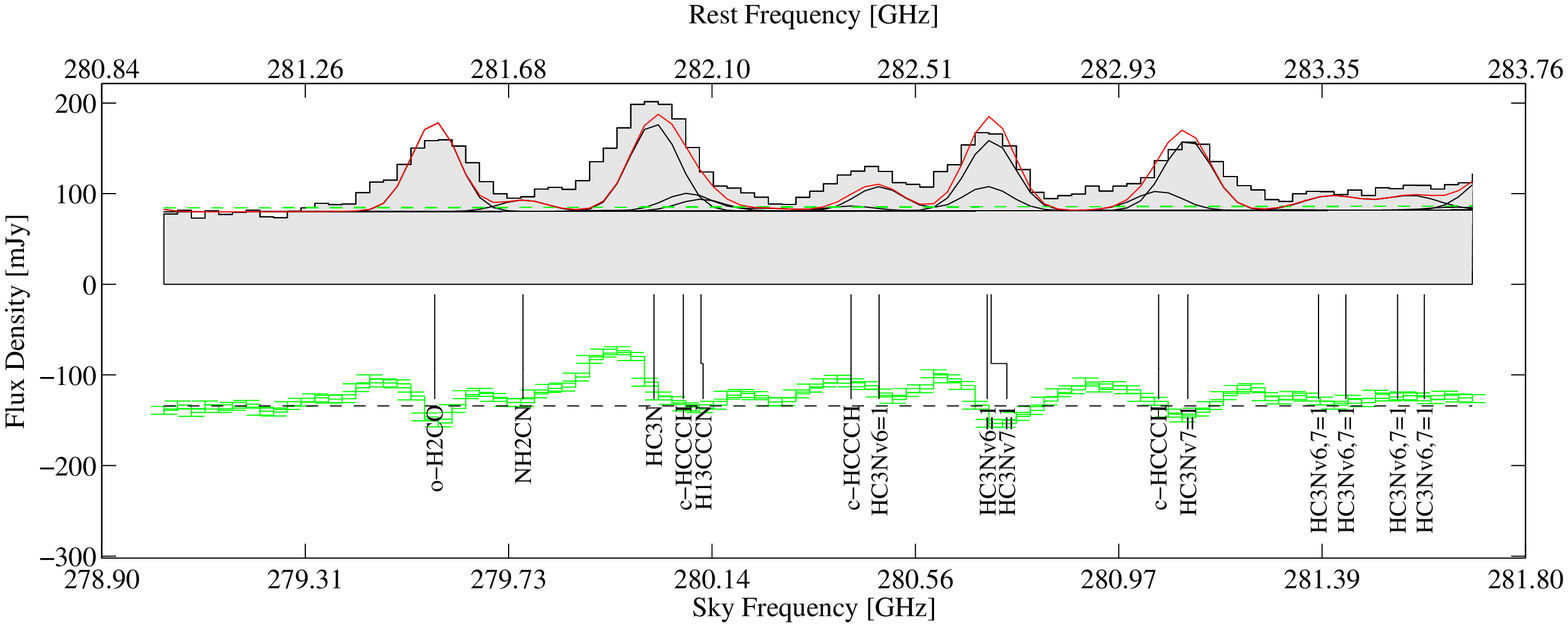}\\
   \includegraphics[width=.8\textwidth,keepaspectratio]{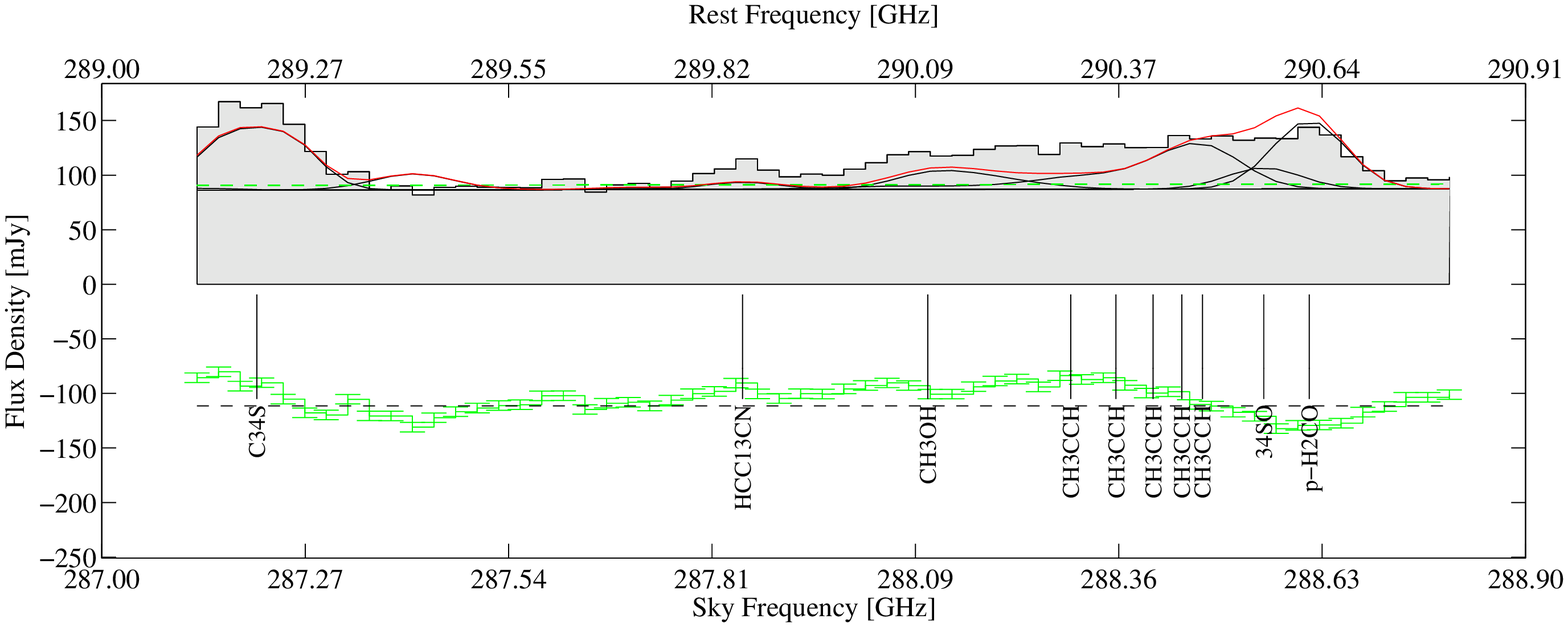}\\
   \includegraphics[width=.8\textwidth,keepaspectratio]{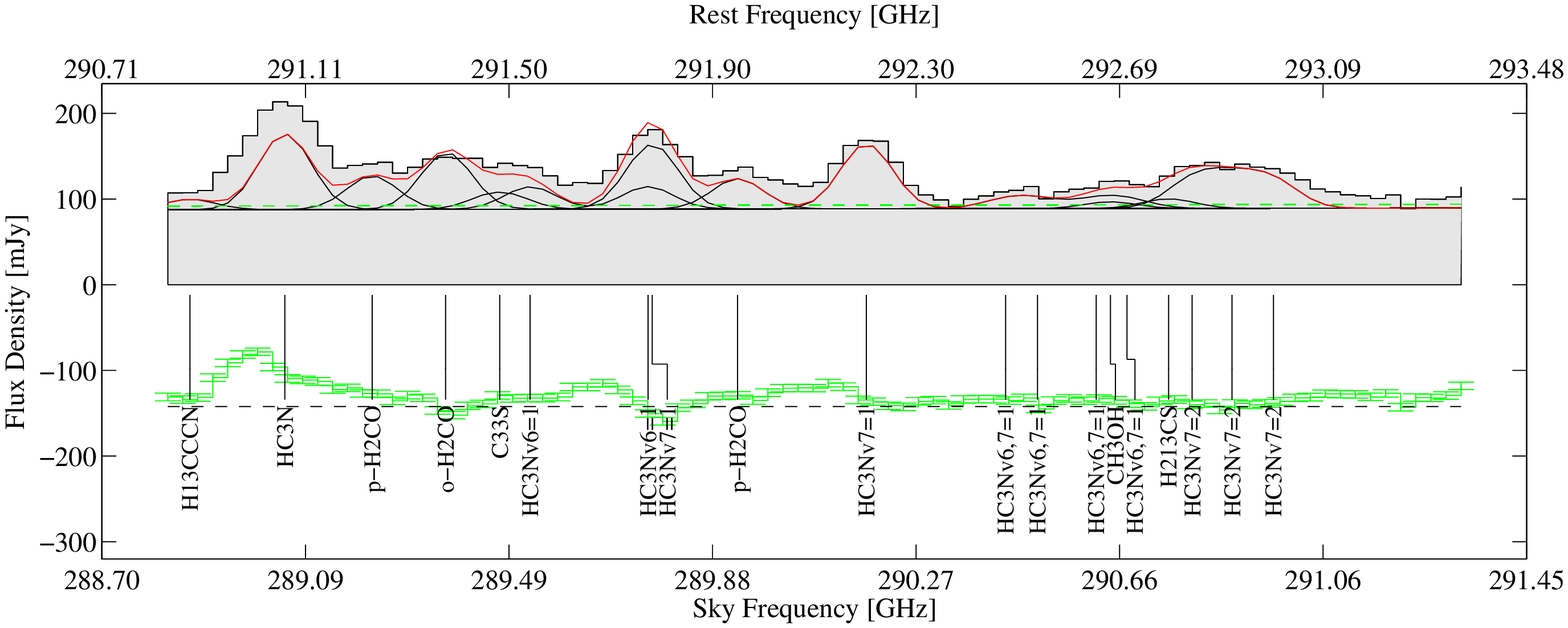}\\
     \caption{\label{fig:fit8} Continues from Fig. \ref{fig:fit1}.}
    \end{figure*}
\clearpage

\begin{figure*}[ph]
   \centering
   \includegraphics[width=.8\textwidth,keepaspectratio]{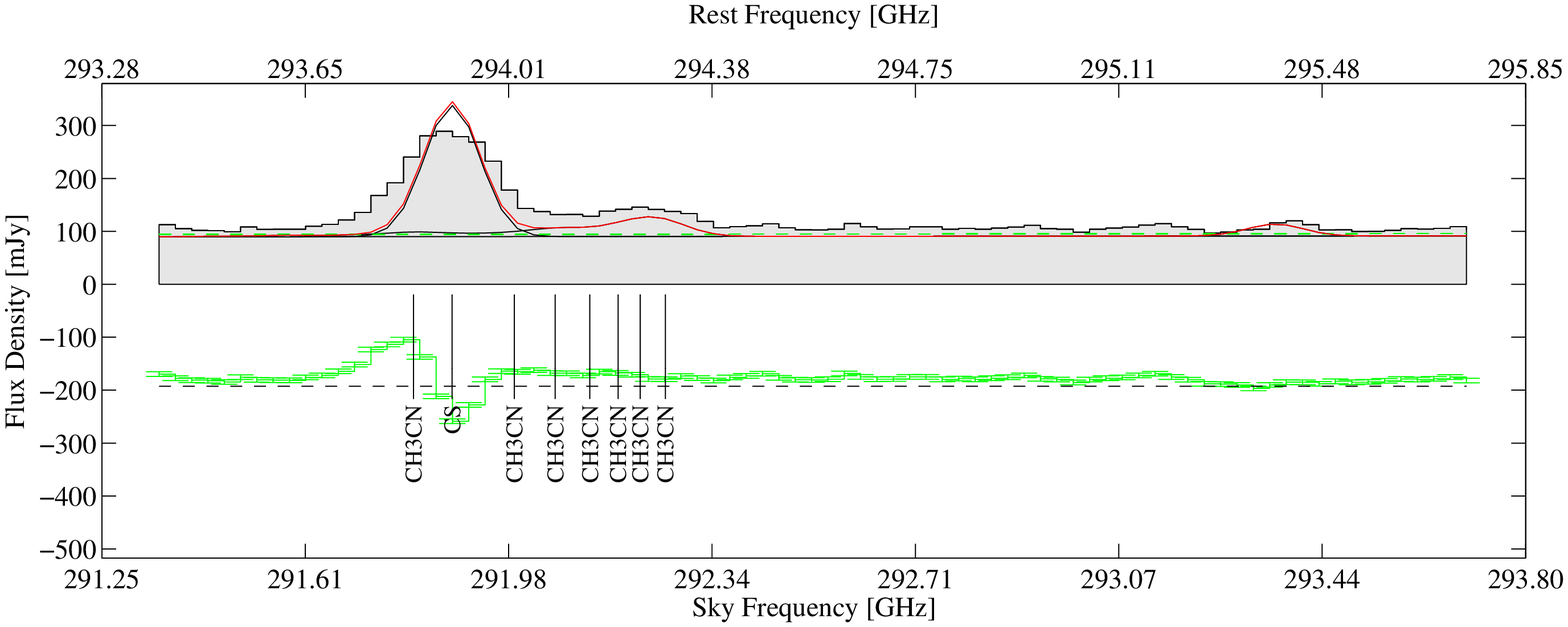}\\
      \caption{\label{fig:fit9} Continues from Fig. \ref{fig:fit1}.}
    \end{figure*}
\clearpage


\begin{table*}
\centering
\begin{tabular}{llllll} 
\hline 
Sky freq. [GHz] & Molecule & Transition & Rest freq. [GHz] & Peak [mJy] & SNR \\ 
\hline 
84.159 & $^{30}$SiO  & 2-1 & 84.746 & 3 & 4 \\  
84.611 & HC$_5$N  & J=32-31 & 85.201 & 3 & 4 \\  
84.757 & HCS$^+$ & 2-1 & 85.348 & 4 & 4 \\  
85.165 & $^{29}$SiO  & 2-1 & 85.759 & 2 & 3 \\  
85.742 & H$^{13}$CN  & J=1-0 & 86.340 & 11 & 13 \\  
86.154 & H$^{13}$CO$^+$ & 1-0 & 86.754 & 4 & 5 \\  
86.246 & SiO  & 2-1 & 86.847 & 13 & 16 \\  
86.712 & CCH  & N=1-0,J=3/2-1/2,F=2-1 & 87.317 & 6 & 7 \\  
86.724 & CCH  & N=1-0,J=3/2-1/2,F=1-0 & 87.329 & 4 & 5 \\  
86.797 & CCH  & N=1-0,J=1/2-1/2,F=1-1 & 87.402 & 4 & 5 \\  
87.255 & HC$_5$N  & J=33-32 & 87.864 & 4 & 4 \\  
88.018 & HCN  & 1-0 & 88.632 & 55 & 68 \\  
88.571 & HCO$^+$  & 1-0 & 89.189 & 41 & 50 \\  
89.899 & HC$_5$N  & J=34-33 & 90.526 & 4 & 5 \\  
90.036 & HNC  & 1-0 & 90.664 & 47 & 58 \\  
90.297 & C$^{34}$S  & 2-1 & 90.926 & 6 & 7 \\  
90.349 & HC$_3$N  & J=10-9 & 90.979 & 22 & 27 \\  
90.571 & HC$_3$N v7=1 & J=10-9,l=1e & 91.203 & 7 & 9 \\  
90.701 & HC$_3$N v7=1 & J=10-9,l=1f & 91.333 & 7 & 9 \\  
91.334 & CH$_3$CN  & 5;3-4;3 & 91.971 & 4 & 5 \\  
91.343 & CH$_3$CN  & 5;2-4;2 & 91.980 & 3 & 3 \\  
91.348 & CH$_3$CN  & 5;1-4;1 & 91.985 & 3 & 3 \\  
91.350 & CH$_3$CN  & 5;0-4;0 & 91.987 & 4 & 5 \\  
91.854 & $^{13}$CS  & 2-1 & 92.494 & 3 & 4 \\  
92.529 & N$_2$H$^+$  & 1-0 & 93.174 & 14 & 17 \\  
92.543 & HC$_5$N  & J=35-34 & 93.188 & 4 & 5 \\  
93.220 & CCS & (7  8 )-(6  7 ) & 93.870 & 3 & 3 \\  
95.187 & HC$_5$N  & J=36-35 & 95.850 & 4 & 5 \\  
95.745 & C$^{34}$S  & 2-1 & 96.413 & 8 & 9 \\  
96.312 & H$^{13}$CCCN  & J=11-10 & 96.983 & 3 & 4 \\  
97.303 & CS  & 2-1 & 97.981 & 47 & 58 \\  
97.830 & HC$_5$N  & J=37-36 & 98.513 & 4 & 5 \\  
98.612 & SO  & 2;3-1;2 & 99.300 & 5 & 6 \\  
98.624 & NH$_2$CN & 5(1,5)-4(1,4),  & 99.311 & 4 & 4 \\  
99.384 & HC$_3$N  & J=11-10 & 100.076 & 30 & 36 \\  
99.547 & HC$_3$N v6=1 & J=11-10,l=1e & 100.241 & 3 & 3 \\  
99.625 & HC$_3$N v6=1 & J=11-10,l=1f & 100.319 & 3 & 3 \\  
99.628 & HC$_3$N v7=1 & J=11-10,l=1e & 100.322 & 10 & 12 \\  
99.771 & HC$_3$N v7=1 & J=11-10,l=1f & 100.466 & 10 & 12 \\  
99.933 & NH$_2$CN & 5(1,4)-4(1,3),  & 100.629 & 4 & 4 \\  
100.012 & HC$_3$N v7=2 & J=11-10,l=0 & 100.709 & 3 & 4 \\  
100.014 & HC$_3$N v7=2 & J=11-10,l=2e & 100.711 & 3 & 4 \\  
100.017 & HC$_3$N v7=2 & J=11-10,l=2f & 100.714 & 3 & 4 \\  
100.474 & HC$_5$N  & J=38-37 & 101.175 & 5 & 6 \\  
103.118 & HC$_5$N  & J=39-38 & 103.837 & 5 & 6 \\  
105.062 & CH$_2$NH & 4(0,4)-3(1,3) & 105.794 & 4 & 5 \\  
105.067 & H$^{13}$CCCN  & J=12-11 & 105.799 & 4 & 5 \\  
105.612 & CCS & (8  9 )-(7  8 ) & 106.348 & 4 & 4 \\  
105.762 & HC$_5$N  & J=40-39 & 106.499 & 5 & 6 \\  
108.405 & HC$_5$N  & J=41-40 & 109.161 & 5 & 6 \\  
108.418 & HC$_3$N  & J=12-11 & 109.174 & 38 & 46 \\  
108.496 & SO  & 3;2-2;1 & 109.252 & 3 & 3 \\  
108.596 & HC$_3$N v6=1 & J=12-11,l=1e & 109.353 & 4 & 5 \\  
108.681 & HC$_3$N v6=1 & J=12-11,l=1f & 109.439 & 4 & 5 \\  
108.684 & HC$_3$N v7=1 & J=12-11,l=1e & 109.442 & 13 & 16 \\  
\hline 
\end{tabular} 

\caption{\label{tab:lines1} Molecular lines detected above the 3-sigma level. A total of 317 lines were detected in our spectral scan. The peak flux densities reported in table were derived with the LTE/NLTE fit of the whole spectrum. The signal to noise ratio reported in column six is the ratio between the peak flux density and the average rms of the ALMA band containing the line. The notation for the transition quantum numbers is the one used by the {\it Splatalogue} database. Tentative detections are shown in italics.}
\end{table*}

\begin{table*}
\centering
\begin{tabular}{llllll} 
\hline 
Sky freq. [GHz] & Molecule & Transition & Rest freq. [GHz] & Peak [mJy] & SNR \\ 
\hline 
108.840 & HC$_3$N v7=1 & J=12-11,l=1f & 109.599 & 13 & 16 \\  
109.102 & HC$_3$N v7=2 & J=12-11,l=0 & 109.863 & 5 & 6 \\  
109.105 & HC$_3$N v7=2 & J=12-11,l=2e & 109.866 & 4 & 5 \\  
109.110 & HC$_3$N v7=2 & J=12-11,l=2f & 109.870 & 4 & 5 \\  
109.438 & $^{13}$CO  & 1-0 & 110.201 & 14 & 18 \\  
109.585 & CH$_3$CN  & 6;4-5;4 & 110.349 & 3 & 4 \\  
109.600 & CH$_3$CN  & 6;3-5;3 & 110.364 & 9 & 12 \\  
109.611 & CH$_3$CN  & 6;2-5;2 & 110.375 & 5 & 6 \\  
109.617 & CH$_3$CN  & 6;1-5;1 & 110.381 & 5 & 7 \\  
109.619 & CH$_3$CN  & 6;0-5;0 & 110.383 & 8 & 9 \\  
111.049 & HC$_5$N  & J=42-41 & 111.823 & 5 & 6 \\  
112.361 & CN  & N=1-0,J=1/2-1/2,F=1/2-3/2 & 113.144 & 10 & 13 \\  
112.387 & CN  & N=1-0,J=1/2-1/2,F=3/2-1/2 & 113.170 & 10 & 12 \\  
112.408 & CN  & N=1-0,J=1/2-1/2,F=3/2-3/2 & 113.191 & 13 & 16 \\  
112.625 & CCS & (9  8 )-(8  7 ) & 113.410 & 3 & 4 \\  
112.702 & CN  & N=1-0,J=3/2-1/2,F=3/2-1/2 & 113.488 & 13 & 16 \\  
112.705 & CN  & N=1-0,J=3/2-1/2,F=5/2-3/2 & 113.491 & 31 & 38 \\  
112.714 & CN  & N=1-0,J=3/2-1/2,F=1/2-1/2 & 113.500 & 10 & 13 \\  
112.723 & CN  & N=1-0,J=3/2-1/2,F=3/2-3/2 & 113.509 & 10 & 12 \\  
214.346 & {\it $^{34}$SO} & 6(5)-5(4) & 215.840 & 11 & 13 \\  
214.373 & c-HCCCH  & 12(6,6)-12(5,7) & 215.867 & 3 & 4 \\  
214.479 & c-HCCCH  & 12(7,6)-12(6,7) & 215.974 & 9 & 10 \\  
214.781 & c-HCCCH  & 3(3,0)-2(2,1) & 216.279 & 5 & 6 \\  
215.210 & H$_2$S & 2(2,0)-2(1,1) & 216.710 & 37 & 45 \\  
215.308 & c-HCCCH  & 11(5,6)-11(4,7) & 216.809 & 8 & 10 \\  
215.344 & c-HCCCH  & 11(6,6)-11(5,7) & 216.846 & 3 & 3 \\  
215.602 & SiO  & 5-4 & 217.105 & 39 & 48 \\  
215.914 & HCC$^{13}$CN  & J=24-23 & 217.420 & 9 & 11 \\  
216.015 & c-HCCCH  & 10(4,6)-10(3,7) & 217.521 & 3 & 3 \\  
216.026 & c-HCCCH  & 10(5,6)-10(4,7) & 217.532 & 8 & 9 \\  
216.314 & c-HCCCH  & 6(1,6)-5(0,5) & 217.822 & 17 & 20 \\  
216.314 & c-HCCCH  & 6(0,6)-5(1,5) & 217.822 & 6 & 7 \\  
216.431 & c-HCCCH  & 5(1,4)-4(2,3) & 217.940 & 11 & 13 \\  
216.545 & c-HCCCH  & 9(3,6)-9(2,7) & 218.055 & 7 & 8 \\  
216.650 & c-HCCCH  & 5(2,4)-4(1,3) & 218.160 & 4 & 4 \\  
216.711 & p-H$_2$CO & 3;0;3-2;0;2 & 218.222 & 52 & 63 \\  
216.813 & HC$_3$N  & J=24-23 & 218.325 & 127 & 155 \\  
216.937 & c-HCCCH  & 8(3,6)-8(2,7) & 218.449 & 5 & 6 \\  
216.949 & NH$_2$CN & 11(1,11)-10(1,10),  & 218.462 & 17 & 21 \\  
216.963 & p-H$_2$CO & 3;2;2-2;2;1 & 218.476 & 26 & 32 \\  
217.169 & HC$_3$N v6=1 & J=24-23,l=1e & 218.683 & 23 & 28 \\  
217.218 & c-HCCCH  & 7(1,6)-7(0,7) & 218.733 & 3 & 4 \\  
217.246 & p-H$_2$CO & 3;2;1-2;2;0 & 218.760 & 24 & 29 \\  
217.339 & HC$_3$N v6=1 & J=24-23,l=1f & 218.854 & 23 & 28 \\  
217.346 & HC$_3$N v7=1 & J=24-23,l=1e & 218.861 & 71 & 87 \\  
217.626 & CCS & (1716)-(1615) & 219.143 & 4 & 5 \\  
217.656 & HC$_3$N v7=1 & J=24-23,l=1f & 219.174 & 71 & 87 \\  
229.443 & H2$^{13}$CS & 7(0,7)-6(0,6) & 231.043 & 5 & 6 \\  
229.620 & $^{13}$CS  & 5-4 & 231.221 & 26 & 32 \\  
230.594 & CCS & (1817)-(1716) & 232.202 & 3 & 4 \\  
231.545 & CCS & (1818)-(1717) & 233.159 & 3 & 4 \\  
232.107 & c-HCCCH  & 23(20,3)-23(19,4) & 233.725 & 2 & 3 \\  
232.295 & c-HCCCH  & 24(18,7)-24(17,8) & 233.914 & 3 & 4 \\  
232.318 & CCS & (1819)-(1718) & 233.938 & 4 & 5 \\  
247.268 & c-HCCCH  & 14(7,7)-14(6,8) & 248.992 & 4 & 5 \\  
247.303 & c-HCCCH  & 14(8,7)-14(7,8) & 249.027 & 12 & 15 \\  
\hline 
\end{tabular} 

\caption{\label{tab:lines2} Continues from Table \ref{tab:lines1}}
\end{table*}

\begin{table*}
\centering
\begin{tabular}{llllll} 
\hline 
Sky freq. [GHz] & Molecule & Transition & Rest freq. [GHz] & Peak [mJy] & SNR \\ 
\hline 
247.330 & c-HCCCH  & 5(2,3)-4(3,2) & 249.054 & 11 & 13 \\  
248.211 & c-HCCCH  & 13(6,7)-13(5,8) & 249.942 & 12 & 15 \\  
248.223 & c-HCCCH  & 13(7,7)-13(6,8) & 249.954 & 4 & 5 \\  
248.385 & c-HCCCH  & 20(19,2)-20(18,3) & 250.117 & 3 & 3 \\  
248.430 & CH$_2$NH & 7(1,6)-7(0,7) & 250.162 & 36 & 44 \\  
248.773 & {\it CH$_3$OH}  & 11;0;0-10;1;0 & 250.507 & 3 & 4 \\  
248.965 & c-HCCCH  & 12(5,7)-12(4,8) & 250.701 & 4 & 5 \\  
248.969 & c-HCCCH  & 12(6,7)-12(5,8) & 250.704 & 12 & 15 \\  
249.562 & c-HCCCH  & 11(4,7)-11(3,8) & 251.302 & 11 & 14 \\  
249.563 & c-HCCCH  & 11(5,7)-11(4,8) & 251.303 & 4 & 5 \\  
249.574 & c-HCCCH  & 7(0,7)-6(1,6) & 251.314 & 29 & 35 \\  
249.574 & c-HCCCH  & 7(1,7)-6(0,6) & 251.314 & 10 & 12 \\  
249.681 & CH$_2$NH & 6(0,6)-5(1,5) & 251.421 & 40 & 49 \\  
250.030 & c-HCCCH  & 10(3,7)-10(2,8) & 251.773 & 3 & 4 \\  
250.030 & c-HCCCH  & 10(4,7)-10(3,8) & 251.773 & 10 & 12 \\  
250.069 & {\it CH$_3$OH}  & 5;-3;0-5;2;0 & 251.812 & 3 & 4 \\  
250.082 & SO  & 6;5-5;4 & 251.825 & 55 & 67 \\  
250.123 & {\it CH$_3$OH}  & 4;-3;0-4;2;0 & 251.867 & 3 & 3 \\  
250.147 & {\it CH$_3$OH}  & 5;3;0-5;-2;0 & 251.891 & 3 & 4 \\  
250.152 & {\it CH$_3$OH}  & 6;3;0-6;-2;0 & 251.896 & 3 & 3 \\  
250.157 & {\it CH$_3$OH}  & 4;3;0-4;-2;0 & 251.900 & 3 & 3 \\  
250.390 & c-HCCCH  & 9(2,7)-9(1,8) & 252.136 & 8 & 9 \\  
250.390 & c-HCCCH  & 9(3,7)-9(2,8) & 252.136 & 3 & 3 \\  
250.662 & c-HCCCH  & 8(2,7)-8(1,8) & 252.410 & 4 & 5 \\  
251.344 & c-HCCCH  & 22(20,3)-22(19,4) & 253.096 & 3 & 4 \\  
251.454 & {\it $^{34}$SO} & 6(6)-5(5) & 253.207 & 14 & 17 \\  
251.815 & NS  & J=11/2-9/2,$\Omega$=1/2,F=13/2-11/2,l=e & 253.570 & 15 & 18 \\  
251.815 & NS  & J=11/2-9/2,$\Omega$=1/2,F=11/2-9/2,l=e & 253.570 & 12 & 15 \\  
251.817 & NS  & J=11/2-9/2,$\Omega$=1/2,F=9/2-7/2,l=e & 253.572 & 10 & 12 \\  
251.887 & HCC$^{13}$CN  & J=28-27 & 253.644 & 8 & 10 \\  
252.210 & NS  & J=11/2-9/2,$\Omega$=1/2,F=13/2-11/2,l=f & 253.968 & 15 & 18 \\  
252.212 & NS  & J=11/2-9/2,$\Omega$=1/2,F=11/2-9/2,l=f & 253.971 & 12 & 15 \\  
252.212 & NS  & J=11/2-9/2,$\Omega$=1/2,F=9/2-7/2,l=f & 253.971 & 10 & 12 \\  
252.457 & $^{30}$SiO  & 6-5 & 254.217 & 34 & 41 \\  
252.774 & c-HCCCH  & 24(21,4)-24(20,5) & 254.536 & 2 & 3 \\  
252.800 & c-HCCCH  & 25(17,8)-25(16,9) & 254.563 & 3 & 4 \\  
252.922 & CH$_2$NH & 4(0,4)-3(0,3) & 254.685 & 89 & 109 \\  
252.936 & HC$_3$N  & J=28-27 & 254.700 & 118 & 144 \\  
253.222 & c-HCCCH  & 5(3,3)-4(2,2) & 254.988 & 4 & 5 \\  
253.351 & HC$_3$N v6=1 & J=28-27,l=1e & 255.117 & 26 & 32 \\  
253.550 & HC$_3$N v6=1 & J=28-27,l=1f & 255.317 & 26 & 32 \\  
253.557 & HC$_3$N v7=1 & J=28-27,l=1e & 255.325 & 79 & 97 \\  
253.711 & H{\it C$^{18}$O}$^+$ & 3-2 & 255.479 & 22 & 27 \\  
253.827 & NS  & J=11/2-9/2,$\Omega$=3/2,F=13/2-11/2,l=f & 255.597 & 6 & 7 \\  
253.827 & NS  & J=11/2-9/2,$\Omega$=3/2,F=13/2-11/2,l=e & 255.597 & 6 & 7 \\  
253.831 & NS  & J=11/2-9/2,$\Omega$=3/2,F=11/2-9/2,l=f & 255.600 & 5 & 6 \\  
253.831 & NS  & J=11/2-9/2,$\Omega$=3/2,F=11/2-9/2,l=e & 255.600 & 5 & 6 \\  
253.833 & NS  & J=11/2-9/2,$\Omega$=3/2,F=9/2-7/2,l=f & 255.603 & 4 & 5 \\  
253.833 & NS  & J=11/2-9/2,$\Omega$=3/2,F=9/2-7/2,l=e & 255.603 & 4 & 5 \\  
253.870 & H$^{13}$CCCN  & J=29-28 & 255.640 & 15 & 19 \\  
253.919 & HC$_3$N v7=1 & J=28-27,l=1f & 255.689 & 79 & 97 \\  
254.069 & CH$_2$NH & 4(2,3)-3(2,2) & 255.840 & 37 & 46 \\  
254.175 & HC$_3$N v6=1, v7=1 & (28 0 0)-(7 0 0) & 255.947 & 10 & 12 \\  
254.216 & HC$_3$N v6=1, v7=1 & (28 0 1)-(7 0 1) & 255.988 & 9 & 12 \\  
254.255 & HCS$^+$ & 6-5 & 256.027 & 35 & 43 \\  
254.299 & HC$_3$N v6=1, v7=1 & (28-2 2)-(7 2 2) & 256.072 & 9 & 11 \\  
\hline 
\end{tabular} 

\caption{\label{tab:lines3} Continues from Table \ref{tab:lines1}}
\end{table*}

\begin{table*}
\centering
\begin{tabular}{llllll} 
\hline 
Sky freq. [GHz] & Molecule & Transition & Rest freq. [GHz] & Peak [mJy] & SNR \\ 
\hline 
254.338 & HC$_3$N v6=1, v7=1 & (28 2 2)-(7-2 2) & 256.111 & 9 & 12 \\  
254.387 & CH$_3$CCH  & 15(6)-14(6) & 256.161 & 5 & 6 \\  
254.392 & CH$_2$NH & 4(3,2)-3(3,1) & 256.165 & 9 & 11 \\  
254.403 & CH$_2$NH & 4(3,1)-3(3,0) & 256.177 & 9 & 11 \\  
254.441 & CH$_3$CCH  & 15(5)-14(5) & 256.214 & 3 & 4 \\  
254.484 & CH$_3$CCH  & 15(4)-14(4) & 256.258 & 4 & 5 \\  
254.485 & HC$_3$N v7=2 & J=28-27,l=0 & 256.260 & 32 & 39 \\  
254.518 & CH$_3$CCH  & 15(3)-14(3) & 256.293 & 10 & 12 \\  
254.537 & HC$_3$N v7=2 & J=28-27,l=2e & 256.311 & 31 & 37 \\  
254.543 & CH$_3$CCH  & 15(2)-14(2) & 256.317 & 6 & 7 \\  
254.557 & CH$_3$CCH  & 15(1)-14(1) & 256.332 & 6 & 7 \\  
254.562 & CH$_3$CCH  & 15(0)-14(0) & 256.337 & 6 & 8 \\  
254.591 & HC$_3$N v7=2 & J=28-27,l=2f & 256.366 & 31 & 37 \\  
255.034 & CH$_3$CN  & 14;12-13;12 & 256.812 & 4 & 4 \\  
255.099 & {\it $^{34}$SO} & 7(6)-6(5) & 256.878 & 19 & 23 \\  
255.148 & CH$_3$CN  & 14;11-13;11 & 256.926 & 4 & 5 \\  
255.251 & CH$_3$CN  & 14;10-13;10 & 257.031 & 4 & 5 \\  
255.333 & CH$_2$NH & 4(2,2)-3(2,1) & 257.113 & 38 & 46 \\  
255.345 & CH$_3$CN  & 14;9-13;9 & 257.125 & 12 & 15 \\  
255.429 & CH$_3$CN  & 14;8-13;8 & 257.209 & 6 & 7 \\  
255.474 & $^{29}$SiO  & 6-5 & 257.255 & 26 & 32 \\  
255.503 & CH$_3$CN  & 14;7-13;7 & 257.284 & 7 & 8 \\  
255.567 & CH$_3$CN  & 14;6-13;6 & 257.349 & 17 & 21 \\  
255.621 & CH$_3$CN  & 14;5-13;5 & 257.403 & 8 & 10 \\  
255.666 & CH$_3$CN  & 14;4-13;4 & 257.448 & 9 & 11 \\  
255.700 & CH$_3$CN  & 14;3-13;3 & 257.483 & 22 & 27 \\  
255.725 & CH$_3$CN  & 14;2-13;2 & 257.507 & 10 & 12 \\  
255.740 & CH$_3$CN  & 14;1-13;1 & 257.522 & 10 & 13 \\  
255.744 & CH$_3$CN  & 14;0-13;0 & 257.527 & 12 & 15 \\  
256.381 & NH$_2$CN & 13(1,13)-12(1,12),  & 258.169 & 15 & 18 \\  
256.467 & SO  & 6;6-5;5 & 258.255 & 63 & 77 \\  
261.966 & HC$_3$N  & J=29-28 & 263.792 & 112 & 137 \\  
262.158 & CH$_2$NH & 3(2,1)-4(1,4) & 263.986 & 8 & 10 \\  
262.162 & H2$^{13}$CS & 8(0,8)-7(0,7) & 263.989 & 5 & 7 \\  
262.395 & HC$_3$N v6=1 & J=29-28,l=1e & 264.224 & 27 & 33 \\  
262.601 & HC$_3$N v6=1 & J=29-28,l=1f & 264.432 & 27 & 33 \\  
262.609 & HC$_3$N v7=1 & J=29-28,l=1e & 264.440 & 79 & 97 \\  
262.621 & H$^{13}$CCCN  & J=30-29 & 264.451 & 15 & 18 \\  
262.636 & c-HCCCH  & 21(20,1)-21(19,2) & 264.467 & 3 & 3 \\  
262.984 & HC$_3$N v7=1 & J=29-28,l=1f & 264.817 & 80 & 97 \\  
263.244 & HC$_3$N v6=1, v7=1 & (29 0 0)-(8 0 0) & 265.079 & 10 & 12 \\  
263.289 & HC$_3$N v6=1, v7=1 & (29 0 1)-(8 0 1) & 265.125 & 10 & 12 \\  
263.380 & HC$_3$N v6=1, v7=1 & (29 2 2)-(8-2 2) & 265.216 & 9 & 12 \\  
263.424 & HC$_3$N v6=1, v7=1 & (29-2 2)-(8 2 2) & 265.260 & 9 & 12 \\  
263.566 & HC$_3$N v7=2 & J=29-28,l=0 & 265.404 & 32 & 39 \\  
263.920 & c-HCCCH  & 4(4,1)-3(3,0) & 265.759 & 16 & 19 \\  
264.012 & HCN v2=1 & J=3-2,l=1e & 265.853 & 47 & 58 \\  
264.046 & HCN  & 3-2 & 265.886 & 379 & 465 \\  
264.347 & c-HCCCH  & 23(15,8)-23(14,9) & 266.190 & 5 & 6 \\  
264.427 & CH$_2$NH & 4(1,3)-3(1,2) & 266.270 & 80 & 98 \\  
265.020 & c-HCCCH  & 24(17,8)-24(16,9) & 266.867 & 4 & 5 \\  
265.195 & c-HCCCH  & 23(21,2)-23(20,3) & 267.044 & 3 & 3 \\  
265.349 & HCN v2=1 & J=3-2,l=1f & 267.199 & 48 & 59 \\  
265.705 & HCO$^+$  & 3-2 & 267.558 & 213 & 261 \\  
266.157 & H2$^{13}$CS & 8(1,7)-7(1,6) & 268.012 & 11 & 14 \\  
268.648 & H$_2$CS & 8(1,8)-7(1,7) & 270.521 & 22 & 27 \\  
\hline 
\end{tabular} 

\caption{\label{tab:lines4} Continues from Table \ref{tab:lines1}}
\end{table*}

\begin{table*}
\centering
\begin{tabular}{llllll} 
\hline 
Sky freq. [GHz] & Molecule & Transition & Rest freq. [GHz] & Peak [mJy] & SNR \\ 
\hline 
269.872 & HCC$^{13}$CN  & J=30-29 & 271.753 & 8 & 9 \\  
269.914 & c-HCCCH  & 22(15,8)-22(14,9) & 271.795 & 7 & 8 \\  
270.042 & HNC v2=1 & J=3-2,l=1e & 271.924 & 86 & 53 \\  
270.098 & HNC  & 3-2 & 271.981 & 328 & 201 \\  
270.863 & C$^{34}$S  & 6-5 & 272.752 & 46 & 28 \\  
270.996 & HC$_3$N  & J=30-29 & 272.885 & 104 & 64 \\  
271.341 & CH$_3$CCH  & 16(6)-15(6) & 273.232 & 6 & 4 \\  
271.344 & c-HCCCH  & 21(13,8)-21(12,9) & 273.236 & 8 & 5 \\  
271.371 & H$^{13}$CCCN  & J=31-30 & 273.263 & 14 & 9 \\  
271.439 & HC$_3$N v6=1 & J=30-29,l=1e & 273.332 & 27 & 16 \\  
271.444 & CH$_3$CCH  & 16(4)-15(4) & 273.337 & 5 & 3 \\  
271.480 & CH$_3$CCH  & 16(3)-15(3) & 273.373 & 12 & 8 \\  
271.506 & CH$_3$CCH  & 16(2)-15(2) & 273.399 & 7 & 4 \\  
271.522 & CH$_3$CCH  & 16(1)-15(1) & 273.415 & 8 & 5 \\  
271.527 & CH$_3$CCH  & 16(0)-15(0) & 273.420 & 8 & 5 \\  
271.653 & HC$_3$N v6=1 & J=30-29,l=1f & 273.546 & 27 & 16 \\  
271.660 & HC$_3$N v7=1 & J=30-29,l=1e & 273.554 & 79 & 48 \\  
275.535 & $^{13}$CS  & 6-5 & 277.455 & 25 & 15 \\  
275.853 & c-HCCCH  & 19(11,8)-19(10,9) & 277.776 & 11 & 7 \\  
276.095 & NH$_2$CN & 14(1,14)-13(1,13),  & 278.020 & 12 & 8 \\  
276.714 & CH$_2$NH & 8(1,7)-8(0,8) & 278.643 & 29 & 18 \\  
276.956 & H$_2$CS & 8(1,7)-7(1,6) & 278.886 & 23 & 14 \\  
277.577 & N$_2$H$^+$  & 3-2 & 279.512 & 144 & 88 \\  
277.604 & c-HCCCH  & 18(11,8)-18(10,9) & 279.539 & 12 & 8 \\  
277.882 & NH$_2$CN & 14(0,14)-13(0,13),  & 279.819 & 7 & 4 \\  
278.864 & HCC$^{13}$CN  & J=31-30 & 280.808 & 7 & 4 \\  
278.922 & c-HCCCH  & 17(9,8)-17(8,9) & 280.867 & 14 & 8 \\  
279.578 & o-H$_2$CO & 4;1;4-3;1;3 & 281.527 & 100 & 61 \\  
279.758 & NH$_2$CN & 14(1,13)-13(1,12),  & 281.708 & 12 & 8 \\  
280.025 & HC$_3$N  & J=31-30 & 281.977 & 97 & 59 \\  
280.085 & c-HCCCH  & 16(8,8)-16(7,9) & 282.037 & 5 & 3 \\  
280.096 & c-HCCCH  & 16(9,8)-16(8,9) & 282.049 & 15 & 9 \\  
280.121 & H$^{13}$CCCN  & J=32-31 & 282.073 & 13 & 8 \\  
280.426 & c-HCCCH  & 4(4,0)-3(3,1) & 282.381 & 5 & 3 \\  
280.483 & HC$_3$N v6=1 & J=31-30,l=1e & 282.439 & 27 & 16 \\  
280.703 & HC$_3$N v6=1 & J=31-30,l=1f & 282.660 & 27 & 16 \\  
280.711 & HC$_3$N v7=1 & J=31-30,l=1e & 282.668 & 77 & 47 \\  
281.053 & c-HCCCH  & 15(7,8)-15(6,9) & 283.012 & 16 & 10 \\  
281.056 & c-HCCCH  & 15(8,8)-15(7,9) & 283.016 & 5 & 3 \\  
281.112 & HC$_3$N v7=1 & J=31-30,l=1f & 283.072 & 78 & 47 \\  
281.378 & HC$_3$N v6=1, v7=1 & (31 0 0)-(0 0 0) & 283.340 & 10 & 6 \\  
281.434 & HC$_3$N v6=1, v7=1 & (31 0 1)-(0 0 1) & 283.396 & 9 & 6 \\  
281.539 & HC$_3$N v6=1, v7=1 & (31 2 2)-(0-2 2) & 283.502 & 9 & 6 \\  
281.594 & HC$_3$N v6=1, v7=1 & (31-2 2)-(0 2 2) & 283.557 & 9 & 6 \\  
287.207 & C$^{34}$S  & 6-5 & 289.209 & 56 & 34 \\  
287.855 & HCC$^{13}$CN  & J=32-31 & 289.862 & 7 & 4 \\  
288.102 & {\it CH$_3$OH}  & 6;0;0-5;0;0 & 290.111 & 13 & 8 \\  
288.293 & CH$_3$CCH  & 17(6)-16(6) & 290.303 & 8 & 5 \\  
288.353 & CH$_3$CCH  & 17(5)-16(5) & 290.364 & 5 & 3 \\  
288.403 & CH$_3$CCH  & 17(4)-16(4) & 290.413 & 6 & 4 \\  
288.441 & CH$_3$CCH  & 17(3)-16(3) & 290.452 & 15 & 9 \\  
288.469 & CH$_3$CCH  & 17(2)-16(2) & 290.480 & 9 & 5 \\  
288.485 & CH$_3$CCH  & 17(1)-16(1) & 290.497 & 9 & 6 \\  
288.491 & CH$_3$CCH  & 17(0)-16(0) & 290.502 & 9 & 6 \\  
288.551 & {\it $^{34}$SO} & 6(7)-5(6) & 290.562 & 19 & 12 \\  
288.611 & p-H$_2$CO & 4;0;4-3;0;3 & 290.623 & 63 & 39 \\  
\hline 
\end{tabular} 

\caption{\label{tab:lines5} Continues from Table \ref{tab:lines1}}
\end{table*}

\begin{table*}
\centering
\begin{tabular}{llllll} 
\hline 
Sky freq. [GHz] & Molecule & Transition & Rest freq. [GHz] & Peak [mJy] & SNR \\ 
\hline 
288.870 & H$^{13}$CCCN  & J=33-32 & 290.884 & 12 & 7 \\  
289.053 & HC$_3$N  & J=32-31 & 291.068 & 88 & 54 \\  
289.221 & p-H$_2$CO & 4;2;3-3;2;2 & 291.238 & 40 & 24 \\  
289.363 & o-H$_2$CO & 4;3;2-3;3;1 & 291.380 & 34 & 21 \\  
289.367 & o-H$_2$CO & 4;3;1-3;3;0 & 291.384 & 33 & 20 \\  
289.468 & {\it C$^{33}$S}  & 6-5 & 291.486 & 20 & 12 \\  
289.527 & HC$_3$N v6=1 & J=32-31,l=1e & 291.545 & 26 & 16 \\  
289.754 & HC$_3$N v6=1 & J=32-31,l=1f & 291.774 & 26 & 16 \\  
289.762 & HC$_3$N v7=1 & J=32-31,l=1e & 291.782 & 75 & 46 \\  
289.927 & p-H$_2$CO & 4;2;2-3;2;1 & 291.948 & 36 & 22 \\  
290.176 & HC$_3$N v7=1 & J=32-31,l=1f & 292.199 & 76 & 46 \\  
290.444 & HC$_3$N v6=1, v7=1 & (32 0 0)-(1 0 0) & 292.469 & 9 & 6 \\  
290.506 & HC$_3$N v6=1, v7=1 & (32 0 1)-(1 0 1) & 292.531 & 9 & 6 \\  
290.619 & HC$_3$N v6=1, v7=1 & (32-2 2)-(1 2 2) & 292.645 & 9 & 6 \\  
290.647 & {\it CH$_3$OH}  & 6;-1;0-5;-1;0 & 292.673 & 8 & 5 \\  
290.679 & HC$_3$N v6=1, v7=1 & (32 2 2)-(1-2 2) & 292.705 & 9 & 6 \\  
290.759 & H2$^{13}$CS & 9(1,9)-8(1,8) & 292.786 & 11 & 7 \\  
290.804 & HC$_3$N v7=2 & J=32-31,l=0 & 292.832 & 31 & 19 \\  
290.881 & HC$_3$N v7=2 & J=32-31,l=2e & 292.909 & 30 & 19 \\  
290.961 & HC$_3$N v7=2 & J=32-31,l=2f & 292.990 & 30 & 19 \\  
291.808 & CH$_3$CN  & 16;9-15;9 & 293.843 & 8 & 5 \\  
291.877 & CS  & 6-5 & 293.912 & 251 & 154 \\  
292.062 & CH$_3$CN  & 16;6-15;6 & 294.098 & 12 & 7 \\  
292.124 & CH$_3$CN  & 16;5-15;5 & 294.161 & 6 & 3 \\  
292.175 & CH$_3$CN  & 16;4-15;4 & 294.212 & 6 & 4 \\  
292.214 & CH$_3$CN  & 16;3-15;3 & 294.251 & 15 & 9 \\  
292.242 & CH$_3$CN  & 16;2-15;2 & 294.280 & 7 & 4 \\  
292.259 & CH$_3$CN  & 16;1-15;1 & 294.297 & 7 & 4 \\  
292.265 & CH$_3$CN  & 16;0-15;0 & 294.302 & 8 & 5 \\  
\hline 
\end{tabular} 

\caption{\label{tab:lines6} Continues from Table \ref{tab:lines1}}
\end{table*}


\begin{table*}
\renewcommand{\tabcolsep}{2pt}
\begin{tabular}{lcccccccccc} 
\hline 
&&&&&&&&&& \\
 & NGC4418 & Arp220 & M82 & NGC253 & NGC1068 & SgrB2(N) & SgrB2(OH) & Orion Bar & TMC-1 & PKS1830 \\ 
SiO &  -9.2...-8.2  &  -9.4...-8.9  & -  &  -10.2...-10.0  &  -9.5...-9.3  &  -10.9...-10.3  & -  &  -10.4...-10.2  & $<$-11.6  &  -9.5...-9.3  \\ 
H$_2$S &  -7.8...-6.7  &  -7.0...-6.7  &  -9.6...-8.5  &  -9.5...-9.1  & -  &  -10.2...-9.5  & -  &  -8.3...-8.1  & $<$-9.3  & -  \\ 
c-HCCCH &  -7.8...-6.7  &  -8.6...-8.2  &  -9.3...-8.8  &  -9.4...-9.1  &  -10.1...-9.8  &  -11.0...-9.9  &  -10.4...-9.5  &  -9.8...-9.6  &  -9.4...-9.2  &  -8.7...-8.5  \\ 
HC$_3$N &  -8.5...-7.2  &  -8.0...-7.7  &  -9.7...-8.9  &  -9.7...-9.3  &  -8.9...-8.1  &  -7.8...-7.0  &  -9.5...-8.9  & -  &  -8.3...-8.1  &  -9.4...-9.2  \\ 
NH$_2$CN &  -9.0...-8.0  & -  &  -9.9...-9.0  &  -10.0...-9.5  & $<$-9.7  &  -11.1...-9.6  &  -10.4...-9.4  & -  & -  & -  \\ 
CH$_3$CN &  -8.4...-7.6  &  -8.5...-8.1  &  -10.2...-9.6  &  -9.7...-9.3  &  -9.6...-9.5  &  -7.0...-6.2  &  -9.9...-9.3  & $<$-10.3  &  -9.1...-8.9  &  -9.3...-9.1  \\ 
CH$_3$OH &  -8.5...-7.4  &  -7.7...-7.2  &  -9.6...-8.2  &  -7.9...-7.7  &  -7.9...-7.7  &  -6.1...-5.4  &  -8.3...-7.2  & $<$-10.1  &  -8.8...-8.6  &  -8.2...-8.0  \\ 
H$_2$CO &  -8.3...-7.5  &  -8.9...-8.5  &  -8.9...-8.4  &  -9.0...-8.7  & $<$-7.1  &  -9.6...-8.9  &  -8.3...-7.6  &  -8.3...-8.1  &  -7.8...-7.6  &  -8.0...-7.8  \\ 
CH$_3$CCH &  -7.8...-6.7  &  -7.5...-7.1  &  -7.8...-7.4  &  -8.2...-8.0  & $<$-8.5  &  -8.6...-7.9  &  -9.1...-8.4  & $<$-8.5  &  -8.3...-8.1  &  -8.6...-8.4  \\ 
\hline 
\end{tabular} 

\caption{\label{tab:abtab} Base-ten logarithm of molecular abundances in Galactic and extragalactic sources. For NGC~4418, the range of values represents the confidence interval of the $\chi^2$ minimization in the best fit model. Data for Sgr~B2(N), Sgr~B2(OH), the Orion Bar, TMC-1, and NGC~253 were taken from \citet{martin06}. Data for Arp~220, NGC~1068, and PKS~1830-211, were taken from \citet{martin2011}, \citet{aladro_1068}, and \citet{muller_pks} respectively.}
\end{table*}

 \begin{figure*}[h]
\centering
\includegraphics[width=.3\textwidth,keepaspectratio]{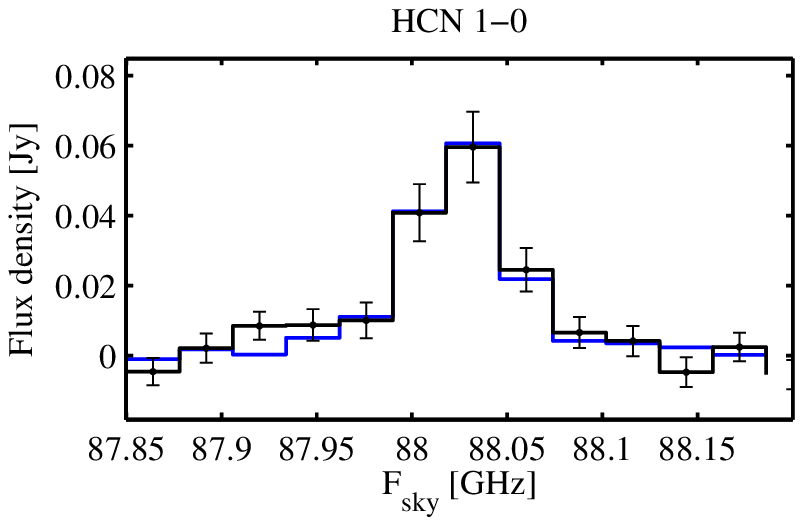}
\includegraphics[width=.3\textwidth,keepaspectratio]{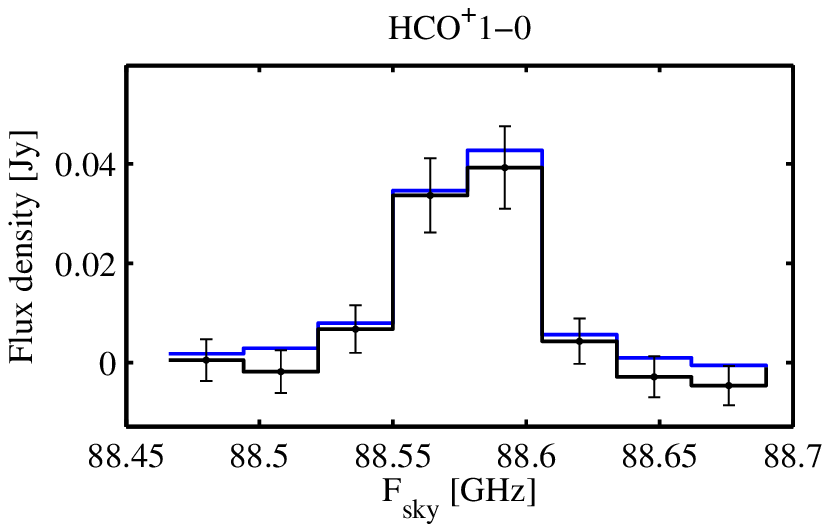}
\includegraphics[width=.3\textwidth,keepaspectratio]{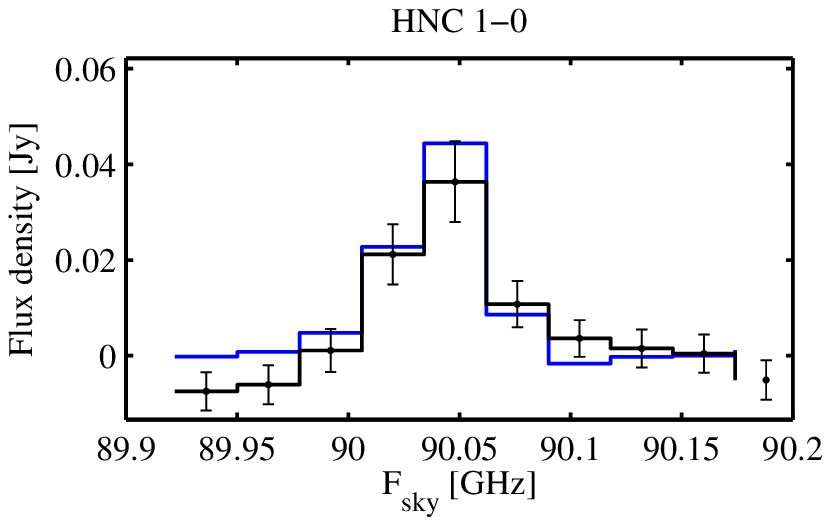}
\includegraphics[width=.3\textwidth,keepaspectratio]{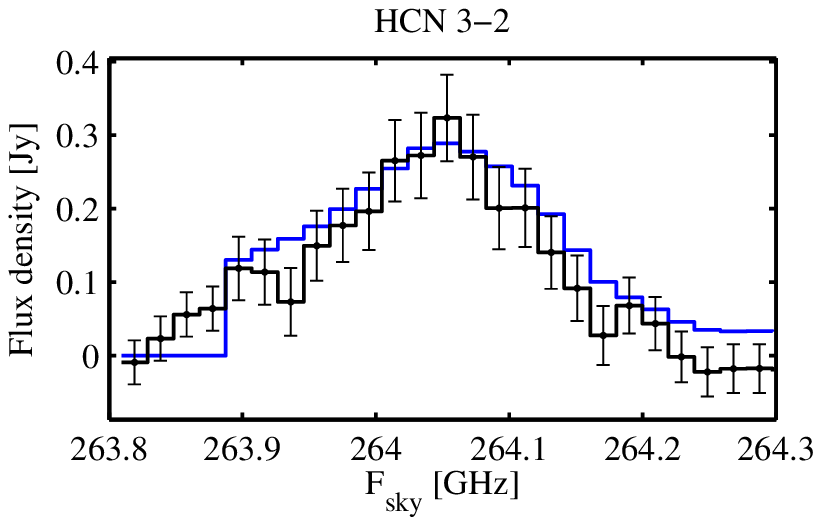}
\includegraphics[width=.3\textwidth,keepaspectratio]{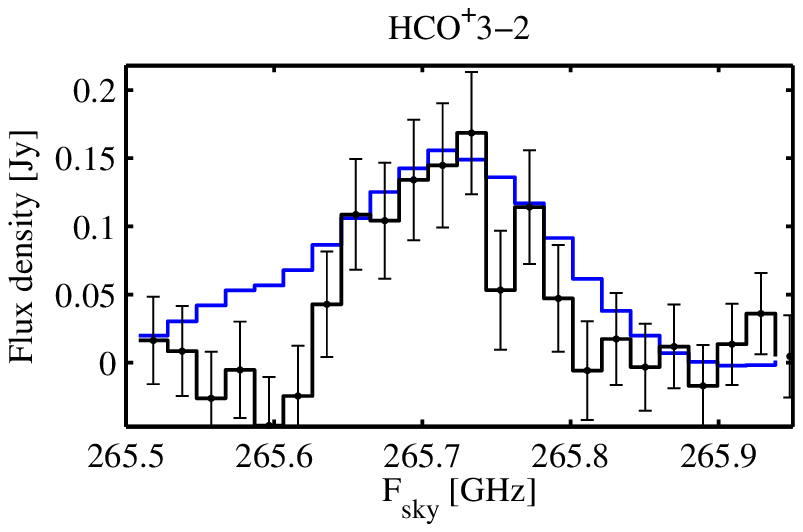}
\includegraphics[width=.3\textwidth,keepaspectratio]{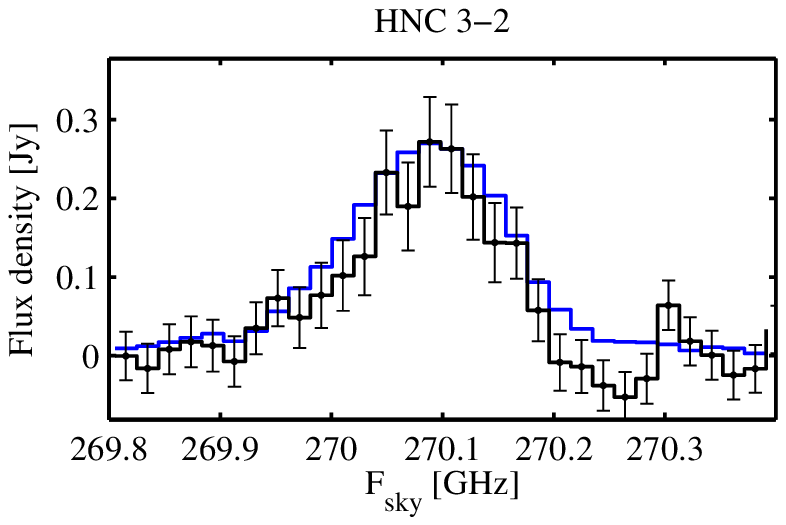}
   \caption{ \label{fig:sdcomp} Comparison of our ALMA data with single-dish spectra of the brightest lines in the scan. The blue line shows the spectrum extracted from the ALMA visibilities by fitting a point source with the CASA routine {\tt uvmultifit}. The black line shows single dish spectra observed by the authors with the IRAM~30m telescope. The error bars show the squared sum of the rms of the two datasets. The spectra were interpolated to a common resolution of 28 and 19.5 MHz at 3 and 1~mm, respectively.}
\end{figure*}

\begin{figure*}[h]
\centering
\includegraphics[width=.4\textwidth,keepaspectratio]{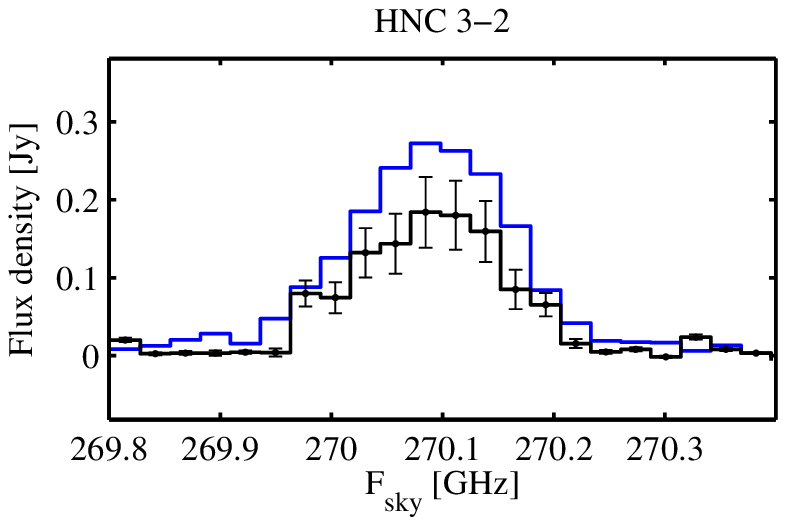}
\includegraphics[width=.4\textwidth,keepaspectratio]{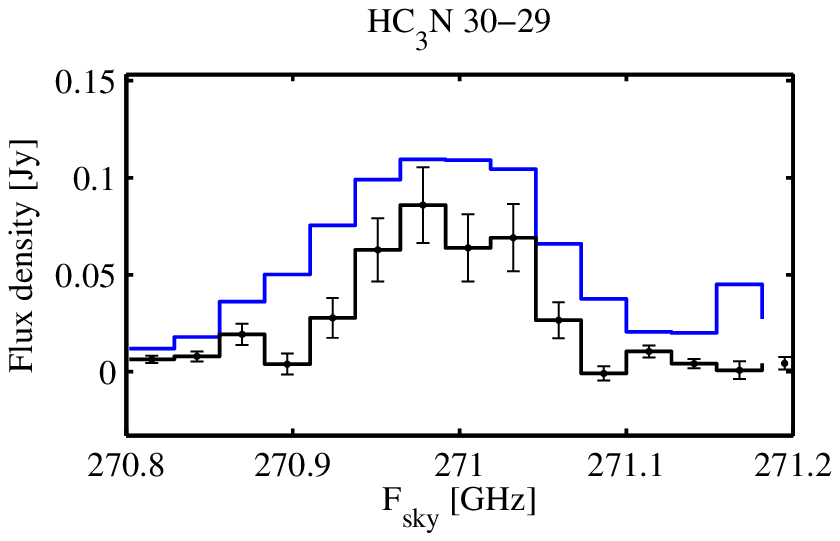}
   \caption{ \label{fig:smacomp} Comparison of our ALMA data with SMA observations by \citet{costagliola2013}. The blue line shows the spectrum extracted from the ALMA visibilities by fitting a point source with the CASA routine {\tt uvmultifit}. The black line shows the flux density measured by the extended configuration of the SMA, with a beam size of 0$''$.4 and a maximum recoverable scale of 2$''$. The error bars show the squared sum of the rms of the two datasets. The spectra were interpolated to the resolution of the ALMA spectral scan.}
\end{figure*}

\begin{table*}
\caption{\label{tab:compflux} Integrated flux densities from pre-ALMA observations}
\begin{center} 
\begin{tabular}{lccc}
\hline
\hline
&&&\\
Molecule & & Integrated flux densities [Jy km/s] & \\
& ALMA$^{\scriptscriptstyle (1)}$ & IRAM 30m$^{\scriptscriptstyle (2)}$ & SMA$^{\scriptscriptstyle (3)}$ \\

 & & & \\
HCN 1-0 & 11.2$\pm$1 & 13$\pm$2 &  -  \\
HCO$^+$ 1-0 & 7$\pm$1 & 8.35$\pm$2 & -  \\
HNC 1-0 & 7$\pm$1 & 6$\pm$2 & -  \\
HC3N 10-9 & 5$\pm$1 & 6$\pm$2 & -  \\
HCN 3-2 & 41$\pm$4 & 43.3$\pm$0.3 &  -  \\
HCO$^+$ 3-2 & 24$\pm$2 & 21$\pm$0.3 & -  \\
HNC 3-2 & 39$\pm$4 & 43$\pm$0.3 & 37$\pm$4 \\
HC3N 30-29 & 15$\pm$2 & - & 11$\pm$1  \\
\hline
\end{tabular}  
\end{center}{\tiny \it (1) This work ; (2) Single-dish fluxes observed with the IRAM~30m telescope ; (3) SMA observations by \citet{costagliola2013}, extended configuration, beam size of~ 0.4$''$.}

\end{table*}

\end{appendix}

\end{document}